\documentclass[twocolumn,floatfix,superscriptaddress,amsmath,showpacs,showkeys,aps,prb,longbibliography]{revtex4-2}
\usepackage[utf8]{inputenc}
\usepackage[final]{graphicx}
\usepackage{bm}
\usepackage[normalem]{ulem}
\usepackage{amsfonts}
\usepackage{amsmath}
\usepackage{amssymb}
\usepackage[usenames]{color}
\usepackage{times}
\usepackage{verbatim}
\usepackage{xcolor}
\usepackage{physics}
\usepackage[hidelinks]{hyperref}
\usepackage{subfigure}
\usepackage{placeins}

\begin{document}
\title{The de Haas--van Alphen effect in a two-dimensional nematic Fermi fluid}
\author{Luís E.  Leite Macêdo}
\affiliation{Departamento de F{\'\i}sica Te\'orica,
Universidade do Estado do Rio de Janeiro, Rua S\~ao Francisco Xavier 524, 20550-013  
Rio de Janeiro, Brazil}
\author{Daniel G. Barci}
\affiliation{Departamento de F{\'\i}sica Te\'orica,
Universidade do Estado do Rio de Janeiro, Rua S\~ao Francisco Xavier 524, 20550-013  
Rio de Janeiro, Brazil}

\date{\today}

\begin{abstract}
Quantum oscillations are widely used to probe the Fermi surface of metals, but their behavior becomes more complex when electronic interactions can spontaneously deform that surface. We investigate this interplay in a two-dimensional model of interacting electrons capable of developing nematic order, using a uniform mean-field approximation in a perpendicular magnetic field. We show that Landau quantization can favor or suppress nematic order as the field varies, producing alternating isotropic and nematic regions near a Pomeranchuk instability, with both continuous and first-order transitions. Conversely, the evolving nematic order modifies the de Haas--van Alphen oscillations in ways that depend on the thermodynamic conditions. At fixed chemical potential, the oscillation frequency can change significantly even when the cyclotron mass varies only modestly. At fixed particle density, conservation of the Fermi-surface area preserves the leading Onsager frequency, while the cyclotron mass and harmonic amplitudes remain sensitive to nematic order. The magnetization and magnetic susceptibility provide additional signatures of the nematic transitions. These results show how quantum oscillations can both reveal and influence electronic order, offering a framework for interpreting experiments in strongly correlated metals.
\end{abstract}

\maketitle
\section{Introduction}
The de Haas--van Alphen (dHvA) effect, discovered in bismuth by de Haas and
van Alphen, is one of the clearest manifestations of quantum mechanics on a
macroscopic scale in a metal~\cite{deHaas1930}.  As a magnetic field is
varied, Landau levels pass successively through the Fermi energy and the
magnetization oscillates periodically in inverse field.  Onsager's
identification of the oscillation frequency with an extremal Fermi-surface
area, $\nu_B=(\hbar/2\pi e)A_{\rm ext}$~\cite{Onsager1952}, transformed this
effect into a precision probe of electronic structure.  Lifshitz and
Kosevich subsequently showed how the field and temperature dependence of
the amplitude encode the cyclotron mass~\cite{Lifshitz1956}.  Together with
the Shubnikov--de Haas effect in magnetotransport~\cite{Shubnikov1930}, these
oscillations became a standard method for mapping Fermi surfaces and
characterizing quasiparticles~\cite{Shoenberg1984}.

For a conventional Landau Fermi liquid~\cite{Pines66}, the oscillatory
magnetization may be organized schematically as
\begin{equation}
M_{\rm osc}=\sum_{p\geq1}{\cal A}_p \; R_T(\chi_p)R_D(p)R_S(p)
\sin\!\left(2\pi p\frac{\nu_B}{B}+\phi_p\right),
\label{eq:I-LK-schematic}
\end{equation}
where $\chi_p$ is the dimensionless thermal argument of harmonic $p$,
${\cal A}_p$ contains geometric prefactors, and $R_T$, $R_D$, and $R_S$
describe thermal damping, scattering, and spin splitting, respectively.  The continuing
observation of quantum oscillations in strongly correlated materials has
greatly enlarged the role of this framework.  Heavy-fermion metals,
cuprates, iron-based superconductors, and ruthenates can display large mass
enhancements, reconstructed Fermi surfaces, competing orders, and quantum
critical behavior~\cite{Continentino1994,Stewart2001,Coleman2005}.  Yet oscillations survive
in remarkably complex regimes, including superconducting and magnetically
ordered phases~\cite{Vignolle2008,DoironLeyraud2007}.  They therefore provide
a rare window onto the evolution of coherent electronic states inside
strongly interacting matter.

The discovery of electronic liquid-crystal phases opened a new chapter in
this development~\cite{KivelsonFradkinEmery-1998,FrKi1999}.  In classical
liquid crystals, orientational or partial translational order lies between
the familiar liquid and crystalline states~\cite{deGPr1998}.  Their quantum
counterparts are collective states of correlated electrons: the electronic
fluid can spontaneously choose a direction, form stripes, or develop more
subtle spatial organization without simply becoming an ordinary crystal.
Such behavior was first discussed prominently in quantum Hall systems and
has since become part of the broader landscape of correlated-electron
physics~\cite{BaFrKiOg2002,Fradkin2010,Fernandes2019}.

The nematic state is perhaps the simplest electronic liquid crystal.  It preserves
translations but breaks rotational symmetry, deforming the electronic fluid
and its Fermi surface.  Nematic behavior has been observed in ruthenates,
iron-based superconductors, and cuprates~\cite{Borzi2007,Kasahara2012,Hinkov2008},
and can coexist or compete with superconductivity and other ordered
states~\cite{Barci-Fradkin-2011, Fernandes2020,MeeseFernandes2026, Meese-PRB-2026, Nei-Barci-Guilherme-2026}.  Its microscopic origin varies among
materials: spin, orbital, lattice, and direct electronic interactions may all
participate.  A simple mechanism for the isotropic/nematic transition is a
quadrupolar Pomeranchuk instability of a Fermi liquid~\cite{Pomeranchuk1959,OgKiFr2001}.

In the presence of a magnetic field, Landau quantization and electronic nematicity become intertwined because the nematic state reshapes the very Fermi surface that controls the quantum-oscillation spectrum. A nematic deformation changes the shape of a Fermi surface but need not change its enclosed area. Consequently, the oscillation phase and local oscillation frequency, the mass inferred from thermal damping, and the distribution of harmonic weight need not evolve in the same way. The order parameter is itself determined self-consistently: even at the static mean-field level, its equilibrium value changes with magnetic field and temperature. The measured waveform can therefore carry information that is absent from a description based on a rigid anisotropic band.

Here we study this problem in a minimal two-dimensional Fermi-surface model
of spinless fermions with an attractive quadrupolar interaction in a perpendicular
magnetic field.  Related models have been investigated using Fermi-liquid
theory~\cite{Barci-Reyes-2013}  and higher-dimensional bosonization~\cite{BarciFradkinRibeiro2018}.  
We derive the mean-field thermodynamic potentials and compare two
experimentally relevant limiting conditions: fixed chemical potential
(grand-canonical ensemble) and fixed particle density (canonical ensemble).  This
comparison is central because the chemical potential oscillates when the
density is constrained, reorganizing the harmonic content of the quantum oscillations even before
nematic order appears.

The first question is how Landau quantization reshapes the nematic
transition. A companion Letter isolates the underlying mechanism: near a
Pomeranchuk instability, the quantizing field can generate reentrant nematic
windows and induce order where the zero-field state is isotropic
~\cite{MacedoBarciLetter}. The present article embeds that mechanism in a full
thermodynamic analysis, resolves local and global stability, and compares the
two ensembles. At fixed chemical potential, the oscillatory modulation of the
quadratic, and especially the quartic, Landau coefficients produces reentrant order and
can divide the equilibrium boundary into continuous and first-order segments.
At fixed density, phase locking suppresses the same feedback, and the
boundaries sampled here remain continuous.

The second question is how the equilibrium nematic deformation reshapes the
dHvA signal. A uniform deformation rescales the Landau-level spacing without
changing the degeneracy per unit  area. At fixed chemical potential,
it changes both the phase and cyclotron mass; when the saddle follows field
and temperature, the phase, and hence the local frequency, can be
parametrically more sensitive than the mass, and spectral weight is
redistributed among harmonics. At fixed density, conservation of
Fermi-surface area locks the leading  Onsager frequency, while the
cyclotron mass, Lifshitz--Kosevich envelope, and harmonic weights evolve. The
magnetic susceptibility acquires a finite mean-field step across a continuous
boundary. The correlated evolution of these observables, rather than any
single anomaly, distinguishes self-consistent nematic feedback from a rigid
anisotropic band or a scalar mass renormalization.

These two effects are complementary manifestations of the same orbital
feedback loop between Landau quantization and nematic order. Sections~II and
III establish the ensemble benchmarks and the uniform nematic Landau
spectrum; Sections~IV and V develop the self-consistent thermodynamics and
phase diagram; and Sec.~VI derives the magnetic and spectral response.
Section~VII discusses the implications and conclusions, with
technical details collected in the appendices.

\section{Free-electron benchmarks and ensemble dependence}
\label{sec:free-benchmarks}
The free two-dimensional electron gas~\cite{Shoenberg2D1984}  provides the natural reference for the ensemble dependence of the nematic problem.  At fixed chemical potential, the particle density oscillates with the magnetic field, whereas at fixed density the chemical potential must adjust accordingly. In realistic systems, reservoir states may partially pin the chemical potential, making these two ensembles useful limiting cases~\cite{ItskovskyManivVagner2000}. Here we retain only the thermodynamic results needed for the subsequent analysis, while the details of the Poisson summation, the canonical convolution, and the exact zero-temperature level counting are deferred to Appendices~\ref{app:free-gas-poisson} and~\ref{app:free-gas-canonical}.

The reference system is a clean, spinless, two-dimensional parabolic band
with sharp Landau levels. The field $\psi({\bf x})$ annihilates a spinless
fermion, and $m$ denotes its band mass.

The Hamiltonian in a uniform perpendicular field is
\begin{equation}
H_0=\int d^2x\,
\psi^\dagger({\bf x})\frac{{\bf D}^2}{2m}\psi({\bf x}),
\label{eq:II-free-Hamiltonian}
\end{equation}
where the covariant derivative is
\begin{equation}
{\bf D}=-i{\boldsymbol\nabla}+{\bf A},
\end{equation}
with \({\boldsymbol\nabla}\times{\bf A}=B\hat{\bf z}\).  We set
\(\hbar=k_B=e=1\), take \(B>0\), and write \(\beta=1/T\).
The one-particle energies form the Landau ladder
\begin{equation}
E_\lambda=\left(\lambda+\frac12\right)\omega_c,
\label{eq:II-free-Landau-levels}
\end{equation}
where $\omega_c=B/m$ and $ \lambda=0,1,2,\ldots$,  with degeneracy per unit  area
\begin{equation}
{\cal D}_B=\frac{B}{2\pi}.
\label{eq:II-free-degeneracy}
\end{equation}
The zero-field density of states is \(\rho_0=m/(2\pi)\), and hence
\({\cal D}_B=\rho_0\omega_c\).

We work in the degenerate, weak-field regime
\(\beta\mu\gg1\) and \(\omega_c/\mu\ll1\), retaining the full
ratio \(T/\omega_c\) in the parabolic level sum. 

\subsection{Fixed chemical potential}
At fixed chemical potential, each Landau level contributes independently to
the grand-potential density,
\begin{equation}
\Omega(T,\mu,B)
=-\frac{{\cal D}_B}{\beta}
\sum_{\lambda=0}^{\infty}
\ln\left[1+e^{-\beta(E_\lambda-\mu)}\right].
\label{eq:II-free-level-sum}
\end{equation}
Poisson summation, neglecting exponentially small band-bottom corrections
in $\mu/T$, separates the potential into zero-field, smooth magnetic and
oscillatory parts:
\begin{equation}
\Omega=\Omega_0+\Omega^{\rm non}+\Omega^{\rm osc} 
\end{equation}
where 
\begin{align}
\Omega_0&=-\frac{m\mu^2}{4\pi}
\left[1+\frac{\pi^2}{3(\beta\mu)^2}\right],
\label{eq:Omega0-free}\\
\Omega^{\rm non}&=\frac{m\omega_c^2}{48\pi},
\label{eq:Omega-non-free}\\
\Omega^{\rm osc}
&=\frac{m\omega_c^2}{4\pi^3}
\sum_{p=1}^{\infty}\frac{(-1)^p}{p^2}
R_T(\chi_p)\cos\left(\frac{2\pi p\mu}{\omega_c}\right).
\label{eq:II-free-grand-potential}
\end{align}
Here $\Omega_0$ includes the Sommerfeld term, $\Omega^{\rm non}$ gives
orbital diamagnetism, and $(-1)^p$ reflects the half-integer level offset.
The Onsager frequency is $\nu_B=m\mu=A_F/(2\pi)$, where $A_F=\pi k_F^2$ is the area of the Fermi surface.  
We use the Lifshitz--Kosevich (LK) function $R_T(z)=z/\sinh z$ throughout.
Its argument includes the harmonic index; in Eq.~\eqref{eq:II-free-grand-potential},
\begin{equation}
R_T(\chi_p)=\frac{\chi_p}{\sinh\chi_p}, 
\label{eq:II-LK-factor}
\end{equation}
where the dimensionless variable
\begin{equation}
\chi_p=\frac{2\pi^2p T}{\omega_c}
\label{eq:Targument}
\end{equation}
measures thermal broadening relative
to the cyclotron energy for harmonic $p$. 

Differentiation of the thermodynamic potential at fixed \(T\) and \(\mu\)
gives the magnetization.  The smooth contribution is 
\(M^d=-\omega_c/(24\pi)\) and the oscillatory part is given by
\begin{align}
M^{\rm osc}={}&-\frac{\mu}{2\pi^2}
\sum_{p=1}^{\infty}\frac{(-1)^p}{p}R_T(\chi_p)
\Bigg\{
\sin\left(\frac{2\pi p\mu}{\omega_c}\right)\nonumber\\
&+\frac{1}{\pi p}\frac{\omega_c}{\mu}
\left[1+\frac{\chi_p}{2}L(\chi_p)\right]
\cos\left(\frac{2\pi p\mu}{\omega_c}\right)
\Bigg\},
\label{eq:Mosc_full}
\end{align}
where 
\begin{equation}
L(x)=\coth x-1/x
\label{eq:Langevin}
\end{equation}
is the Langevin function.  The cosine quadrature is smaller by
\(O(\omega_c/\mu,T/\mu)\).  It arises from differentiating the smooth
prefactor and the thermal factor, whereas the leading sine term comes from
the rapidly varying phase.  The dominant dHvA waveform is therefore the
usual LK sine series.  Equation~\eqref{eq:Mosc_full} will be the
grand-canonical isotropic benchmark for the interacting problem.

\subsection{Fixed density}
Because the degeneracy of every Landau level is
proportional to \(B\), the chemical potential must oscillate with the field
in order to keep the density unchanged.  The appropriate potential is the
Helmholtz free-energy density,
\begin{equation}
{\cal F}(T,n,B)=\Omega[T,\mu(n,B),B]+\mu(n,B)n,
\label{eq:II-Legendre}
\end{equation}
where  \(\mu(n,B)\) is obtained by inverting 
\begin{equation}
n=-\left.\frac{\partial\Omega}{\partial \mu}\right|_{T,B} .
\end{equation}

We denote by
\(\mu_0=2\pi n/m\) the zero-temperature Fermi energy for the
prescribed density; the smooth finite-temperature chemical potential differs
only by exponentially small corrections in this degenerate two-dimensional
benchmark.  The equation of state separates into its smooth and
oscillatory contributions,
\begin{equation}
n=\frac{m\mu}{2\pi}+n_{\rm osc}(T,\mu,B).
\label{eq:II-free-density}
\end{equation}
In the thermally damped regime, the oscillatory part is small and the
constraint can be solved perturbatively by writing
\(\mu=\mu_0+\delta\mu\).  To leading order,
\begin{equation}
\delta\mu=-\frac{2\pi}{m}n_{\rm osc}+\cdots .
\end{equation}
Here, to the same order,
\begin{align}
n_{\rm osc}(\mu_0)&=\frac{m\omega_c}{2\pi}
\sum_{p=1}^{\infty}\frac{(-1)^p}{\pi p}
R_T(\chi_p)\sin\left(\frac{2\pi p\mu_0}{\omega_c}\right),
\end{align}

Expansion of the Legendre transform around \(\mu_0\) has an important
simplification: all terms linear in \(\delta\mu\) cancel by the equation of
state.  The first nontrivial correction is quadratic:
\begin{equation}
{\cal F}
={\cal F}_0+\Omega^{\rm non}
+\Omega^{\rm osc}(\mu_0)
+\Delta{\cal F}_N+\cdots,
\label{eq:Fcan-correction}
\end{equation}
where ${\cal F}_0(T,n)\equiv\Omega_0(T,\mu_0)+\mu_0n$ is the smooth
zero-field Helmholtz free-energy density at the prescribed density, and
\begin{equation}
\Delta{\cal F}_N=\frac{\pi}{m}n_{\rm osc}^2 .
\end{equation}
The last term is the leading correction generated by the density constraint.
Although formally second order in the oscillatory density, it need not be
negligible for the waveform.  A product of harmonics $p$ and $j$ generates harmonics at $p+j$ and
$|p-j|$. Equal indices also give a phase-independent contribution, included
in $\Delta{\cal F}_N$ but excluded from the oscillatory series below.  The result can still
be organized as a Fourier series because all frequencies are integer
multiples of the same fundamental.  Defining
\(\theta=2\pi\mu_0/\omega_c\) and
\(\chi=2\pi^2T/\omega_c\), the oscillatory canonical free energy can be
written, through this quadratic mixing order, as
\begin{equation}
{\cal F}^{\rm osc}
=\frac{m\omega_c^2}{4\pi^3}
\sum_{\ell=1}^{\infty}(-1)^\ell
{\cal C}^{(N)}_\ell(\chi)\cos(\ell\theta),
\label{eq:II-free-canonical-spectrum}
\end{equation}
where
\begin{align}
{\cal C}^{(N)}_\ell(\chi)
&=\frac{R_T(\ell\chi)}{\ell^2}+K_\ell(\chi),
\label{eq:II-free-canonical-kernel}
\end{align}
with
\begin{align}
K_\ell(\chi)&=
\chi^2\frac{\coth(\ell\chi)-\ell}{\sinh(\ell\chi)}.
\label{eq:Kell}
\end{align}
The first term in \({\cal C}^{(N)}_\ell\) is the direct LK contribution
evaluated at \(\mu_0\); \(K_\ell\), generated by the density constraint,
mixes harmonics already in the isotropic gas and establishes the canonical
reference for the nematic response. For example, in the thermally
damped limit,
\begin{equation}
{\cal C}^{(N)}_2\simeq
\frac{R_T(\chi_2)}{4}-\frac{R_T^2(\chi_1)}{2}.
\label{eq:II-free-second-harmonic}
\end{equation}
The second term originates from the product of two fundamental harmonics and
can dominate the direct LK second harmonic even while
\(|\delta\mu|/\omega_c\) remains small.

The harmonic construction requires $|\delta\mu|/\omega_c\ll1$ and
cannot be extrapolated uniformly to $T=0$. At zero temperature a generic
fixed density partially fills a Landau level, pinning the chemical
potential to it. Direct level counting gives the exact piecewise canonical
free energy and magnetization. The derivation is given in Appendix~\ref{app:free-gas-canonical}.
The sawtooth orientation differs from the pinned-$\mu$ limit, and integer
fillings give singular susceptibility features while temperature or level
broadening rounds the free-gas features~\cite{VagnerManiv1985,ItskovskyManivVagner1996,ItskovskyManivVagner2000}.
The exact zero-temperature waveform contains the full harmonic series,
whereas thermal damping restricts the appreciable harmonics at finite temperature. 

\section{Nematic model and uniform Landau spectrum}
\label{sec:nematic-model-spectrum}
We use a magnetic-field extension of the continuum quadrupolar model
introduced by Oganesyan, Kivelson, and Fradkin~\cite{OgKiFr2001}, which
describes a Fermi-surface Pomeranchuk instability in the $l=2$ channel.
Our goal is to isolate the low-energy interplay between quadrupolar
Fermi-surface order and Landau quantization, without constructing a
material-specific microscopic band structure. We retain a rotationally
invariant continuum description of the relevant Fermi-surface degrees of
freedom and couple it to the magnetic field through covariant derivatives.
The reference parabolic dispersion provides the analytic Landau spectrum;
departures from it enter later as controlled corrections to the mean-field
thermodynamics.

\subsection{Model Hamiltonian}
\label{sec:model}
The simplest Hamiltonian describing local quadrupolar interactions in a magnetic field can be written as
\begin{align}
H&=H_0+H_I,  
\label{eq:H}   \\
H_0&=\int d^2x\;\psi^\dagger({\bf x})\varepsilon({\bf D})\psi({\bf x}),
\label{eq:H0}\\
H_I&=\frac{f_2}{4}\int d^2x\;\Tr\!\left[\hat Q^2({\bf x})\right].
\label{eq:HI}
\end{align}
$\psi$ is the spinless fermion field; as in the free gas example,
${\bf D}=-i{\boldsymbol\nabla}+{\bf A}$ and
${\boldsymbol\nabla}\times{\bf A}=B\hat{\bf z}$.  We use
$e=\hbar=k_B=1$, as in the preceding section.  The dispersion is kept general at this stage to separate the
Fermi-surface structure relevant to the nematic instability from
nonuniversal details of the band away from it. For the reference parabolic
dispersion used to obtain the analytic Landau spectrum,
$\varepsilon({\bf D})={\bf D}^2/(2m)$.  In Eq.~\eqref{eq:HI}, $\hat Q$
denotes the quadrupolar momentum operator defined below.  With the convention
adopted there, the attractive channel relevant to the nematic instability has
$f_2<0$.

The local quadrupolar density is the symmetric, traceless tensor
\begin{equation}
Q_{ij}=\psi^\dagger\left[\frac{1}{2}\{D_i,D_j\}
-\frac{\delta_{ij}}{2}{\bf D}^2\right]\psi.
\label{eq:Quadrupole}
\end{equation}
Because $[D_x,D_y]\neq0$ in a magnetic field, the anticommutator makes the
symmetry $Q_{ij}=Q_{ji}$ explicit.  Under a gauge transformation,
$\psi\to e^{-i\chi}\psi$ and
${\bf A}\to{\bf A}+{\boldsymbol\nabla}\chi$, the covariant derivative
transforms as $D_i\psi\to e^{-i\chi}D_i\psi$.  It follows that $Q_{ij}$ is
locally gauge invariant.  Its two independent components may be collected in
the matrix representation
\begin{equation}
\hat Q=\begin{pmatrix}Q^+&Q^-\\Q^-&-Q^+\end{pmatrix},
\end{equation}
where
\begin{align}
Q^+&=\frac{1}{2}\psi^\dagger(D_x^2-D_y^2)\psi, 
\label{eq:Q+}\\
Q^-&=\frac{1}{2}\psi^\dagger\{D_x,D_y\}\psi.
\label{eq:Q-}
\end{align}
This definition reduces continuously to the usual quadrupolar density as $B\to0$.

The quadrupole probes the anisotropic distribution of kinetic momentum.
The perpendicular field enters through the covariant derivatives and changes
the orbital dynamics and quadrupolar response.  It preserves rotations in
the plane and does not itself select a nematic director.

\subsection{Bosonic effective action}
We formulate the interacting problem in the grand-canonical ensemble and
decouple the attractive quadrupolar interaction with a Hubbard--Stratonovich
field.  

For the attractive case, \(f_2=-|f_2|\), the Euclidean partition function is
\begin{align}
Z=\int{\cal D}\Phi{\cal D}\psi^\dagger{\cal D}\psi\;
e^{-S[\psi^\dagger,\psi,\Phi]},
\label{eq:Zpsiphi}
\end{align}
where
\begin{align}
S[\psi^\dagger,\psi,\Phi]
&=\int_x\Bigl\{\psi^\dagger
[\partial_\tau+\varepsilon({\bf D})-\mu]\psi
\nonumber\\
&\hspace{1. cm}-\frac{1}{2}\Tr(\Phi\hat Q)
+\frac{1}{4|f_2|}\Tr(\Phi^2)\Bigr\}.
\label{eq:SPhi-f}
\end{align}
Here and below, the compact integration measure is
$\int_x=\int d^2x\int_0^\beta d\tau$.

In Eqs.~\eqref{eq:Zpsiphi} and \eqref{eq:SPhi-f}, the real, symmetric,
traceless auxiliary field
\begin{equation}
\Phi=\begin{pmatrix}\phi^+&\phi^-\\ \phi^-&-\phi^+\end{pmatrix},
\label{eq:III-Phi}
\end{equation}
linearizes the quadrupolar interaction.  The explicit construction and
normalization of the transformation are given in
Appendix~\ref{app:HS-transformation}.

The fermions can now be integrated out exactly, leaving a partition function
for the bosonic auxiliary field $\Phi$,
\begin{equation}
Z=\int{\cal D}\Phi\,e^{-S_{\rm eff}[\Phi]},
\label{eq:ZPhi}
\end{equation}
with the bosonic effective action
\begin{equation}
S_{\rm eff}[\Phi]
=\frac{1}{4|f_2|}\int_x\operatorname{Tr}(\Phi^2)
-\operatorname{Tr}\ln\left(\beta\hat O[\Phi]\right).
\label{eq:Seff}
\end{equation}

For the parabolic dispersion, the operator $\hat O[\Phi]$ is given by
\begin{align}
\hat O[\Phi]&=\partial_\tau-\mu+{\cal H}_{\rm eff}[\Phi],
\label{eq:O}
\end{align}
where the effective Hamiltonian can be written as 
\begin{align}
{\cal H}_{\rm eff}[\Phi]
&=\frac{1}{2m}G_{ij}\frac{1}{2}\{D_i,D_j\}.
\label{eq:Heff}
\end{align}
For a spatially uniform auxiliary field $\Phi$,
\begin{equation}
G_{ij}=\delta_{ij}-m\Phi_{ij}
\end{equation}
is a dimensionless inverse-mass tensor (equivalently, a metric-like tensor for the quadratic kinetic energy) and deforms
the quadratic dispersion while preserving gauge covariance.  This uniform
interpretation is sufficient for the spectrum calculated below; a
Hermitian extension to nonuniform fields is specified in Appendix~\ref{app:HS-transformation}.

In the effective action of Eq.~\eqref{eq:Seff}, 
the trace of $\Phi^2$ contracts tensor indices, while the logarithmic trace is
a functional trace over space and imaginary time, evaluated in a
one-particle basis.

\subsection{Uniform static saddle}
We evaluate the functional integral in Eq.~\eqref{eq:ZPhi} at a static,
spatially uniform Hubbard--Stratonovich field,
\begin{equation}
\Phi(\tau,{\bf x})=\bar\Phi,
\end{equation}
and approximate
\begin{equation}
Z\simeq e^{-S_{\rm eff}[\bar\Phi]}.
\label{eq:MF-saddle-Z}
\end{equation}
The stationary configuration satisfies
\begin{equation}
\left.\frac{\delta S_{\rm eff}}{\delta\Phi_{ij}}
\right|_{\Phi=\bar\Phi}=0.
\label{eq:MF-saddle-eq}
\end{equation}
The variation of
Eq.~\eqref{eq:SPhi-f} gives the mean-field relation
\begin{equation}
\bar\Phi_{ij}=|f_2|\,\langle Q_{ij}\rangle_{\rm MF},
\label{eq:Phi-Q-MF}
\end{equation}
which directly relates the saddle-point solution $\bar\Phi$ to the nematic
order parameter $\langle Q_{ij}\rangle_{\rm MF}$ at this level of
approximation.

The uniform static saddle thus provides the basis for determining the equilibrium thermodynamic properties and the stability of the nematic phase in a magnetic field.

\subsection{Landau levels in a uniform nematic background}
A uniform nematic deformation rescales the Landau-level spacing while
preserving the degeneracy per  area. For the parabolic part of the
dispersion, the one-particle Hamiltonian at the saddle is
\begin{equation}
{\cal H}_{\rm eff}[\bar\Phi]
=\frac{1}{2m}G_{ij}\frac{1}{2}\{D_i,D_j\},
\label{eq:Heff-MF}
\end{equation}
with
\begin{equation}
G_{ij}=\delta_{ij}-m\bar\Phi_{ij}.
\label{eq:G-MF}
\end{equation}
At this point it is useful to introduce the rotationally invariant amplitude of
the uniform traceless saddle,
\begin{equation}
q^2\equiv \frac{m^2}{2}\Tr(\bar\Phi^2),
\qquad 0\le q<1.
\label{eq:q-def}
\end{equation}
The two eigenvalues of $m\bar\Phi$ are $\pm q$, and therefore
\begin{equation}
\det G=1-q^2.
\end{equation}
We shall denote
\begin{equation}
\Delta(q)\equiv\sqrt{\det G}=\sqrt{1-q^2}.
\label{eq:Delta-q}
\end{equation}
The orientation of the nematic director drops out of the uniform thermodynamic
potential, as required by rotational invariance, and $q<1$ is precisely the
domain in which the quadratic kinetic form is positive definite.

The eigenvalue problem can be mapped exactly onto the isotropic Landau problem
by an auxiliary linear change of coordinates with a nonunit Jacobian, as shown
in Appendix~\ref{app:spectrum-uniform}.  The energies are
\begin{equation}
E_\lambda(q)=\left(\lambda+\frac12\right)\omega_c\Delta(q),
\label{eq:ciclotron}
\end{equation}
with $\lambda=0,1,2,\ldots,$
whereas the degeneracy per unit  area remains
\begin{equation}
{\cal D}_B=\frac{B}{2\pi}.
\label{eq:degeneracy}
\end{equation}
The degeneracy counts magnetic flux quanta through the area.
Uniform anisotropy changes the cyclotron dynamics but preserves the
guiding-center state count at fixed flux.

Equivalently, the cyclotron mass of the quadratic saddle is
\begin{equation}
m_c(q)=\frac{m}{\Delta(q)}.
\label{eq:mc-nematic}
\end{equation}

This result is exact for the quadratic dispersion at a static uniform saddle.
It will also be the unperturbed spectrum when corrections to the  dispersion are included perturbatively.

\section{Mean-field thermodynamics and Landau expansion}
\label{sec:mean-field}
We determine the uniform nematic background self-consistently from the
mean-field thermodynamic potential. After establishing the zero-field
instability and its stabilization, we derive the field-dependent Landau
coefficients at fixed chemical potential and fixed density.

\subsection{Grand potential at fixed chemical potential}
We first work at fixed chemical potential.  We keep
the parabolic part of the dispersion, for which
Eqs.~\eqref{eq:ciclotron} and \eqref{eq:degeneracy} imply that the fermionic
determinant follows from the free level sum of Sec.~II with spacing
$\omega_c\Delta$ and the unchanged degeneracy $B/(2\pi)$.  We have 
\begin{equation}
\Omega_{\rm MF}(q)
=\Omega_{f,{\rm par}}(q)
+\frac{1}{2m^2|f_2|}\; q^2
+\delta\Omega_\gamma(q),
\label{eq:OmegaMF-general}
\end{equation}
where $\Omega_{f,{\rm par}}(q)$ is the grand potential of the parabolic
dispersion with the renormalized cyclotron frequency
$\omega_c\to\omega_c\Delta(q)$.  The second term in
Eq.~\eqref{eq:OmegaMF-general} comes from the Hubbard--Stratonovich
transformation, while $\delta\Omega_\gamma$ denotes the contribution of the
radial dispersion operator introduced below. This separation is useful
because the Landau spectrum of the parabolic saddle is exact, while corrections can be treated by ordinary perturbation theory in the
same Landau-level basis.

For the quadratic spectrum, in the degenerate regime $\beta\mu\gg1$, the
parabolic fermionic part separates into smooth, nonoscillatory magnetic, and
oscillatory terms,
\begin{equation}
\Omega_{f,{\rm par}}(q)=\Omega_f^0(q)+\Omega_f^{\rm non}(q)
+\Omega_f^{\rm osc}(q),
\label{eq:OmegaMF-decomp}
\end{equation}
with
\begin{align}
\Omega_f^0(q)
&=-\frac{m\mu^2}{4\pi\Delta}
\left[1+\frac{\pi^2}{3(\beta\mu)^2}\right],
\label{eq:calF0}\\[4pt]
\Omega_f^{\rm non}(q)
&=\frac{m\mu^2}{48\pi}
\left(\frac{\omega_c}{\mu}\right)^2\Delta,
\label{eq:calFnon}\\[4pt]
\Omega_f^{\rm osc}(q)
&=\frac{m\mu^2}{4\pi}
\left(\frac{\omega_c}{\mu}\right)^2\Delta
\sum_{n=1}^{\infty}\frac{(-1)^n}{\pi^2n^2}
\nonumber\\
&\quad\times
R_T\!\left(\frac{\chi_n}{\Delta}\right)
\cos\!\left(\frac{2\pi n\mu}{\omega_c\Delta}\right),
\label{eq:Fosc-MF}
\end{align}
where $\chi_n=2\pi^2n/(\beta\omega_c)$ and
$R_T(x)=x/\sinh x$.  Equations~\eqref{eq:calF0}--\eqref{eq:Fosc-MF} keep the
full dependence on $q$ generated by the quadratic saddle.  The assumptions are
$\beta\mu\gg1$ and $\omega_c/\mu\ll1$; no expansion in $\beta\omega_c$ has
been made.

The nematic order modifies both the Lifshitz--Kosevich damping scale and the
phase of each harmonic through $\Delta(q)$. The oscillatory determinant in
turn feeds back into the equilibrium order parameter.  The irrelevant radial operator adds perturbative
nonuniversal corrections to this structure, derived below and in
Appendix~\ref{app:irrelevant-term}.

\subsection{Zero-field instability and stabilization of the nematic phase}
The zero-field limit establishes the nematic instability and the role of
the stabilizing dispersion correction, connecting the calculation to
previous studies~\cite{OgKiFr2001,BaOx2003,Lawler2006}.

We first isolate the zero-field, zero-temperature instability.  We define
\begin{equation}
a\equiv\frac{m\mu^2}{4\pi}\qquad\mbox{and}
\qquad
F_2\equiv m^2 a f_2.
\label{eq:A-F2-def}
\end{equation}
While $a$ sets the grand-potential density scale, $F_2$ is the dimensionless
quadrupolar coupling.  For the attractive channel, $F_2<0$, we obtain
\begin{align}
\Omega_{\rm MF}=\frac{a}{2}\left(\frac{q^2}{|F_2|}
-\frac{2}{\sqrt{1-q^2}}\right)+\delta\Omega_\gamma(q).
\end{align}
Expanding this expression in powers of $q$, we find for the quadratic term
\begin{equation}
\Delta\Omega_{\rm MF}^{(2)}
=\frac{a}{2}\left(\frac{1}{|F_2|}-1\right)q^2,
\label{eq:quadratic-zero}
\end{equation}
and the isotropic saddle loses stability at
\begin{equation}
F_2^c=-1.
\label{eq:F2-critical-zero}
\end{equation}
The isotropic state is locally unstable for $F_2<-1$, in agreement with
the Pomeranchuk criterion.   However, the strictly parabolic dispersion does not by itself stabilize the
ordered state because the quartic term in the $q$ expansion is negative.
Studies of the zero-field quantum nematic transition in the random-phase
approximation~\cite{OgKiFr2001} and bosonized formulations~\cite{BaOx2003}
showed that
radial curvature near the Fermi surface can stabilize the nematic phase.  In
the corresponding low-energy description, the free fermionic dispersion is approximated by
\begin{equation}
\varepsilon(k)-\mu=v_F(k-k_F)+\frac{\gamma}{m k_F}(k-k_F)^3+\ldots
\label{eq:curvature}
\end{equation}
where the normalization is chosen so that $\gamma$ is dimensionless. 
This expression displays the usual linear approximation that defines the Fermi velocity $v_F$, and a cubic correction that preserves particle--hole symmetry.   This low-energy form isolates the cubic radial correction studied in
bosonized descriptions of the isotropic--nematic quantum phase
transition~\cite{BaOx2003,Lawler2006}, where it stabilizes the ordered state. 

In a magnetic field, in which the momentum is not a good quantum number, we can represent this type of radial curvature by the isotropic,
gauge-covariant higher-order dispersion term
\begin{equation}
\delta\varepsilon_\gamma({\bf D})
=\frac{\gamma}{8mk_F^4}({\bf D}^2-k_F^2)^3,
\qquad \gamma>0.
\label{eq:FS-irrelevant}
\end{equation}
Here $k_F$ is the Fermi momentum of the isotropic zero-field reference state; at
fixed chemical potential, $k_F^2=2m\mu$.  Near that surface,
Eq.~\eqref{eq:FS-irrelevant} reproduces the cubic radial term in
Eq.~\eqref{eq:curvature}.  We now evaluate its contribution to the local
Landau coefficients.  The quadratic Landau spectrum remains the unperturbed
problem, while this correction is treated at leading order in perturbation theory.

This term does not change the location of the Fermi surface, the Fermi velocity, or
the quadratic nematic coefficient at $T=B=0$; its first contribution to the
uniform Landau expansion is quartic.  An explicit calculation, given in
Appendix~\ref{app:irrelevant-term}, gives
\begin{equation}
\Delta\Omega_{\rm MF}
=\frac{a}{2}r_0q^2+\frac{a}{4}u_0q^4+O(q^6),
\label{eq:Landau-zero}
\end{equation}
with
\begin{equation}
r_0=\frac{1}{|F_2|}-1,
\qquad
u_0=\frac{3}{16}(\gamma-8).
\label{eq:r0-u0}
\end{equation}
Hence, $\gamma>8$
makes the local quartic coefficient positive without moving the isotropic
instability $F_2^c=-1$.    Other irrelevant radial corrections can provide the
same local stabilization while changing nonuniversal higher-order
coefficients. The particular form in Eq.~\eqref{eq:FS-irrelevant} is a
simple representation of the band curvature normal to the Fermi surface
and is not intended as a global microscopic dispersion.
Local quartic stability does not by itself determine the potential at finite
amplitude.  The thermodynamic potential and its dependence on higher-order
stabilizing terms are examined in Sec.~\ref{sec:full-potential}.

For $u_0>0$ the saddle close to the zero-field transition is
\begin{equation}
q_{\rm sp}^2=
\begin{cases}
0, & r_0>0,\\[3pt]
-r_0/u_0+O(r_0^2), & r_0<0.
\end{cases}
\label{eq:qsp-zero}
\end{equation}
Thus $q_{\rm sp}\propto(-r_0)^{1/2}$ on the ordered side.  The irrelevant
operator fixes the nonuniversal amplitude of the order parameter but not the
critical coupling or the mean-field exponent.

\subsection{Magnetic field: oscillatory renormalization of the saddle}
At finite magnetic field, Landau quantization modulates the nematic
coefficients. Expanding the fixed-$\mu$ grand potential near the isotropic
saddle while retaining the magnetic oscillations gives
\begin{equation}
\Delta\Omega_{\rm MF}
=\frac{a}{2}r(T,B)q^2+\frac{a}{4}u(T,B)q^4+O(q^6),
\label{eq:Landau-TB}
\end{equation}
and it is convenient to write the quadratic coefficient as
\begin{equation}
r(T,B)=\frac{1}{|F_2|}-{\cal X}_{2}(T,B),
\label{eq:r-X2}
\end{equation}
where ${\cal X}_2$ denotes the dimensionless irreducible electronic response
in the quadrupolar channel. We decompose this response into the parabolic
contribution and the correction generated by the radial dispersion operator:
\begin{equation}
{\cal X}_2(T,B)
={\cal X}_{2,{\rm par}}(T,B)+\delta{\cal X}_{2,\gamma}(T,B)\; .
\label{eq:X2-full-gamma}
\end{equation}
The former is the parabolic fixed-point result, while
$\delta{\cal X}_{2,\gamma}$ contains the higher-dispersion corrections.

The parabolic contribution reads
\begin{align}
{\cal X}_{2,{\rm par}}(T,B)
={}&1+\frac{\pi^2}{3(\beta\mu)^2}
+\frac{1}{12}\left(\frac{\omega_c}{\mu}\right)^2
+{\cal X}_{2,{\rm par}}^{\rm osc}(T,B),
\label{eq:X2-total}
\end{align}
where the oscillatory component is given by
\begin{align}
{\cal X}_{2,{\rm par}}^{\rm osc}
={}&\left(\frac{\omega_c}{\mu}\right)^2
\sum_{n=1}^{\infty}\frac{(-1)^n}{\pi^2n^2}\!R_T\!(\chi_n)\!
\Bigg\{
\left[1+\chi_n L(\chi_n)\right]
\cos\theta_n
\nonumber\\
&\hspace{2.5cm}+\theta_n\sin\theta_n
\Bigg\},
\label{eq:X2-osc}
\end{align}
with 
\begin{equation}
\theta_n=\frac{2\pi n\mu}{\omega_c} .
\label{eq:theta-L}
\end{equation}
The thermal argument $\chi_n$ and the Langevin function $L(x)$ are defined in
Eqs.~\eqref{eq:Targument} and~\eqref{eq:Langevin}. These expressions are derived in
Appendix~\ref{app:Landau-expansion}.  

The term proportional to
$\theta_n\sin\theta_n$ is  of order $\omega_c/\mu$ and is the
leading oscillatory contribution to the quadratic coefficient of the
parabolic theory; the remaining oscillatory term is of order
$(\omega_c/\mu)^2$.

To evaluate the correction $\delta{\cal X}_{2,\gamma}$, we perform
a Poisson resummation of its leading grand-potential contribution (see
Appendix~\ref{app:irrelevant-term}). Although the radial operator in
Eq.~\eqref{eq:FS-irrelevant} leaves the quadratic coefficient unchanged at
$T=B=0$, it produces controlled nonuniversal corrections at nonzero $T$ or
$B$. In the degenerate, weak-field regime
$T/\mu\ll1$ and $\omega_c/\mu\ll1$, combining the retained low-energy
terms with the parabolic contribution, we obtain for the quadrupolar response
\begin{align}
{\cal X}_2(T,B)={}&1+\pi^2\left(\frac13-\frac{\gamma}{4}\right)
\left(\frac{T}{\mu}\right)^2
\nonumber\\
&+\left(\frac1{12}-\frac{7\gamma}{16}\right)
\left(\frac{\omega_c}{\mu}\right)^2
\nonumber\\
&+\left(\frac{\omega_c}{\mu}\right)^2
\sum_{n=1}^{\infty}\frac{(-1)^n}{\pi^2n^2}R_T(\chi_n)
\Bigg\{\theta_n\sin\theta_n
\nonumber\\
&\;+\left(1-\frac{3\gamma}{4}\right)
\left[1+\chi_nL(\chi_n)\right]\cos\theta_n\Bigg\}
+\cdots .
\label{eq:X2-complete-lowenergy}
\end{align}
The ellipsis denotes subleading terms beyond the retained low-energy expansion.
Equation~\eqref{eq:X2-complete-lowenergy}
shows that the radial operator leaves the leading
$O(\omega_c/\mu)$ sine quadrature unchanged at this order of approximation; its
relative size is of order $\gamma\omega_c/\mu$.

The loss of local isotropic stability at fixed $T$ and $B$ is determined by $r(T,B)=0$
or, equivalently
\begin{equation}
F_2^c(T,B)=-\frac{1}{{\cal X}_2(T,B)}.
\label{eq:F2-critical-TB}
\end{equation}
For $u>0$ and without preemption, this is the continuous equilibrium
boundary.  Landau quantization modulates the nematic susceptibility.
The leading quadratic-model contribution is
contained in Eqs.~\eqref{eq:X2-total}--\eqref{eq:X2-osc}, while the precise
subleading coefficients depend on the higher-order dispersion terms.  In
the $T\to0$, $B\to0$ reference limit,
$\delta{\cal X}_{2,\gamma}\to0$ and ${\cal X}_2\to1$, recovering
Eq.~\eqref{eq:F2-critical-zero}.

For the quartic coefficient, we define
\begin{equation}
\epsilon_B\equiv\frac{\omega_c}{\mu},
\qquad
L_n\equiv L(\chi_n),
\qquad
L'_n\equiv\left.\frac{dL(x)}{dx}\right|_{x=\chi_n}.
\end{equation}
We decompose
\begin{equation}
u(T,B)=u_{f,{\rm par}}^{\rm sm}+u_{f,{\rm par}}^{\rm osc}
+u_{\gamma}^{\rm sm}+\delta u_{\gamma}^{\rm osc},
\label{eq:u-decomp}
\end{equation}
where the parabolic smooth contribution is
\begin{equation}
u_{f,{\rm par}}^{\rm sm}
=-\frac{3}{2}\left[1+\frac{\pi^2}{3}\left(\frac{T}{\mu}\right)^2\right]
-\frac{1}{24}\epsilon_B^2,
\label{eq:u-smooth}
\end{equation}
and the oscillatory contribution is
\begin{align}
u_{f,{\rm par}}^{\rm osc}
={}&\sum_{n=1}^{\infty}(-1)^n R_T(\chi_n)
\Bigg\{-2\cos\theta_n
\label{eq:u-par-osc}\\
&+\frac{\epsilon_B}{\pi n}\left(2\chi_nL_n-1\right)\sin\theta_n
\nonumber\\
&+\frac{\epsilon_B^2}{2\pi^2n^2}
\left[\chi_n^2\left(L_n^2-L'_n\right)-1-\chi_nL_n\right]
\cos\theta_n\Bigg\}.
\nonumber 
\end{align}
Equation~\eqref{eq:u-par-osc} follows directly by differentiating the
unexpanded dHvA term twice with respect to $q^2$; the derivation is given in
Appendix~\ref{app:Landau-expansion}.  Its leading term is of order unity,
\begin{equation}
u_{f,{\rm par}}^{\rm osc}
=-2\sum_{n=1}^{\infty}(-1)^nR_T(\chi_n)\cos\theta_n
+O(\epsilon_B),
\label{eq:uosc-leading}
\end{equation}
in contrast with the leading oscillatory correction to $r(T,B)$, which is
$O(\epsilon_B)$.  The reason is that two derivatives of the rapidly varying
phase $\theta_n/\Delta(q)$ generate two powers of $\mu/\omega_c$ and cancel
the explicit $\epsilon_B^2$ prefactor of the oscillatory grand potential.

For the representative irrelevant operator of Eq.~\eqref{eq:FS-irrelevant},
the smooth leading contribution is
\begin{align}
u_{\gamma}^{\rm sm}
={}&\frac{3\gamma}{16}
+\frac{15\pi^2\gamma}{4}\left(\frac{T}{\mu}\right)^2
+\frac{71\gamma}{16}\epsilon_B^2
+\ldots\label{eq:u-gamma-smooth}
\end{align}
The ellipsis represents terms of order $\gamma T^4/\mu^4$, $
\gamma\epsilon_B^2T^2/\mu^2$ and
$\gamma\epsilon_B^4$.
The leading oscillatory correction is
\begin{align}
\delta u_{\gamma}^{\rm osc}
={}&\frac{3\gamma\epsilon_B}{2\pi}
\sum_{n=1}^{\infty}\frac{(-1)^n}{n}R_T(\chi_n)
\left[1-\chi_nL_n\right]\sin\theta_n
\nonumber\\
&+O(\gamma\epsilon_B^2).
\label{eq:u-gamma-osc}
\end{align}
The exact Landau-level representations, from which
Eqs.~\eqref{eq:u-gamma-smooth} and~\eqref{eq:u-gamma-osc} follow, are given in
Appendix~\ref{app:irrelevant-term}.  In particular,
$u_{\gamma}^{\rm sm}(0,0)=3\gamma/16$, and combining it with
Eq.~\eqref{eq:u-smooth} gives again
$u_0=3(\gamma-8)/16$ at strictly $B=T=0$.  The finite-$T$ and smooth finite-$B$
coefficients in Eq.~\eqref{eq:u-gamma-smooth} are nonuniversal properties of
the representative dispersion correction, whereas the leading $O(1)$ term in
Eq.~\eqref{eq:uosc-leading} is fixed by the parabolic Fermi-surface theory.

In the thermally damped regime, where the first harmonic dominates,
\begin{equation}
u(T,B)\simeq u_{f,{\rm par}}^{\rm sm}+u_\gamma^{\rm sm}+2R_T(\chi_1)\cos\theta_1+O(\epsilon_B),
\label{eq:u-first-harmonic}
\end{equation}
while, to the same accuracy, the dominant first-harmonic correction to the quadratic
coefficient is
\begin{equation}
r_{\rm osc}\simeq
\frac{2\epsilon_B}{\pi}R_T(\chi_1)\sin\theta_1.
\label{eq:r-first-harmonic}
\end{equation}
Thus the leading oscillations of $r$ and $u$ are approximately in quadrature:
the phase boundary is displaced mainly by a sine harmonic, whereas its local
quartic stability is modulated mainly by a cosine harmonic.  In particular,
near a Landau level crossing, $\theta_1=\pi$ modulo $2\pi$, the leading
oscillatory displacement of $r$ vanishes while the reduction of $u$ is
maximal.

The magnetic field can therefore drive the quartic coefficient through zero
even when its zero-field value is positive.  The continuous line becomes
marginal where $r=u=0$, defining candidate tricritical points.  Their
classification requires the higher-order and global tests performed in
Sec.~\ref{sec:full-potential}; beyond them,
higher-order terms determine the continuation of the phase boundary.

Within the local continuous regime, with $u>0$, small amplitude and no
preempting minimum, the saddle is
\begin{equation}
q_{\rm sp}^2(T,B)=
\begin{cases}
0, & r(T,B)>0,\\[3pt]
-r(T,B)/u(T,B)+O(r^2), & r(T,B)<0.
\end{cases}
\label{eq:qsp-TB}
\end{equation}
The oscillatory quadratic term shifts the continuous boundary, while the
quartic term controls the local stability and the ordered amplitude.
Because the dHvA phase contains $1/\Delta=1+q^2/2+\cdots$, truncating the
Landau expansion of an individual harmonic also requires
\begin{equation}
q^2\,\frac{\mu}{\omega_c}\ll1
\label{eq:Landau-phase-control}
\end{equation}
for the harmonics that are not thermally suppressed.  For harmonic $n$, the phase increment is proportional to
$n q^2\mu/\omega_c$.  Since the thermal factor suppresses harmonics only
when $nT/\omega_c$ becomes large, the thermally relevant harmonics extend
 to $n\sim\omega_c/T$.    Near Landau-level crossings, the local expansion is controlled for $\mu q^2\ll T$, so its range of validity shrinks as $T\to0$. The finite-amplitude analysis in Sec.~\ref{sec:full-potential} therefore uses the unexpanded oscillatory expression of Eq.~\eqref{eq:Fosc-MF}.

The consequences of the self-consistent saddle for the Landau-level
spectrum and crossing fields are discussed in
Appendix~\ref{app:selfconsistent-spectrum}. 

\subsection{Fixed density in the thermally damped regime}
\label{sec:MF-fixed-density}
At fixed density, the leading dHvA frequency is constrained even before performing the explicit canonical calculation.  For a single closed Fermi surface in two dimensions, the particle density fixes the area enclosed by the Fermi surface,
\begin{equation}
n=\frac{A_F}{(2\pi)^2},
\end{equation}
in the present spinless convention.  A nematic distortion can therefore change the \emph{shape} of the Fermi surface, but at fixed density it cannot change its enclosed area.  Since the Onsager frequency is determined by this area,
$
\nu_B=A_F/2\pi
$,
 the fundamental oscillation frequency is locked at fixed density.  This conclusion does not rely on the specific form of  the dispersion: it follows from state counting together with the Onsager relation, provided the deformation does not change the  topology of the Fermi surface. 

The role of the canonical calculation is then to determine how the chemical potential adjusts as the nematic deformation changes while the Fermi-surface area remains fixed.  For the quadratic saddle this compensation can be shown  explicitly.  We first keep the nematic amplitude \(q\) fixed and perform the Legendre transformation with respect to \(\mu\); only afterwards  we allow the equilibrium nematic saddle to shift.  The perturbative harmonic construction below is carried out in the thermally damped regime introduced in Sec.~\ref{sec:free-benchmarks}.

For the quadratic saddle, the Landau-level spacing is
$
\omega_c\Delta(q)
$,
whereas the degeneracy per unit area,
$
{\cal D}_B=B/2\pi
$, 
is independent of the nematic deformation.  Consequently, the smooth density at fixed \(q\) is
\begin{equation}
n_{\rm sm}(q,\mu)
=
{\cal D}_B\,
\frac{\mu}{\omega_c\Delta(q)}
=
\frac{m\mu}{2\pi\Delta(q)}.
\label{eq:nsm-nematic-fixedN}
\end{equation}
Equivalently, the combination
$\nu=
\mu/(\omega_c\Delta(q))
$
measures the smooth Landau-level filling.  Since the level degeneracy is unchanged by the nematic deformation, fixing the density fixes this filling.

Defining the zero-field chemical potential associated with the prescribed density,
\begin{equation}
\mu_0\equiv\frac{2\pi n}{m},
\label{eq:mu0-fixedN-nematic}
\end{equation}
the smooth density constraint \(n_{\rm sm}(q,\bar\mu)=n\) gives
\begin{equation}
\bar\mu(q)=\mu_0\Delta(q).
\label{eq:mubar-fixedN-nematic}
\end{equation}
Thus, within the quadratic model, the compensation required by fixed density has a simple form: the nematic deformation changes the Landau-level spacing by the factor \(\Delta(q)\), while the chemical potential changes by exactly the same factor.  The rapidly varying phase therefore becomes
\begin{equation}
\frac{2\pi p\bar\mu(q)}
{\omega_c\Delta(q)}
=
\frac{2\pi p\mu_0}{\omega_c}
\equiv p\theta_0,
\label{eq:phase-lock-fixedN}
\end{equation}
and is independent of the nematic amplitude.

This explicit cancellation is the quadratic realization of the more general fixed-area argument above.  
Once the irrelevant radial curvature is included, this particular cancellation is no longer exact: the chemical potential must instead be determined from the more general condition that the enclosed Fermi-surface area remain fixed.  The curvature can consequently modify the cyclotron mass, the thermal damping, the oscillation amplitudes, and the detailed relation between $\bar\mu$ and $q$, while the leading Onsager frequency remains fixed by the density.  Corrections to the simple harmonic description associated with the curvature and with the oscillatory part of the chemical potential are subleading in the regime considered here and will be treated accordingly below.

At fixed $q$, the oscillatory density evaluated at the smooth chemical
potential is
\begin{align}
n_{\rm osc}(q)={}&\frac{m\omega_c}{2\pi}
\sum_{p=1}^{\infty}\frac{(-1)^p}{\pi p}
\widetilde R_T(p)
\sin(p\theta_0),
\label{eq:nosc-nematic-fixedN}
\end{align}
where $\widetilde R_T(p)\equiv R_T(p\widetilde\chi)$ and
\begin{equation}
\widetilde R_T(p)=\frac{p\widetilde\chi}{\sinh(p\widetilde\chi)},
\qquad
\widetilde\chi=\frac{2\pi^2}{\beta\omega_c\Delta(q)}.
\label{eq:RT-tilde-fixedN}
\end{equation}
The smooth compressibility at fixed $q$ is
\begin{equation}
\kappa(q)=\left.\frac{\partial n_{\rm sm}}{\partial\mu}\right|_q
=\frac{m}{2\pi\Delta(q)}.
\label{eq:kappa-nematic-fixedN}
\end{equation}
Consequently, the oscillating chemical potential is
\begin{align}
\delta\mu
&=-\frac{n_{\rm osc}}{\kappa}
\nonumber \\
&=-\omega_c\Delta
\sum_{p=1}^{\infty}\frac{(-1)^p}{\pi p}
\widetilde R_T(p)\sin(p\theta_0)
+O(n_{\rm osc}^2).
\label{eq:dmu-nematic-fixedN}
\end{align}

Expanding the Legendre transformation about $\bar\mu(q)$ gives, to second
order in the oscillatory amplitude,
\begin{equation}
\begin{aligned}
{\cal F}_N(q,n,B)
={}&{\cal F}_{N,{\rm sm}}(q,n,B)
+\Omega_{\rm osc}(q,\bar\mu) \\ 
&+\frac{n_{\rm osc}^2(q)}{2\kappa(q)}+\ldots,
\end{aligned}
\label{eq:FN-nematic-fixedN}
\end{equation}
where the ellipsis denotes terms of third and higher order in the
thermally damped oscillatory contribution.  The
canonical correction is therefore
\begin{equation}
\Delta{\cal F}_N^{(2)}(q)
=\frac{\pi\Delta(q)}{m}\,n_{\rm osc}^2(q).
\label{eq:FN-correction-nematic}
\end{equation}
This is the direct nematic counterpart of Eq.~\eqref{eq:Fcan-correction}.

As in the free gas, the quadratic term mixes the original harmonics through
sum and difference channels.  Absorbing its zero-frequency part into the
smooth free energy, the harmonic spectrum at a fixed nematic background can
be written
\begin{align}
{\cal F}_{N,{\rm harm}}^{\rm osc}(q)
={}&\frac{m\omega_c^2\Delta(q)}{4\pi^3}
\sum_{\ell=1}^{\infty}(-1)^\ell
{\cal C}^{(N)}_\ell(\widetilde\chi)
\cos(\ell\theta_0),
\label{eq:FN-spectrum-nematic}
\end{align}
where
\begin{align}
{\cal C}^{(N)}_\ell(x)
&=\frac{R_T(\ell x)}{\ell^2}+K_\ell(x),
\label{eq:Cell-nematic-fixedN}
\end{align}
Here $K_\ell$ is the canonical convolution kernel defined in
Eq.~\eqref{eq:Kell}, now evaluated at the nematically renormalized
thermal variable $x=\widetilde\chi=\chi/\Delta$.  In particular, in the
thermally damped limit,
\begin{equation}
{\cal C}^{(N)}_2
\simeq\frac{\widetilde R_T(2)}{4}
-\frac{\widetilde R_T^2(1)}{2},
\label{eq:C2-nematic-fixedN}
\end{equation}
so that the fixed-density constraint can strongly reorganize the higher
harmonics even when the oscillation of the chemical potential remains small.

Equation~\eqref{eq:FN-spectrum-nematic} describes the canonical oscillatory
spectrum at a fixed nematic background.  At equilibrium, however, the
oscillatory free energy also shifts the nematic saddle.  This oscillation-induced shift of the equilibrium saddle generates an additional second-order contribution to the free
energy, distinct from the canonical mixing $n_{\rm osc}^2/(2\kappa)$.  At
fixed density, phase locking means that derivatives with respect to the
nematic amplitude act only on the smooth prefactor and on the LK factor, and
therefore do not acquire the large $\mu_0/\omega_c$ factors associated with a
rapidly varying phase.  Provided the smooth curvature remains finite, the
saddle-shift contribution is consequently smaller than the canonical
harmonic mixing by an additional factor of order
$(\omega_c/\mu_0)^2$ in the dHvA regime.  The explicit perturbative expansion
and the resulting harmonic convolution are given in
Appendix~\ref{app:saddle-readjustment}.  The oscillatory correction to the
prefactor $\Delta$ multiplying $n_{\rm osc}^2$ enters only at third order in
the oscillatory contribution and is likewise discussed there.

\paragraph*{Canonical Landau coefficients.}
Expanding the canonical free energy about the isotropic state gives the
local coefficients entering the phase diagram. We define
\begin{equation}
\begin{gathered}
A_0=\frac{m\mu_0^2}{4\pi},\qquad
\epsilon_0=\frac{\omega_c}{\mu_0},\\
\chi=\frac{2\pi^2T}{\omega_c},\qquad
\theta_0=\frac{2\pi\mu_0}{\omega_c},
\end{gathered}
\end{equation}
and use the fixed-density normalization
$F_2=m^3\mu_0^2f_2/(4\pi)$.  Writing
\begin{equation}
{\cal F}_N(q)-{\cal F}_N(0)
=\frac{A_0}{2}r_Nq^2+\frac{A_0}{4}u_Nq^4+O(q^6),
\label{eq:Landau-fixedN-explicit}
\end{equation}
we use the same microscopic reference in both ensembles.
The comparison with the grand-canonical ensemble refers to the same isotropic
microscopic state, with $\mu=\mu_0$ and hence the same reference momentum
$k_{F0}^2=2m\mu_0$.  The change of ensemble is implemented by the Legendre
constraint and keeps the microscopic irrelevant operator fixed.
The parabolic smooth contribution is
\begin{align}
r_{N,{\rm par}}^{\rm sm}
&=\frac1{|F_2|}-1-\frac{\pi^2T^2}{3\mu_0^2}-\frac{\epsilon_0^2}{12},
\label{eq:rN-par-smooth}\\
u_{N,{\rm par}}^{\rm sm}
&=-\frac12\left(1+\frac{\epsilon_0^2}{12}\right)-\frac{\pi^2T^2}{2\mu_0^2}.
\label{eq:uN-par-smooth}
\end{align}
The density of states of the quadratic nematic saddle is proportional
to $1/\Delta$.  Its canonical Sommerfeld term is therefore
$-m\pi T^2/(12\Delta)$ and contributes to both Landau coefficients.  In particular, the Legendre
transform changes the zero-field parabolic quartic coefficient from the
grand-canonical value $-3/2$ to $-1/2$.

The direct oscillatory term and the canonical harmonic mixing can be kept
together by defining
\begin{align}
C_\ell(z)&=\frac{R_\ell(z)}{\ell^2}+K_\ell(z),
\label{eq:Cell-explicit-fixedN}
\end{align}
Here $C_\ell(z)\equiv{\cal C}^{(N)}_\ell(z)$,
$R_\ell(z)\equiv R_T(\ell z)$, and $K_\ell$ is given by
Eq.~\eqref{eq:Kell}. The first term is the direct first-order oscillatory
contribution and $K_\ell$ contains the second-order Legendre mixing derived
above. Direct expansion of
$\Delta C_\ell(\chi/\Delta)$ gives
\begin{align}
r_{N,{\rm par}}^{\rm osc}
={}&-\frac{\epsilon_0^2}{\pi^2}
\sum_{\ell=1}^{\infty}(-1)^\ell
\left[C_\ell-\chi C_\ell'\right]\cos(\ell\theta_0),
\label{eq:rN-par-osc-explicit}\\
u_{N,{\rm par}}^{\rm osc}
={}&\frac{\epsilon_0^2}{2\pi^2}
\sum_{\ell=1}^{\infty}(-1)^\ell
\left[\chi^2C_\ell''-C_\ell+\chi C_\ell'\right]
\cos(\ell\theta_0),
\label{eq:uN-par-osc-explicit}
\end{align}
where $C_\ell,C_\ell'$, and $C_\ell''$ are evaluated at $z=\chi$.
Equations~\eqref{eq:rN-par-osc-explicit} and
\eqref{eq:uN-par-osc-explicit} exhibit the explicit weak-field factor
$(\omega_c/\mu_0)^2$ multiplying the direct oscillatory contribution and
its second-order canonical mixing.

For the numerical phase diagrams we evaluate the canonical coefficients
directly from the finite-temperature Landau-level sums, solving the full
parabolic density constraint at each nematic amplitude. The radial
dispersion correction is Legendre transformed with the reference momentum
$k_{F0}^2=2m\mu_0$ held fixed. This procedure retains the complete parabolic
thermal occupations and all mixed oscillatory contributions at the
retained order of the dispersion correction. The explicit level-sum
formulas and density derivatives are given in
Appendices~\ref{app:fixedN-nematic} and~\ref{app:canonical-radial};
the numerical checks are given in Appendix~\ref{app:completion-convergence}.

The local coefficients combine the parabolic contribution and the radial
dispersion correction,
\begin{align*}
r_N&=r_{N,{\rm par}}+\delta r_{N,\gamma},\\
u_N&=u_{N,{\rm par}}+\delta u_{N,\gamma}.
\end{align*}
In the zero-field, zero-temperature limit these coefficients reduce to
\begin{equation}
r_N(0,0)=\frac1{|F_2|}-1,\qquad
u_N(0,0)=\frac{3\gamma-8}{16}.
\label{eq:rN-uN-zero}
\end{equation}
Thus the representative radial dispersion correction stabilizes the local
canonical saddle for $\gamma>8/3$. Compared with fixed chemical potential,
the density constraint both locks the rapid oscillation phase and
suppresses the oscillatory modulation of the local Landau coefficients.
These results provide the starting point for the direct ensemble
comparison in Sec.~\ref{sec:phase-diagram}.

The Landau expansion is a local diagnostic of stability near the isotropic
saddle. At fixed chemical potential, equilibrium competition at finite
amplitude is determined by global minimization of the unexpanded
finite-temperature potential specified in Sec.~\ref{sec:full-potential}.

\section{Mean-field phase diagram}
\label{sec:phase-diagram}
The equilibrium phase diagrams at fixed chemical potential and fixed density are obtained by minimizing the corresponding mean-field potentials. The Landau coefficients derived in Sec.~IV characterize the isotropic instability and its local nonlinear stability. At fixed chemical potential, Landau quantization can also produce competing finite-amplitude minima and first-order transitions, which are captured by the global minimization of the finite-temperature potential. At fixed density, phase locking suppresses this competition in the regime considered here.

\subsection{Fixed chemical potential: global equilibrium and local interpretation}
\label{sec:full-potential}
At fixed chemical potential we determine the equilibrium state from the
global minimum of
$\Delta\Omega_\mu(q)=\Omega_{\rm MF}(q)-\Omega_{\rm MF}(0)$.
The potential contains the parabolic fermionic determinant, the
Hubbard--Stratonovich term, and the radial dispersion correction introduced
in Sec.~IV. With $a=m\mu^2/(4\pi)$ and
$\epsilon_B=\omega_c/\mu$, the dimensionless potential is
\begin{align}
V_\mu(q)\equiv&\frac{\Delta\Omega_\mu(q)}{a}
 \nonumber \\
=&\frac{q^2}{2|F_2|}
-A_T\left(\frac{1}{\Delta}-1\right)
+\frac{\epsilon_B^2}{12}(\Delta-1)\nonumber\\
+&\epsilon_B^2[S(\Delta)-S(1)]
+\delta\omega_\gamma(q),
\label{eq:full-potential-mu}
\end{align}
with
\begin{align}
A_T&=1+\frac{\pi^2}{3}\left(\frac{T}{\mu}\right)^2,
\\
S(\Delta)&=\Delta\sum_{n=1}^{n_{\max}}
\frac{(-1)^n}{\pi^2n^2}
R_T(\chi_n/\Delta)\cos(\theta_n/\Delta).
\label{eq:full-potential-S}
\end{align}
Here $\Delta=\sqrt{1-q^2}$, and $n_{\max}$ is the numerical cutoff on the
dHvA harmonic sum. Calculations based on Eq.~\eqref{eq:full-potential-S} use
$n_{\max}=40$, for which the result is converged as documented in
Appendix~\ref{app:numerical-completion}. Neither
the rapidly varying dHvA phase nor the Lifshitz--Kosevich thermal factor is
expanded in $q$. The last term,
$\delta\omega_\gamma(q)\equiv
[\delta\Omega_\gamma(q)-\delta\Omega_\gamma(0)]/a$, is the dimensionless
matched finite-amplitude contribution of the radial dispersion correction, constructed in
Appendix~\ref{app:numerical-completion}.

It is useful to formulate the global minimization in a way that isolates
the dependence on the quadrupolar coupling. We write
\begin{equation}
V_\mu(q)=\frac{q^2}{2|F_2|}+K(q),
\label{eq:phase-V-K-decomposition}
\end{equation}
where $K(q)$ contains all contributions to the deformation energy that are
independent of $F_2$: the fermionic contribution, the smooth magnetic term,
the dHvA contribution, and the radial correction. We choose
$K(0)=0$, so that $V_\mu(0)=0$. 

For a given trial amplitude $q>0$, coexistence with the isotropic state
occurs when $V_\mu(q)=0$. Solving this condition for the interaction
strength gives
\begin{equation}
\frac{1}{|F_2|}=-\frac{2K(q)}{q^2}.
\end{equation}

This motivates defining the finite-amplitude threshold function
for $q>0$
\begin{equation}
\mathcal R(q)\equiv-\frac{2K(q)}{q^2},
\label{eq:phase-V-R}
\end{equation}
in such a way that, 
\begin{equation}
\mathcal R(0)\equiv
\lim_{q\to0^+}\mathcal R(q)={\cal X}_2.
\label{eq:phase-V-R-0}
\end{equation}
For $q>0$, $\mathcal R(q)$ is the value of $1/|F_2|$
at which a state with deformation $q$ becomes degenerate
with the isotropic state. 

The isotropic state is globally stable only if
$V_\mu(q)\geq0$ for every allowed deformation. Equivalently,
$1/|F_2|$ must lie above the threshold $\mathcal R(q)$ for every $q$.
As the attraction is increased, the first competing state to become
degenerate with the isotropic one is therefore the amplitude at which
$\mathcal R(q)$ is maximal. The equilibrium boundary is
\begin{equation}
\frac{1}{|F_{2,c}|}
=\sup_{0\leq q<1}\mathcal R(q).
\label{eq:full-potential-boundary}
\end{equation}
The location $q_*$ of this maximum simultaneously identifies the character
of the transition. If $q_*=0$, the isotropic minimum loses stability
continuously. If $q_*>0$, a finite-amplitude nematic minimum becomes
degenerate with the isotropic state before the latter loses local
stability, and the transition is first order.

\begin{figure}[t]
\centering
\includegraphics[width=\columnwidth]{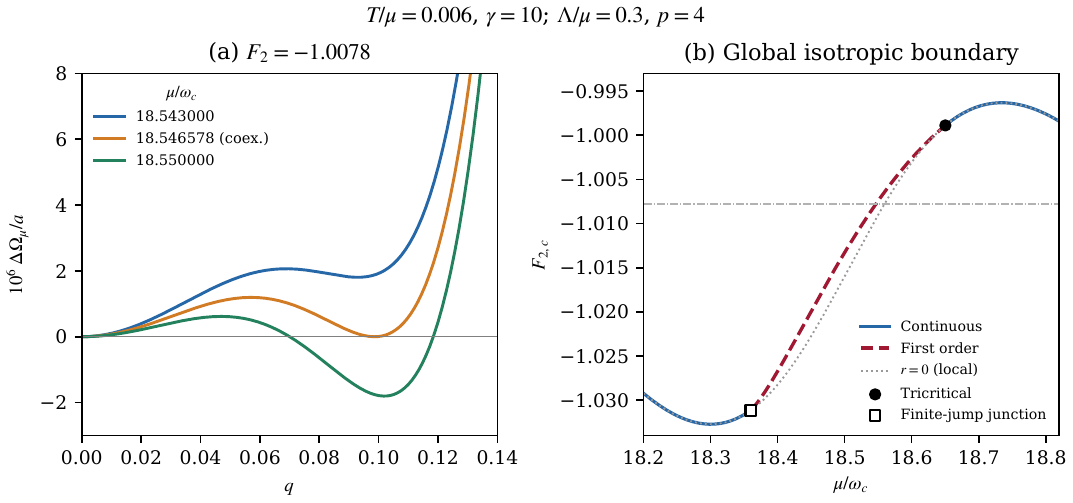}
\caption{Global fixed-$\mu$ phase structure at $T/\mu=0.006$ and
$\gamma=10$, with $\Lambda/\mu=0.3$ and $p=4$.
(a) Finite-amplitude potential at $F_2=-1.0078$ before, at, and after
coexistence at a weakly first-order transition. At coexistence the isotropic
and nematic minima are degenerate while the isotropic minimum remains
locally stable.
(b) Equilibrium isotropic boundary obtained from the global minimization:
solid blue segments are continuous and dashed red segments are first order;
the dotted curve is the local $r_\mu=0$ locus. The filled circle marks the
tricritical endpoint of the representative finite-amplitude continuation and the
open square the finite-jump junction on the left. The horizontal
dash-dotted line is the coupling used in panel (a).}
\label{fig:full-potential-first-order}
\end{figure}

Figure~\ref{fig:full-potential-first-order} illustrates
of this situation. At $T/\mu=0.006$, $\gamma=10$, and
$F_2=-1.0078$, the finite-amplitude potential develops a competing nematic
minimum. Coexistence occurs at $\mu/\omega_c\simeq18.5466$, with
$q_{\rm jump}\simeq0.0986$. The isotropic state is still a
local minimum at coexistence. 
Coexistence lies between the spinodal at which the metastable nematic
minimum appears and the spinodal at which the isotropic minimum loses local
stability.

Repeating the global minimization over successive inverse-field periods gives
the fixed-$\mu$ equilibrium boundary shown in
Fig.~\ref{fig:ensemble-comparison-cuts}(a). Landau quantization produces an
oscillatory modulation of $F_{2,c}$ around its zero-field value, so that a
horizontal cut at fixed $F_2$ can enter and leave the nematic phase repeatedly
as the magnetic field is varied. Parts of the boundary are continuous,
whereas at sufficiently low temperature finite-amplitude coexistence replaces
portions of the local instability line by first-order segments. 

The local Landau expansion of Sec.~\ref{sec:mean-field} explains the
structure of the global phase boundary. Close to the isotropic saddle, the
dimensionless potential in either ensemble has the form
\begin{align}
{\cal V}(q;T,B)-{\cal V}(0;T,B)
=&\frac{r(T,B)}{2}q^2+\frac{u(T,B)}{4}q^4 \nonumber\\
&+\frac{v(T,B)}{6}q^6+\cdots .
\label{eq:phase-V-Landau}
\end{align}
At fixed chemical potential,
\begin{equation}
r_\mu(T,B)=\frac{1}{|F_2|}-{\cal X}_2(T,B),
\end{equation}
so a locally continuous instability satisfies
\begin{equation}
r_\mu(T,B)=0,\qquad u_\mu(T,B)>0,
\label{eq:phase-V-continuous}
\end{equation}
or equivalently
\begin{equation}
F_{2,c}^{(\mu)}(T,B)=-\frac{1}{{\cal X}_2(T,B)}
\qquad\text{provided}\qquad u_\mu(T,B)>0.
\label{eq:phase-V-F2c-mu}
\end{equation}
On the ordered side, sufficiently close to such a continuous boundary,
$q_{{\rm sp},\mu}^2=-r_\mu/u_\mu+O(r_\mu^2)$.

Combining
Eq.~\eqref{eq:phase-V-K-decomposition} with the local expansion above and
using $r_\mu=1/|F_2|-{\cal X}_2$, the $F_2$-independent part of the
potential behaves for small $q$ as
\[
K(q)=-\frac{{\cal X}_2}{2}q^2
+\frac{u_\mu}{4}q^4
+\frac{v_\mu}{6}q^6+\cdots .
\]
Substitution into Eq.~\eqref{eq:phase-V-R} then gives
\[
\mathcal R(q)
={\cal X}_2-\frac{u_\mu}{2}q^2
-\frac{v_\mu}{3}q^4+\cdots ,
\]
which explicitly confirms the continuous extension at the
origin in Eq.~\eqref{eq:phase-V-R-0}.

The usual local condition $r_\mu=0$ is therefore the $q\to0$ limit of the global criterion
Eq.~\eqref{eq:full-potential-boundary}.  On the other hand, 
the quartic coefficient also acquires a simple global interpretation.
For $u_\mu>0$, $\mathcal R(q)$ initially decreases away from the origin, so
the first instability is continuous unless a more distant finite-$q$
maximum overtakes the one at $q=0$. For $u_\mu<0$, by contrast,
$\mathcal R(q)$ initially increases, signaling that a finite-amplitude
state can become competitive before the isotropic curvature vanishes.
Thus a negative $u_\mu$ is a local diagnostic of the finite-amplitude
competition seen in the global minimization.

The first-harmonic expressions in Eqs.~\eqref{eq:r-first-harmonic}
and~\eqref{eq:u-first-harmonic} explain this competition. Near a Landau-level
crossing, the leading oscillatory displacement of $r_\mu$ vanishes while
the reduction of $u_\mu$ is maximal. The field can thus make finite-amplitude
nematic configurations competitive while the isotropic state remains locally
stable, as in Fig.~\ref{fig:full-potential-first-order}.

The simultaneous local conditions
\begin{equation}
r_\mu(T,B)=u_\mu(T,B)=0
\label{eq:phase-V-tricritical-candidate}
\end{equation}
identify candidate tricritical points. These local conditions are necessary
but not sufficient for tricriticality. The sixth-order term
must stabilize the local expansion, and the representative finite-amplitude
potential must contain no lower competing minimum that preempts the local
instability.

Landau quantization strongly modulates the quartic coefficient $u_\mu$, producing regions with $u_\mu<0$ that favor first-order transitions when the potential is stabilized at higher order. These transitions persist under the finite-amplitude continuations tested. For the representative continuation used in the figures, the radial term $\delta\omega_\gamma$ provides the stabilization without additional higher-order terms. At larger $\mu/\omega_c$, the nematic jump vanishes at a tricritical endpoint with a positive sixth-order coefficient. At smaller $\mu/\omega_c$, a negative sixth-order coefficient allows a finite-amplitude minimum to preempt the local transition, producing a junction with a finite jump. The detailed endpoint structure depends on the chosen continuation (Appendix~\ref{app:global-endpoints}).

\subsection{Fixed density: phase locking and the canonical minimum}
At fixed density the equilibrium state minimizes the canonical free energy.
In the thermally damped regime, Eq.~\eqref{eq:FN-nematic-fixedN} includes
the direct oscillatory term and the second-order canonical mixing. For the
numerical phase boundary, the local coefficients are evaluated directly from
the finite-temperature Landau-level sums described in Sec.~IV and
Appendices~\ref{app:fixedN-nematic} and~\ref{app:canonical-radial}.

The phase locking in Eq.~\eqref{eq:phase-lock-fixedN} removes the dependence
of the rapid dHvA phase on the nematic amplitude.
Derivatives with respect to $q^2$ act on the smooth prefactor and the LK
factor, without the rapid-phase amplification present at fixed $\mu$.
The oscillatory corrections to $r_N$ and $u_N$ consequently carry the
explicit weak-field factor $(\omega_c/\mu_0)^2$ in
Eqs.~\eqref{eq:rN-par-osc-explicit} and
\eqref{eq:uN-par-osc-explicit}. Canonical mixing and the small
oscillation-induced shift of the saddle do not restore the order-one
quartic modulation of the grand-canonical ensemble.
The change of ensemble also modifies the smooth quartic coefficient. For
the representative radial dispersion correction,
Eq.~\eqref{eq:rN-uN-zero} gives
$u_N(0,0)=(3\gamma-8)/16$, and $u_N$ remains positive throughout the
regime displayed below. Its weak variation is shown in
Fig.~\ref{fig:ensemble-comparison-cuts}(d).

The phase transition therefore evolves continuously through the line defined by
\begin{equation}
r_N(T,B)=0,\qquad u_N(T,B)>0.
\label{eq:phase-V-canonical-condition}
\end{equation}
On the ordered side,
$q_{{\rm sp},N}^2=-r_N/u_N+O(r_N^2)$ within the local domain of control.
A direct numerical extremization of the canonical free energy finds no
competing lower uniform minimum over the parameter range used in the phase
diagrams. The details of this check are given in
Appendix~\ref{app:canonical-preemption-check}.

The canonical boundary still oscillates with inverse field, but phase locking
removes the strong grand-canonical amplification of the quartic sector and
produces the much smoother equilibrium boundary shown in
Fig.~\ref{fig:ensemble-comparison-cuts}(b).

\subsection{Direct comparison of the ensembles}
\begin{figure}
\centering
\includegraphics[width=8.5 cm]{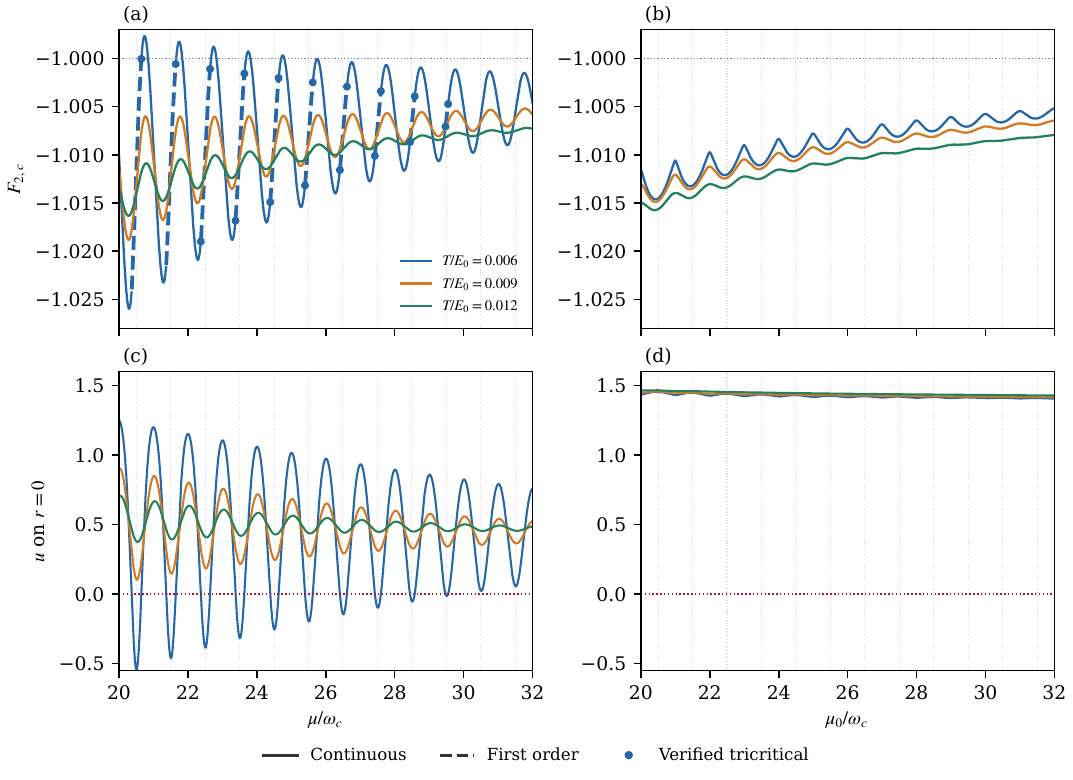}
\caption{Comparison of the equilibrium phase boundary and its local quartic
diagnostic for $\gamma=10$, with fixed chemical potential in the left
column and fixed density in the right column. The reference energy $E_0$
is defined in the text.
Panels (a),(b): $F_{2,c}$ as a function of inverse field for
$T/E_0=0.006$, $0.009$, and $0.012$, plotted with a common vertical scale.
At fixed $\mu$, solid segments are continuous and dashed segments are
first-order coexistence lines obtained from the global minimization. At
fixed density the boundary is continuous in the sampled regime, and the
finite-amplitude search finds no preempting uniform minimum. Filled circles
mark the $r_\mu=u_\mu=0$ candidates that also pass the positive-sixth-order
and global-minimum tests; preempted candidates are not marked as
tricritical. The dashed horizontal line marks the zero-field value
$F_{2,c}=-1$.
Panels (c),(d): the quartic coefficient evaluated on the local $r=0$ locus,
again on common scales. The red dashed line marks $u=0$. Vertical dotted
lines show the isotropic Landau-level crossings. The lower panels expose
the strong grand-canonical modulation of $u_\mu$ and the comparatively
smooth, strictly positive canonical coefficient $u_N$.}
\label{fig:ensemble-comparison-cuts}
\end{figure}

The phase diagrams reveal a pronounced ensemble dependence
that goes beyond a simple shift of the critical coupling.
Figure~\ref{fig:ensemble-comparison-cuts} compares the phase
boundaries on common scales, with $E_0=\mu$ at fixed chemical
potential and $E_0=\mu_0$ at fixed density. The fixed-$\mu$
boundary exhibits strong inverse-field modulation, including
first-order segments at low temperatures, whereas the
fixed-density boundary is substantially smoother and remains
continuous throughout the sampled regime. The lower panels
expose the corresponding contrast in nonlinear stability:
$u_\mu$ oscillates strongly and changes sign, while
$u_N>1.40$ in all three temperature cuts. These quartic
coefficients are evaluated along the local $r=0$ loci and
serve as stability diagnostics.

\begin{figure}[t]
\centering
\includegraphics[width=\columnwidth]{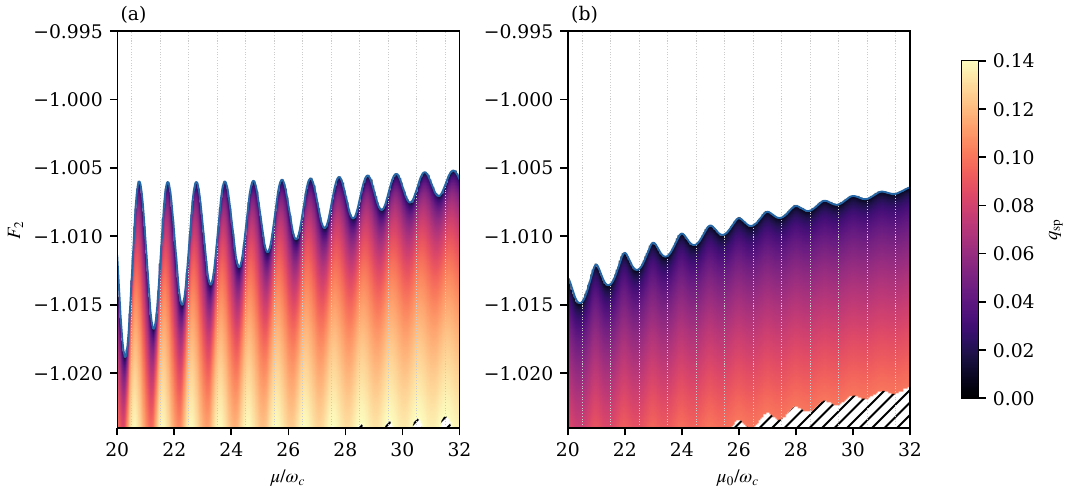}
\caption{Mean-field phase maps at $T/E_0=0.009$ and $\gamma=10$, at fixed
chemical potential in (a) and fixed density in (b). White denotes the
isotropic phase and the color gives the equilibrium nematic amplitude
within the corresponding controlled construction. In (a),
$q_{{\rm sp},\mu}$ is obtained from global minimization of the
finite-amplitude grand potential. In (b), the finite-amplitude check finds
no competing uniform minimum, so the canonical solution is represented by
$q_{{\rm sp},N}=\sqrt{-r_N/u_N}$ in the ordered region. Blue curves mark
the local $r=0$ loci. Diagonal hatching masks
$q_{{\rm sp},\mu}>0.14$ in (a), outside the stated small-deformation window
of the effective grand potential, and $q_{{\rm sp},N}^2>0.01$ in (b),
outside the local canonical bound. The fixed-$\mu$ phase-expansion
condition $q^2\mu/\omega_c\ll1$ is not imposed on the unexpanded global
potential. Vertical dotted lines indicate isotropic Landau-level crossings.
Identical axes and color scales expose the stronger grand-canonical
corrugation and the smoother canonical phase structure.}
\label{fig:phase-maps-ensembles}
\end{figure}

The difference extends to the structure of the ordered
regions, as shown by the phase maps in
Fig.~\ref{fig:phase-maps-ensembles}. At $T/E_0=0.009$,
the fixed-$\mu$ phase exhibits a more strongly corrugated
pattern of field-induced nematic order, whereas at fixed
density the ordered regions are smoother and more regularly
connected. The transitions at fixed $\mu$ are continuous
in this temperature cut; first-order segments emerge at
lower temperatures, as seen in
Fig.~\ref{fig:ensemble-comparison-cuts}. Thus, the
thermodynamic constraint controls not only the location
and order of the transition, but also the geometry and
connectivity of the reentrant nematic regions.

\subsection{Range of validity}
\label{sec:phase-validity}
The phase diagrams are obtained in the degenerate weak-field regime, $T/E_0,\omega_c/E_0\ll1$. The radial correction is treated perturbatively, requiring $\gamma[\max(T,\omega_c)/E_0]^2\ll1$, while the finite-amplitude analysis focuses on small nematic deformations, $\gamma q^2\lesssim1$. The masked regions in Fig.~\ref{fig:phase-maps-ensembles} impose more conservative bounds, which are satisfied by the representative results for $\gamma=10$.

The equilibrium phase boundaries are determined from the full finite-temperature potential at fixed chemical potential and from Landau-level sums at fixed density. The resulting phase diagrams provide the equilibrium nematic backgrounds for the magnetic and spectral responses analyzed in the following section.

\section{de Haas--van Alphen effect in the nematic phase}
\label{sec:dhva-nematic}
The equilibrium nematic deformation evolves with magnetic field and
temperature, modifying the dHvA response. We examine this feedback through
the magnetization and susceptibility at fixed chemical potential and fixed
density, distinguishing oscillation phase, cyclotron mass, thermal damping,
and harmonic content. Figures~\ref{fig:VI-cyclotron-mass-map}--\ref{fig:VI-selfconsistent-susceptibility}
compare the equilibrium response with the isotropic reference at the same
temperature, field window, and thermodynamic constraint.

\subsection{Magnetization at the equilibrium saddle}
Let ${\cal V}_X(q;T,B)$ denote the dimensional thermodynamic
potential density, with $X=\mu$ for the grand potential and
$X=N$ for the Helmholtz free energy. The canonical subscript
$N$ denotes fixed density $n$ at fixed physical area.
The dimensionless deformation potentials of Sec.~V are
multiplied by $A_\mu=a$ or $A_N=A_0$ when computing magnetic
responses. The equilibrium saddle satisfies
\begin{equation}
\left.\partial_q{\cal V}_X\right|_{q=q_{{\rm sp},X}}=0\; , 
\label{eq:VI-stationarity}
\end{equation}
and the potential at the saddle is 
\begin{equation}
{\cal V}_{X,{\rm eq}}={\cal V}_X(q_{{\rm sp},X}(T,B);T,B).
\end{equation}
Consequently,
\begin{align}
M_X
&=-\left.\frac{d{\cal V}_{X,{\rm eq}}}{dB}\right|_X\nonumber\\
&=-\left.\frac{\partial{\cal V}_X}{\partial B}\right|_{q,X}
-\left.\frac{\partial{\cal V}_X}{\partial q}\right|_{B,X}
\frac{dq_{{\rm sp},X}}{dB}\nonumber\\
&=-\left(\left.\frac{\partial{\cal V}_X}{\partial B}\right|_{q,X}
\right)_{q=q_{{\rm sp},X}} .
\label{eq:VI-envelope-M}
\end{align}
Stationarity eliminates the explicit saddle-derivative term
from the magnetization. The field dependence of the equilibrium
saddle nevertheless modulates the response through its
appearance in the remaining partial derivative. At a
first-order transition, the competing minima define separate
magnetization branches, which may exhibit a jump at coexistence.

To isolate the changes due to nematic order, we introduce an
isotropic reference evaluated at $q=0$,
with the same temperature, magnetic field, thermodynamic
constraint, and approximation:
\begin{equation}
M_X^{\rm iso}\equiv
-\left.\partial_B{\cal V}_X\right|_{q=0},
\label{eq:VI-isotropic-benchmarks}
\end{equation}
and
\begin{equation}
\delta M_X^{\rm nem}\equiv M_X-M_X^{\rm iso}.
\label{eq:deltaM}
\end{equation}
At fixed density, this subtraction separates the nematic contribution
from the harmonic mixing already present in the isotropic Fermi gas.

\subsection{Fixed chemical potential}
Using Eq.~\eqref{eq:VI-envelope-M}, the magnetization of the parabolic
model with the dispersion correction  is
\begin{equation}
M_\mu=M_{\mu,{\rm par}}^d(q_{{\rm sp},\mu})
+M_{\mu,{\rm par}}^{\rm osc}(q_{{\rm sp},\mu})
+\delta M_{\mu,\gamma}(q_{{\rm sp},\mu}).
\label{eq:VI-Mmu-final}
\end{equation}
The three terms are the smooth diamagnetic response, the dHvA oscillations,
and the leading radial-dispersion correction, respectively. They are given by
\begin{equation}
M_{\mu,{\rm par}}^d(q)=-\frac{\omega_c\Delta(q)}{24\pi},
\label{eq:VI-Mmu-diamagnetic}
\end{equation}
and
\begin{align}
M_{\mu,{\rm par}}^{\rm osc}(q)
={}&-\frac{\mu}{2\pi^2}\sum_{p=1}^{\infty}
\frac{(-1)^p}{p}R_T(z_p)
\Bigg\{\sin\phi_p\nonumber\\
&+\frac{\epsilon_q}{\pi p}
\left[1+\frac{z_p}{2}L(z_p)\right]\cos\phi_p\Bigg\}.
\label{eq:VI-Mmu-par}
\end{align}
Here
\begin{equation}
z_p=\frac{\chi_p}{\Delta},\qquad
\phi_p=\frac{\theta_p}{\Delta},\qquad
\epsilon_q=\frac{\omega_c\Delta}{\mu}.
\label{eq:VI-fixedmu-variables}
\end{equation}
The correction $\delta M_{\mu,\gamma}$ is given in
Appendix~\ref{app:fixedmu-magnetization}; the Hubbard--Stratonovich term
has no explicit field dependence.

The dominant sine term gives the dHvA oscillations, while the cosine term
retains the field derivatives of the prefactor and thermal damping.
For a field-independent saddle, the parabolic response is the free-gas
result with cyclotron mass $m_c=m/\Delta$.  At equilibrium,
$q_{{\rm sp},\mu}(B)$ modulates both the oscillation phase and the thermal
damping. Setting $q_{{\rm sp},\mu}=0$ gives the isotropic benchmark.
The nematic contribution evolves continuously across a continuous boundary;
first-order coexistence can instead produce a magnetization jump.

\subsection{Fixed density}
At fixed density the chemical potential adjusts with the magnetic field.
Using the equilibrium derivative in Eq.~\eqref{eq:VI-envelope-M}, the
canonical magnetization at the retained order in the radial dispersion
correction is
\begin{equation}
M_N=\left[M_{N,{\rm par}}^{\rm levels}(q)
+\delta M_{N,\gamma}(q)\right]_{q=q_{{\rm sp},N}}
+O(\gamma^2).
\label{eq:VI-MN-final}
\end{equation}
The two terms are evaluated at the equilibrium saddle of the same canonical
free energy, with the density constraint imposed.  Their explicit level-sum
forms are given in
Appendix~\ref{app:canonical-magnetization}.

In the thermally damped regime, the ensemble dependence is explicit in the harmonic representation.  The
parabolic response contains the nonoscillatory contribution
\begin{align}
M_{N,{\rm par}}^{\rm non}(q)
={}&-\frac{\omega_c\Delta(q)}{24\pi}\nonumber\\
&-\frac{\omega_c\Delta(q)}{8\pi^3}
\left[2D(\widetilde\chi)-\widetilde\chi D'(\widetilde\chi)\right].
\label{eq:VI-MN-non}
\end{align}
Its oscillatory contribution is
\begin{align}
M_{N,{\rm par}}^{\rm harm}(q)
={}&-\frac{\omega_c\Delta(q)}{4\pi^3}
\sum_{\ell=1}^{\infty}(-1)^\ell\Bigg\{
\left[2C_\ell-\widetilde\chi C_\ell'\right]\nonumber\\
&\quad\times\cos(\ell\theta_0)
+\ell\theta_0C_\ell\sin(\ell\theta_0)\Bigg\},
\label{eq:VI-MN-harm}
\end{align}
Here $C_\ell$ is the canonical harmonic coefficient defined in
Eq.~\eqref{eq:Cell-explicit-fixedN}, evaluated at the nematically
renormalized thermal argument $\widetilde\chi=\chi/\Delta(q)$, while
$\theta_0=2\pi\mu_0/\omega_c$ is the $q$-independent canonical phase
introduced in Sec.~\ref{sec:MF-fixed-density} and
Eq.~\eqref{eq:phase-lock-fixedN}.  The phase-independent part of the
second-order canonical mixing is denoted by
$D(z)\equiv z^2\sum_{p\geq1}\sinh^{-2}(pz)$.

Equations~\eqref{eq:VI-MN-non} and~\eqref{eq:VI-MN-harm} make the
fixed-density physics explicit.  The rapid phase $\ell\theta_0$ is
independent of the nematic amplitude, whereas $q$ enters the thermal factor
and the canonical harmonic coefficients through $\widetilde\chi$.
Nematicity therefore changes the thermal damping and redistributes harmonic
weight without shifting the  Onsager frequency.  The appropriate
isotropic canonical benchmark is obtained by setting $q=0$ while keeping the
same density constraint, so that the phase-independent term $D$ and the
canonical kernel $K_\ell$ are present on both sides of the comparison.

The harmonic form above applies in the thermally damped regime; the numerical
equilibrium response is evaluated from the full finite-temperature level
sums, with the density constraint and the radial dispersion correction
treated consistently as described in
Appendix~\ref{app:canonical-magnetization}.

Table~\ref{tab:VI-ensemble-comparison} summarizes the resulting contrast
between the two thermodynamic constraints.

\begin{table*}[t]
\caption{Quantum-oscillation signatures of a uniform parabolic nematic
background. Rows involving the equilibrium saddle incorporate its field
and/or temperature dependence. All statements refer to the thermally damped,
weak-field regime used in Secs.~IV and V.}
\label{tab:VI-ensemble-comparison}
\begin{ruledtabular}
\begin{tabular}{lcc}
Quantity & Fixed chemical potential & Fixed density \\
\hline
Rapid phase & $2\pi\ell\mu/[\omega_c\Delta(q)]$
& $2\pi\ell\mu_0/\omega_c$ \\
Onsager frequency (fixed deformation) & $\nu_{B,\mu}=m\mu/\Delta(q)$
& $\nu_{B,N}=m\mu_0$ \\
LK cyclotron mass & $m_c=m/\Delta(q)$
& $m_c=m/\Delta(q)$ \\
Equilibrium thermal factor &
$R_T\!\left[2\pi^2\ell Tm/(B\Delta(q_{{\rm sp},\mu}))\right]$
& $R_T\!\left[2\pi^2\ell Tm/(B\Delta(q_{{\rm sp},N}))\right]$
inside $C_\ell$ \\
Isotropic harmonic mixing & Grand-canonical LK harmonics
& Canonical kernel $K_\ell$ \\
Effect of $q_{\rm sp}(B)$ & Phase and amplitude modulation
& Predominantly amplitude modulation \\
Most distinctive signature & \shortstack{Local oscillation-frequency drift\\and spectral transfer}
& Frequency locking with modified LK damping \\
\end{tabular}
\end{ruledtabular}
\end{table*}

\subsection{Phase, local frequency, cyclotron mass, and harmonic content}
The nematic deformation affects different quantum-oscillation observables in
different ways.  We therefore distinguish the Onsager frequency of a rigid
background, the local oscillation frequency of a field-dependent equilibrium
state, and the cyclotron mass governing thermal damping; the harmonic content
provides a further ensemble-dependent diagnostic.

For a fixed deformation, the Onsager frequency is the coefficient conjugate
to \(1/B\) in the oscillation phase.  In the parabolic model,
\begin{equation}
\nu_{B,\mu}(q)=\frac{m\mu}{\Delta(q)},\qquad
\nu_{B,N}=m\mu_0 .
\label{eq:VI-frequency-comparison}
\end{equation}
We define the shift from the isotropic reference by
$\delta\nu_{B,X}\equiv\nu_{B,X}-\nu_{B,X}^{\rm iso}$, with
$\nu_{B,\mu}^{\rm iso}=m\mu$ and $\nu_{B,N}^{\rm iso}=m\mu_0$. For small
order parameter,
\begin{equation}
\frac{\delta\nu_{B,\mu}}{\nu_{B,\mu}^{\rm iso}}
=\frac{q^2}{2}+O(q^4),\qquad
\delta\nu_{B,N}=0,
\label{eq:VI-smallq-frequency}
\end{equation}
whereas the cyclotron mass in both ensembles is
\begin{equation}
m_c(q)=\frac{m}{\Delta(q)}.
\label{eq:VI-cyclotron-mass-comparison}
\end{equation}
Thus fixed density separates the oscillation frequency from the cyclotron
mass: nematicity changes the thermal scale while the  Onsager
frequency remains locked.  At fixed chemical potential, by contrast, the
same factor \(1/\Delta(q)\) enters both quantities.

The distinction becomes sharper when the nematic deformation follows the
magnetic field.  With \(y=1/B\), the fixed-\(\mu\) phase of harmonic \(\ell\)
and its local oscillation frequency are
\begin{align}
\Phi_{\mu,\ell}(y)
&=\frac{2\pi\ell m\mu y}{\Delta[q_{{\rm sp},\mu}(y)]},
\nonumber\\
\nu_{B,\mu}^{\rm loc}(y)
&=\frac{1}{2\pi\ell}\frac{d\Phi_{\mu,\ell}}{dy}
=m\mu\left[
\frac{1}{\Delta(q_{{\rm sp},\mu})}
+y\frac{d}{dy}\frac{1}{\Delta(q_{{\rm sp},\mu})}
\right].
\label{eq:VI-local-frequency}
\end{align}
Because \(q_{{\rm sp},\mu}\) is field dependent, this phase is not generally
linear in \(1/B\).  The second term in Eq.~\eqref{eq:VI-local-frequency} is
absent for a rigid background and measures the field dependence of the
equilibrium nematic deformation.  At fixed density,
\begin{equation*}
\Phi_{N,\ell}=\frac{2\pi\ell\mu_0}{\omega_c},\qquad
\nu_{B,N}^{\rm loc}=m\mu_0
\end{equation*}
within the parabolic canonical theory.  The saddle can still modulate
amplitudes and harmonic weights, but it does not generate the corresponding
phase drift.  A local frequency defined from the phase should not be
identified with the maximum of a finite-window Fourier spectrum, which can
also move because of amplitude modulation, interference, and the finite
analysis window.  Further phase and Landau-fan details are given in
Appendices~\ref{app:magnetization} and~\ref{app:selfconsistent-spectrum}.

At equilibrium, the order parameter also changes with temperature.  The LK
factor therefore becomes
\begin{align}
R_{\ell,X}^{\rm eq}(T,B)
&=R_T\!\left[z_{\ell,X}^{\rm eq}(T,B)\right],
\nonumber\\
z_{\ell,X}^{\rm eq}(T,B)
&=\frac{2\pi^2\ell Tm}
{B\Delta[q_{{\rm sp},X}(T,B)]},
\qquad X=\mu,N .
\label{eq:VI-selfconsistent-LK}
\end{align}
For \(R_T(z)=z/\sinh z\), the logarithmic temperature derivative at fixed
field and fixed thermodynamic constraint is
\begin{equation}
\begin{aligned}
\frac{d\ln R_{\ell,X}^{\rm eq}}{d\ln T}
={}&\left[1-z_{\ell,X}^{\rm eq}\coth z_{\ell,X}^{\rm eq}\right]\\
&\times\left[1-\frac{d\ln\Delta[q_{{\rm sp},X}(T,B)]}{d\ln T}\right].
\end{aligned}
\label{eq:VI-selfconsistent-LK-slope}
\end{equation}
The first factor is the usual rigid-background LK slope, while the second
reflects the thermal evolution of the nematic state.  Consequently, a
temperature fit across a region in which \(q_{{\rm sp},X}\) changes
appreciably need not measure a field-independent mass \(m/\Delta\): the
temperature dependence of the order parameter also enters the oscillation
amplitude.  A continuous boundary can therefore produce a kink in the
LK-factor slope.

\begin{figure}[t]
\centering
\includegraphics[width=\columnwidth]
{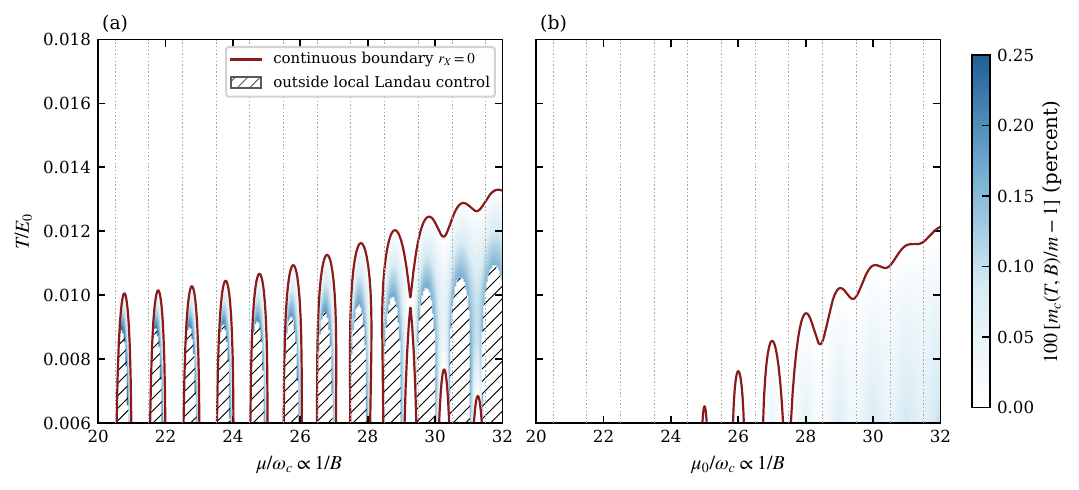}
\caption{Self-consistent cyclotron-mass enhancement in the
temperature--inverse-field plane at \(F_2=-1.008\) and \(\gamma=10\):
(a) fixed chemical potential and (b) fixed density.  The color scale shows
\(100[m_c(T,B)/m-1]\), with
\(m_c/m=1/\sqrt{1-q_{{\rm sp},X}^2(T,B)}\).
Here $q_{{\rm sp},X}^2=-r_X/u_X$ in the ordered phase, so this local-Landau
map uses the phase-expansion bound at fixed $\mu$, unlike the unexpanded
global map in Fig.~\ref{fig:phase-maps-ensembles}(a). Red curves mark the continuous
boundaries \(r_X=0\), and vertical dotted lines mark the isotropic
Landau-level crossings.  Hatched regions are ordered solutions outside the
conservative local-Landau bounds
\(q_{{\rm sp},\mu}^2\mu/\omega_c\leq0.1\) and
\(q_{{\rm sp},N}^2\leq0.01\); they are not assigned a color.  White regions
inside the control domain are isotropic and have \(m_c=m\).  The plotted mass
is the parabolic-saddle LK mass; explicit perturbative corrections from the
irrelevant operator are omitted.}
\label{fig:VI-cyclotron-mass-map}
\end{figure}

Figure~\ref{fig:VI-cyclotron-mass-map} shows the resulting equilibrium
cyclotron-mass enhancement in the two ensembles.

The leading parabolic harmonics provide the corresponding amplitude and
spectral diagnostics:
\begin{align}
M_{\mu,\ell}^{\rm osc}(q)
&=-\frac{\mu}{2\pi^2}\frac{(-1)^\ell}{\ell}
R_T\!\left(\frac{\chi_\ell}{\Delta}\right)
\sin\!\left(\frac{\theta_\ell}{\Delta}\right)
+O\!\left(\frac{\omega_c}{\mu}\right),
\label{eq:VI-Mmu-leading-comparison}\\
M_{N,\ell}^{\rm harm}(q)
&=-\frac{\mu_0\Delta}{2\pi^2}(-1)^\ell\ell
C_\ell(\widetilde\chi)\sin(\ell\theta_0)
+O\!\left(\frac{\omega_c}{\mu_0}\right).
\label{eq:VI-MN-leading-comparison}
\end{align}
These expressions make explicit the phase-and-amplitude modulation at fixed
$\mu$ and the amplitude-side nematic response at fixed density. In the latter
case, $C_\ell$ already contains the canonical isotropic kernel $K_\ell$, so
the reference must be the canonical $q=0$ spectrum.

\begin{figure}[t]
\centering
\includegraphics[width=8.5 cm]
{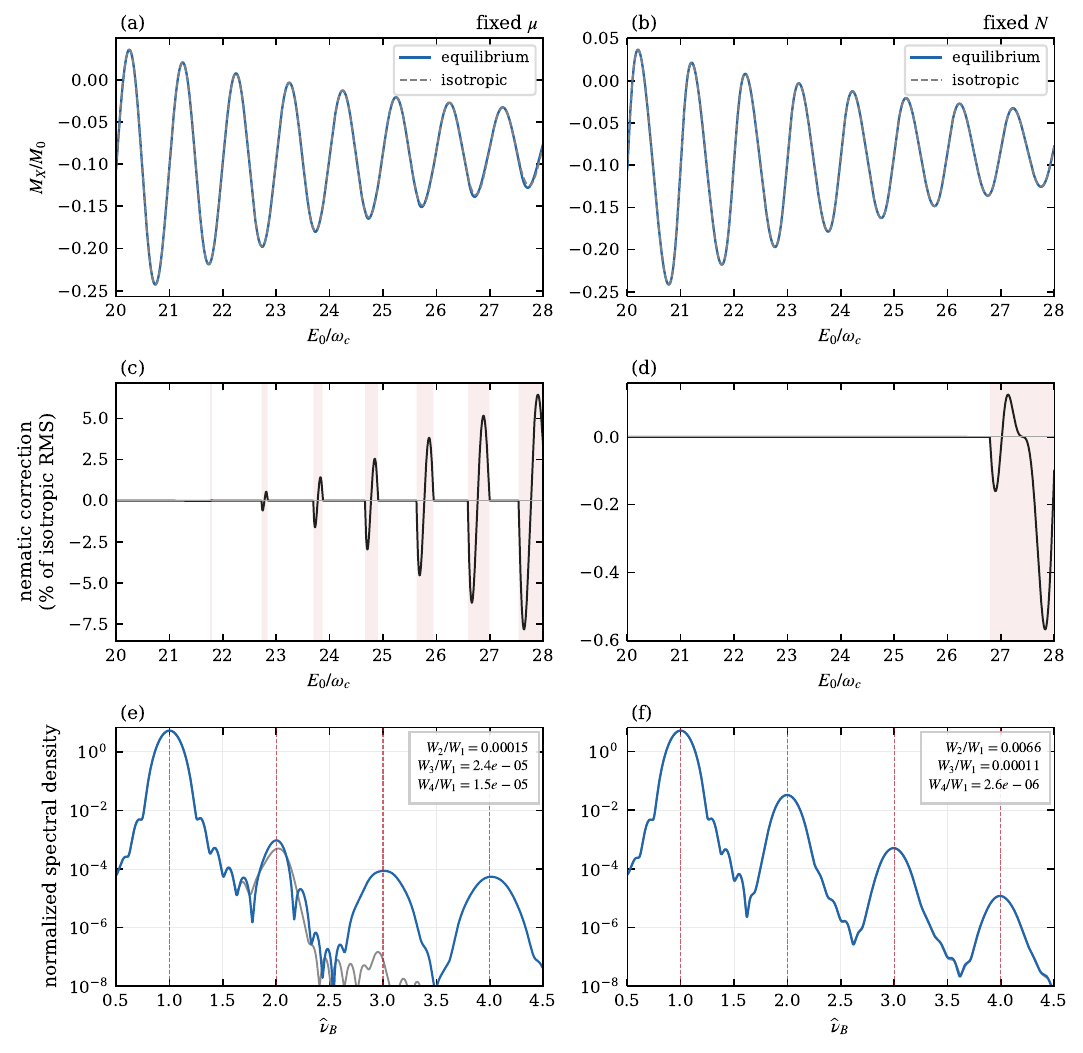}
\caption{Equilibrium magnetization and finite-window Fourier spectra
from the thermodynamic potential. We use $20\leq E_0/\omega_c\leq28$,
$T/E_0=0.010$, $\gamma=10$, $F_2=-1.0078$ at fixed chemical potential, and
$F_2=-1.009$ at fixed density; the different couplings keep both examples
inside their control domains.  The left and right columns show fixed $\mu$
and fixed density, respectively.  (a),(b) Full equilibrium magnetization
(blue) and isotropic reference (gray dashed), in units of
$M_0=E_0/(2\pi^2)$.  (c),(d) Direct nematic correction, normalized to the
isotropic RMS reference defined in Appendix~\ref{app:numerical-spectral}; it
vanishes identically in the isotropic regions.  Shading marks ordered
intervals.  (e),(f) Power spectra of the equilibrium and isotropic
magnetizations.  Here $\widehat\nu_B$ denotes the Fourier frequency normalized
to the isotropic fundamental, and $W_\ell$ denotes the integrated spectral
weight in the window around the $\ell$th harmonic.  Dashed lines mark the
first four harmonics, and boxes give the equilibrium ratios $W_\ell/W_1$.
The response includes the isotropic level sum and field derivatives of the
minimized deformation potential.  Detailed numerical and spectral conventions
are given in Appendix~\ref{app:numerical-spectral}.}
\label{fig:VI-spectral-transfer}
\end{figure}

The equilibrium spectra in Fig.~\ref{fig:VI-spectral-transfer} are obtained
by differentiating the same thermodynamic potential whose minimum determines
\(q_{{\rm sp},X}(B)\).  The numerical differentiation, normalization,
windowing procedure, and definitions of the integrated harmonic weights are
given in Appendix~\ref{app:numerical-spectral}.

Figure~\ref{fig:VI-spectral-transfer} illustrates the resulting difference in
the equilibrium waveform.  At fixed chemical potential, the field-dependent
nematic saddle produces visible phase modulation and transfers spectral
weight to higher harmonics.  At fixed density, the Onsager phase remains
locked and the equilibrium spectrum stays much closer to its canonical
isotropic reference; most of the higher-harmonic background in this example
is already generated by the density constraint itself.  The numerical harmonic
ratios are reported in Appendix~\ref{app:numerical-spectral}.

These examples lie in the thermally damped continuous regime of
Sec.~\ref{sec:full-potential} and Appendix~\ref{app:canonical-preemption-check},
within the control conditions of Sec.~\ref{sec:phase-validity}; they do not
cross the low-temperature first-order segments.  At a fixed-\(\mu\)
first-order boundary, a jump in \(q_{{\rm sp},\mu}\) produces finite changes
in the parabolic phase, LK mass, and harmonic amplitudes.

\subsection{Magnetic susceptibility and boundary response}
The magnetic susceptibility probes an additional aspect of the
self-consistent response.  Stationarity removes an explicit
\(dq_{{\rm sp},X}/dB\) term from the magnetization, but the field dependence
of the equilibrium saddle reappears upon differentiating once more:
\begin{equation*}
\chi_X=\left.\frac{dM_X}{dB}\right|_X
=-\left.\frac{d^2{\cal V}_{X,{\rm eq}}}{dB^2}\right|_X,
\qquad X=\mu,N.
\end{equation*}
Differentiating the stationarity condition gives
\begin{equation}
\chi_X=
-\partial_B^2{\cal V}_X
+\frac{(\partial_B\partial_q{\cal V}_X)^2}
{\partial_q^2{\cal V}_X}
\bigg|_{q=q_{{\rm sp},X}} .
\label{eq:VI-chi-envelope}
\end{equation}
The first term is the fixed-saddle response.  The second describes the
field-induced shift of the equilibrium saddle; it is nonnegative on a
stable branch and vanishes at the isotropic saddle because \({\cal V}_X\) is
even in \(q\).  Equation~\eqref{eq:VI-chi-envelope} applies on a smooth stable
branch with \(\partial_q^2{\cal V}_X>0\).  Its derivation is given in
Appendix~\ref{app:equilibrium-susceptibility}.

At a continuous nematic boundary, this gives a finite susceptibility step.
The local ordered contribution to the equilibrium potential is
\begin{equation*}
\delta{\cal V}_{X,{\rm eq}}^{\rm ord}
=-\frac{A_Xr_X^2}{4u_X}\Theta(-r_X),
\end{equation*}
and hence
\begin{equation}
\delta\chi_X^{\rm ord}
=A_X\partial_B^2\left(\frac{r_X^2}{4u_X}\right)
\Theta(-r_X).
\label{eq:VI-chi-boundary}
\end{equation}
Although the corresponding nematic contribution to the magnetization
vanishes continuously as \(r_X\to0\), its field derivative does not.  For a
generic crossing with \(u_X>0\), the ordered-side limit is
\(A_X(\partial_Br_X)^2/(2u_X)\), whereas the isotropic-side contribution
vanishes.  The equilibrium susceptibility therefore exhibits a finite
mean-field step at a continuous nematic transition, understood as a one-sided
limit of Eq.~\eqref{eq:VI-chi-envelope}.  At a first-order boundary that
equation applies separately to each smooth stable branch; the transition
itself is characterized by the magnetization jump between the competing
minima.

\begin{figure}[t]
\centering
\includegraphics[width=8.5 cm]
{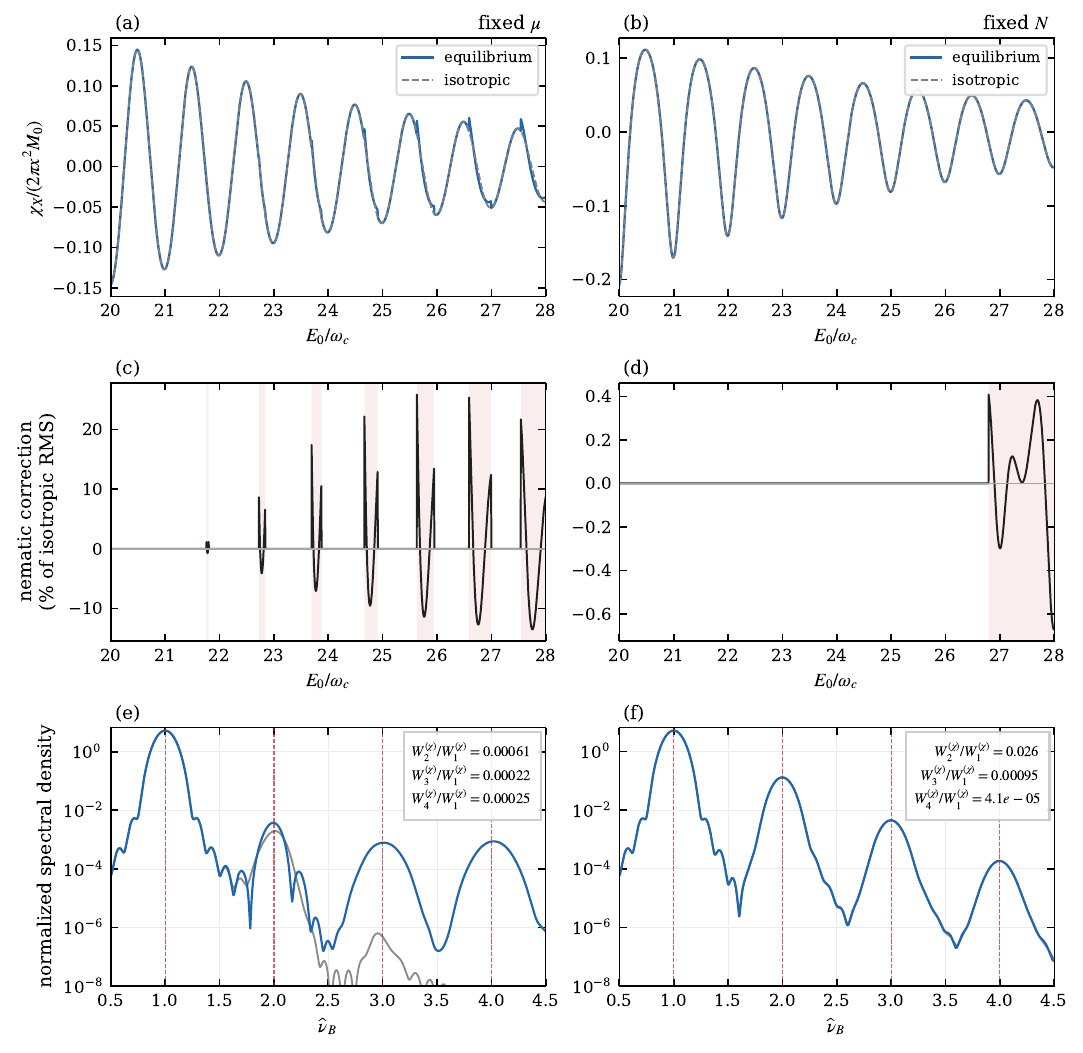}
\caption{Equilibrium scaled susceptibility and its Fourier spectrum from the
same thermodynamic potential as Fig.~\ref{fig:VI-spectral-transfer}.
The left and right columns show fixed $\mu$ and fixed density, respectively.
(a),(b) Full equilibrium susceptibility (blue) and isotropic reference (gray
dashed), divided by $2\pi x^2M_0$, with $x=E_0/\omega_c$ and
$M_0=E_0/(2\pi^2)$.
(c),(d) Direct equilibrium-minus-isotropic difference of these scaled
responses, expressed as a percentage of the RMS of the scaled isotropic
response over the displayed field window.  The correction is zero in the
isotropic regions and turns on at the continuous nematic boundary; shading
marks ordered intervals.  (e),(f) Fourier spectra of the equilibrium and
isotropic scaled susceptibilities.  Here $\widehat\nu_B$ denotes the Fourier
frequency normalized to the isotropic fundamental, and $W_\ell^{(\chi)}$
denotes the integrated susceptibility spectral weight in the window around
the $\ell$th harmonic.  Dashed lines mark the first four harmonics, and boxes
give $W_\ell^{(\chi)}/W_1^{(\chi)}$ for the equilibrium response.  Parameters
and ensemble-specific couplings are those of
Fig.~\ref{fig:VI-spectral-transfer}.  The susceptibility includes both the
explicit field dependence of the thermodynamic potential and the field
dependence of the equilibrium saddle.  Detailed differentiation,
normalization, and spectral conventions are given in
Appendix~\ref{app:numerical-spectral}.}
\label{fig:VI-selfconsistent-susceptibility}
\end{figure}

Figure~\ref{fig:VI-selfconsistent-susceptibility} displays this boundary
response together with the complete equilibrium susceptibility and its
finite-window spectrum.

Beyond the boundary anomaly, the additional field derivative also enhances
the oscillatory response.  For the parabolic harmonics, the leading rigid
contribution is
\begin{equation}
\chi_{X,\ell}^{\rm osc,rig}(q)
=-\frac{(-1)^\ell}{B}\Phi_{X,\ell}S_{X,\ell}
\cos\Phi_{X,\ell}
+O\!\left(\frac{S_{X,\ell}}{B}\right).
\label{eq:VI-chi-leading}
\end{equation}
Here \(S_{X,\ell}\) is the corresponding sine coefficient of the
magnetization harmonic and \(\Phi_{X,\ell}\) is the oscillation phase defined
in the preceding subsection.  The field derivative rotates the dominant
harmonic by approximately \(\pi/2\) and amplifies its rapidly varying part by
the large semiclassical factor \(\Phi_{X,\ell}\sim\nu_B/B\).

For a locally periodic magnetization harmonic with amplitude
\({\cal A}_{X,\ell}\),
\begin{equation}
{\cal A}^{(\chi)}_{X,\ell}\simeq
2\pi y^2\ell\nu_B\,{\cal A}_{X,\ell},
\qquad y=\frac{1}{B}.
\label{eq:VI-chi-spectral-enhancement}
\end{equation}
Here $\nu_B$ denotes the corresponding local fundamental frequency. Thus
susceptibility power ratios acquire approximately an additional factor
\(\ell^2\) relative to the corresponding magnetization ratios, apart from slow
envelope derivatives and the field dependence of the saddle.  The harmonic
form has the same thermally damped scope as the magnetization expansion; the
nonuniform low-temperature limit is discussed in
Appendices~\ref{app:free-gas-canonical} and~\ref{app:Landau-expansion}.

The numerical results in Fig.~\ref{fig:VI-selfconsistent-susceptibility}
display both the predicted boundary step and the derivative-enhanced higher
harmonics. The spectral redistribution is stronger at fixed chemical
potential, whereas the fixed-density correction remains smaller relative to
the canonical isotropic background. Numerical differentiation, normalization,
and Fourier conventions are given in Appendix~\ref{app:numerical-spectral}.

\section{Discussion and Conclusions}
\label{sec:discussion}
Landau quantization and nematic order are coupled through a
self-consistent orbital feedback mechanism. Near a Pomeranchuk
instability, the magnetic field can stabilize reentrant nematic
regions, while the resulting Fermi-surface deformation modifies
the de Haas--van Alphen response. The phase diagram and the
quantum oscillations are thus two manifestations of the same
underlying mechanism.

The equilibrium phase diagram depends strongly on the
thermodynamic constraint. At fixed chemical potential, Landau
quantization can produce both continuous and first-order
nematic transitions. At fixed density, phase locking suppresses
the nonlinear amplification, and the transitions remain
continuous in the regime studied. These transitions have
distinct magnetic signatures: the susceptibility exhibits a
finite mean-field step at continuous boundaries, while
first-order coexistence can produce a jump in magnetization.

The thermodynamic constraint also determines how nematicity
affects quantum oscillations. At fixed chemical potential,
the field-dependent order parameter modulates the oscillation
phase, producing local-frequency drift and spectral-weight
redistribution without a comparable enhancement of the
cyclotron mass. At fixed density, conservation of the
Fermi-surface area locks the leading  Onsager
frequency, while the cyclotron mass, thermal damping, and
harmonic amplitudes continue to evolve. In this case,
nematic spectral changes must be compared with the canonical
isotropic response, which already contains harmonic mixing.

These results suggest that the combined evolution of
oscillation frequencies, cyclotron masses, harmonic content,
and magnetic response provides a useful signature of
self-consistent electronic order. The ruthenate
Sr$_3$Ru$_2$O$_7$ is a relevant example: pronounced
field-dependent oscillation frequencies have been observed
near the metamagnetic regime~\cite{Mercure2010}. Our results
offer a possible mechanism for such behavior, while multiband
effects may produce responses intermediate between the two
thermodynamic limits considered here. Similar investigations
may be useful in other correlated metals with nematic
tendencies, including FeSe$_{1-x}$S$_x$~\cite{Coldea2019}.

The uniform mean-field approach captures the essential orbital
feedback mechanism. In two dimensions, orientational
fluctuations and lattice anisotropy can modify the critical
behavior~\cite{MeWa1966,KoTh1973,BaSt2009,JoKaKiNe1977}.
Extending the present framework to include these effects and
realistic multiband structures offers a natural route toward
understanding the interplay between field-induced electronic
order and quantum oscillations in correlated metals.

\section*{ACKNOWLEDGMENTS}
We acknowledge partial financial support from the Brazilian agencies
{\em Funda\c c\~ao Carlos Chagas Filho de Amparo \`a Pesquisa do Estado do Rio
de Janeiro} (FAPERJ), {\em Conselho Nacional de Desenvolvimento Cient\'\i fico
e Tecnol\'ogico} (CNPq), and {\em Coordena\c c\~ao de Aperfei\c coamento de
Pessoal de N\'\i vel Superior} (CAPES), Finance Code 001.  We also acknowledge
support from the INCT project Advanced Quantum Materials through CNPq
(Proc. 408766/2024-7), FAPESP (Proc. 2025/27091-3), and CAPES.  L.E.L.M.
acknowledges a CAPES Ph.D. fellowship.

\vspace*{0.2 cm}
\noindent\textit{Data availability.---}
A reproducibility package containing the numerical code and data required
to reproduce the results of this work is publicly available on
Zenodo~\cite{MacedoBarciCode}.

\appendix

\section{Grand-canonical free gas and Poisson summation}
\label{app:free-gas-poisson}
This appendix gives the derivation of the fixed-\(\mu\) benchmark summarized
in Sec.~\ref{sec:free-benchmarks}.

The grand-canonical partition function is given by the coherent-state functional integral
\begin{align}
Z_0 = \int \mathcal{D}\psi^\dagger \mathcal{D}\psi\;
e^{-\int d^2x \int_{0}^{\beta} d\tau\;
\psi^\dagger\left[\partial_{\tau}+\varepsilon\left({\bf D}\right)-\mu\right]\psi},
\label{eq:PF-Gas}
\end{align}
where $\varepsilon({\bf D}) = {\bf D}^2/(2m)$ and ${\bf A}$ is taken in the symmetric gauge.
Integrating out the fermionic fields formally yields
\begin{align}
Z_0 = \det\bigl(\partial_{\tau}+\varepsilon\left({\bf D}\right)-\mu\bigr)
= e^{\operatorname{Tr}\ln\left(\partial_{\tau}+\varepsilon\left({\bf D}\right)-\mu\right)} .
\end{align}

The grand-potential density is defined as
\begin{align}
\Omega(\beta, B, \mu) &= -\frac{1}{\beta L^2}\ln Z_0 \nonumber \\
&= -\frac{1}{\beta L^2}\operatorname{Tr}\ln\bigl(\partial_{\tau}+\varepsilon\left({\bf D}\right)-\mu\bigr).
\label{eq:Omega-free}
\end{align}
Here $L^2$ is the  area of the two-dimensional sample.
The trace can be evaluated exactly by expanding in the basis of Landau-level eigenstates, leading to
\begin{align}
\Omega(\beta, B, \mu)
&= -\frac{{\cal D}_B}{\beta}\sum_{\lambda=0}^{\infty}
\ln\!\Bigl(1+e^{-\beta \left(E_{\lambda}-\mu\right) }\Bigr),
\label{eq:FED-summed}
\end{align}
where the single-particle spectrum consists of highly degenerate Landau levels with energies
\begin{equation}
E_\lambda = \left(\lambda+\frac{1}{2}\right)\omega_c,
\label{eq:Landau-Spectrum}
\end{equation}
where $\omega_c=B/m$ is the cyclotron frequency. The degeneracy per unit area of each Landau level is
\begin{equation}
{\cal D}_B = \frac{B}{2\pi}.
\label{eq:Degeneracy}
\end{equation}
It is also useful to define the zero-field density of states per unit area,
\begin{equation}
\rho_0=\frac{m}{2\pi},
\label{eq:rho0}
\end{equation}
so that ${\cal D}_B=\rho_0\omega_c$.

Equation~\eqref{eq:FED-summed} can be conveniently rewritten as an energy integral involving the density of states:
\begin{equation}
\Omega(\beta, B, \mu) = -\frac{1}{\beta } \int_{0}^{\infty} d\epsilon \; \rho(\epsilon)\;
\ln\!\Bigl( 1+e^{-\beta\left(\epsilon-\mu \right)} \Bigr),
\label{eq:FED-rho}
\end{equation}
where the density of states for the clean system is
\begin{equation}
\rho(\epsilon) = {\cal D}_B\sum_{\lambda=0}^\infty \delta\!\left(\epsilon-E_\lambda\right).
\label{eq:rho-free}
\end{equation}

Applying the Poisson summation formula we obtain the exact representation
\begin{equation}
\rho(\epsilon) = \frac{m}{2\pi} \left[ 1 + 2 \sum_{k=1}^{\infty} (-1)^k \cos\!\left( \frac{2\pi k}{\omega_c } \epsilon \right) \right],
\label{eq:DE}
\end{equation}
which isolates the smooth ($k=0$) and oscillatory ($k \neq 0$) contributions.

Substituting Eq.~\eqref{eq:DE} into Eq.~\eqref{eq:FED-rho} and integrating by parts, we find, in the degenerate Fermi regime $\beta\mu\gg1$ and up to corrections exponentially small in $\beta\mu$,
\begin{align}
\Omega = \Omega_0 + \Omega^{\rm non} + \Omega^{\rm osc},
\label{eq:Free-Energy-0}
\end{align}
with the three distinct contributions given by
\begin{align}
\Omega_0 &= -\frac{m\mu^2}{4\pi}\left[1+\frac{\pi^2}{3} \frac{1}{(\beta\mu)^2}\right], \label{eq:Omega0} \\[4pt]
\Omega^{\rm non} &= \frac{m\mu^2}{48\pi}\left(\frac{\omega_c}{\mu}\right)^2,
\label{eq:Omeganon} \\[4pt]
\Omega^{\rm osc} &= \frac{m\mu^2}{4\pi}\left(\frac{\omega_c}{\mu}\right)^2\sum_{n=1}^\infty
\frac{(-1)^n}{n^2\pi^2}\, R_T(\chi_n)\,
\cos\!\left(\frac{2n\pi \mu}{\omega_c}\right).
\label{eq:Omegaosc}
\end{align}
Equation~\eqref{eq:Omega0} is the standard grand potential of a two-dimensional free Fermi gas; it includes the quadratic temperature correction that guarantees a linear-in-$T$ specific heat at low temperatures.
The second term, $\Omega^{\rm non}$ [Eq.~\eqref{eq:Omeganon}], is the
nonoscillatory diamagnetic contribution, which is quadratic in the magnetic
field.
The last term, Eq.~\eqref{eq:Omegaosc}, encodes the quantum oscillations of the grand potential.
In Eq.~\eqref{eq:Omegaosc}, $R_T(\chi_n)$ is the familiar
Lifshitz--Kosevich thermal damping factor,
\begin{equation}
R_T(\chi_n) = \frac{\chi_n}{\sinh(\chi_n)},
\end{equation}
with
\begin{equation}
\chi_n = \frac{2n\pi^2}{\omega_c\beta}.
\end{equation}

The phase in Eq.~\eqref{eq:Omegaosc} gives the Onsager frequency
$\nu_B=A_F/(2\pi)=m\mu$ of the present two-dimensional parabolic band, where
$A_F$ is the zero-field Fermi-surface area, while $R_T(\chi_n)$ gives the
cyclotron mass $m_c=m$.  The relative amplitudes and phases determine the
harmonic content of the waveform.  Disorder and spin splitting, which would
introduce additional damping and phase factors, are absent from the clean
spinless model considered here.

Equations~\eqref{eq:Omega0}--\eqref{eq:Omegaosc} require the degenerate
regime $\beta\mu\gg1$, in which corrections from the lower band edge are
exponentially small.  Throughout the dHvA analysis we also assume
$\omega_c/\mu\ll1$, so that many Landau levels are occupied.  The ratio
$T/\omega_c$ remains arbitrary at this stage and enters through
$\chi_n=2\pi^2n/(\beta\omega_c)$ and $R_T(\chi_n)$.

From the grand potential we compute the magnetization at fixed chemical potential,
\begin{equation}
M = -\left.\frac{\partial \Omega}{\partial B}\right|_{\beta,\mu}.
\end{equation}
It naturally decomposes into two parts, $M = M^d + M^{\rm osc}$.
The smooth diamagnetic contribution is
\begin{equation}
M^d = -\frac{\omega_c}{24\pi},
\end{equation}
while the oscillatory part can be written as
\begin{align}
M^{\rm osc} &= -\frac{\mu}{2\pi^2}\sum_{n=1}^{\infty} \frac{(-1)^n}{ n}\, R_T(\chi_n)\,
\Bigg\{
\sin\!\left(\frac{2n\pi \mu}{\omega_c} \right) \nonumber \\
&\quad +
\frac{1}{\pi n} \left(\frac{\omega_c}{\mu}\right)\left[1+\frac{1}{2}\chi_n L(\chi_n) \right]
\cos\!\left(\frac{2n\pi \mu}{\omega_c} \right)
\Bigg\},
\end{align}
where $L(x) = \coth(x) - 1/x$.

This expression can be combined exactly into a single cosine with a shifted phase. Defining
\begin{equation}
c_n=\frac{1}{\pi n}\left(\frac{\omega_c}{\mu}\right)
\left[1+\frac{1}{2}\chi_n L(\chi_n)\right],
\label{eq:cn}
\end{equation}
we obtain
\begin{align}
M^{\rm osc} &= -\frac{\mu}{2\pi^2}\sum_{n=1}^{\infty} \frac{(-1)^n}{ n}\,
R_T(\chi_n)\, A_n\,
\cos\!\left(\frac{2n\pi \mu}{\omega_c}-\varphi_n \right),
\label{eq:Mosc_compact}
\end{align}
where
\begin{equation}
A_n=\sqrt{1+c_n^2},
\qquad
\varphi_n=\arctan\!\left(\frac{1}{c_n}\right).
\end{equation}
For $c_n\ll1$,
\begin{equation}
A_n=1+\frac{c_n^2}{2}+\mathcal O(c_n^4),
\qquad
\varphi_n=\frac{\pi}{2}-c_n+\mathcal O(c_n^3).
\end{equation}
The quadrature ratio $c_n$ interpolates between $c_n\sim\omega_c/\mu$ when $\chi_n\ll1$ and $c_n\sim\pi/(\beta\mu)$ when $\chi_n\gg1$. Thus, under the two basic conditions $\omega_c/\mu\ll1$ and $1/(\beta\mu)\ll1$, the cosine term in Eq.~\eqref{eq:Mosc_full} is subleading for all thermal regimes. To leading order,
\begin{equation}
M^{\rm osc} = -\frac{\mu}{2\pi^2}\sum_{n=1}^{\infty} \frac{(-1)^n}{ n}\,
R_T(\chi_n)\,
\cos\!\left(\frac{2n\pi \mu}{\omega_c}-\frac{\pi}{2} \right).
\label{eq:Mosc_simple}
\end{equation}
This approximation captures the dominant de Haas--van Alphen oscillations and will serve as a benchmark for the interacting theory in Sec.~\ref{sec:mean-field}.

\section{Fixed-density free gas and the exact $T=0$ limit}
\label{app:free-gas-canonical}
This appendix collects the perturbative Legendre transformation, the
canonical harmonic convolution, and the separate exact treatment required at
zero temperature.

At fixed density, the appropriate thermodynamic potential is the Helmholtz
free-energy density $\mathcal{F}(T,n,B)$, obtained from the grand potential
by the Legendre transformation
\begin{equation}
\mathcal{F}(\beta,B,n) = \Omega(\beta,B,\mu[n]) + \mu[n]\, n,
\label{eq:Legendre}
\end{equation}
where the chemical potential as a function of density, $\mu[n]$, is obtained by inverting the relation
\begin{equation}
n = -\left.\frac{\partial \Omega}{\partial \mu}\right|_{\beta,B}.
\label{eq:density_from_omega}
\end{equation}

In the degenerate Fermi regime $1/(\beta\mu)\ll1$, the leading behavior is captured by the expansion
\begin{equation}
n = \frac{m\mu}{2\pi} + n_{\rm osc}(\beta,B,\mu),
\label{eq:n_expansion}
\end{equation}
where the oscillatory correction to the density is
\begin{align}
n_{\rm osc}(\beta,B,\mu) &= \frac{m}{2\pi}\omega_c\sum_{p=1}^\infty
\frac{(-1)^p}{\pi p}\,
R_T(\chi_p)\,
\sin\!\bigl(2\pi p\mu/\omega_c\bigr).
\label{eq:nosc}
\end{align}

We write the chemical potential as
\begin{equation}
\mu = \mu_0 + \delta\mu,
\label{eq:mu_expansion}
\end{equation}
with $\mu_0 = 2\pi n/m$ the zero-field Fermi energy.
Imposing the condition of constant density translates into an implicit equation for $\delta\mu$:
\begin{align}
\delta\mu &= -\frac{2\pi}{m}\, n_{\rm osc}(\beta,B,\mu_0+\delta\mu) \nonumber \\
&= -\omega_c\sum_{p=1}^\infty \frac{(-1)^p}{\pi p}\,
R_T(\chi_p)\,
\sin\!\bigl(2\pi p(\mu_0+\delta\mu)/\omega_c\bigr).
\label{eq:dmu_osc}
\end{align}
Equation~\eqref{eq:dmu_osc} has no general closed-form solution.  We therefore
analyze it separately in the limiting regimes relevant below.

\subsection{Thermally damped regime}
The thermally damped (high-temperature or weak-field) regime is characterized by $\chi_1\gg1$, namely $\beta\omega_c\ll 2\pi^2$; parametrically, the temperature is large compared with the Landau-level spacing while still satisfying $\beta\mu\gg1$. In this regime the factor $R_T(\chi_p)$ strongly suppresses the oscillatory terms. A perturbative inversion of Eq.~\eqref{eq:dmu_osc} is justified when
\begin{equation}
\left|\frac{1}{\rho_0}\frac{\partial n_{\rm osc}}{\partial\mu}\right|\ll1,
\qquad
\frac{|\delta\mu|}{\omega_c}\ll1.
\end{equation}
These conditions are fulfilled when the thermally damped oscillatory contribution is small. Solving Eq.~\eqref{eq:dmu_osc} iteratively then gives, to leading order,
\begin{align}
\delta\mu = -\frac{2\pi}{m}\, n_{\rm osc}(\beta,B,\mu_0).
\end{align}

Expanding the free energy of Eq.~\eqref{eq:Legendre} around $\mu = \mu_0$ yields
\begin{align}
\mathcal{F} &= \Omega_0(T,\mu_0) + \frac{\partial\Omega_0}{\partial\mu}\bigg|_{\mu_0} \delta\mu
+ \frac{1}{2} \frac{\partial^2\Omega_0}{\partial\mu^2}\bigg|_{\mu_0} (\delta\mu)^2 + \cdots \nonumber\\
&\quad + \Omega^{\rm non}(B) + \Omega^{\rm osc}(T,\mu_0,B)
+ \frac{\partial\Omega^{\rm osc}}{\partial\mu}\bigg|_{\mu_0} \delta\mu + \cdots \nonumber\\
&\quad + \mu_0 n + \delta\mu\, n.
\label{eq:F_expansion}
\end{align}
Using $\partial\Omega_0/\partial\mu = -n$, the terms linear in $\delta\mu$ cancel exactly.
The remaining contributions are
\begin{align}
\mathcal{F}(T,n,B)={}&\Omega_0(T,\mu_0)+\mu_0 n+\Omega^{\rm non}(B)
\nonumber\\
&+\Omega^{\rm osc}(\beta,B,\mu_0)
+\frac{\partial\Omega^{\rm osc}}{\partial\mu}\bigg|_{\mu_0}\delta\mu
\nonumber\\
&+\frac{1}{2}\frac{\partial^2\Omega_0}{\partial\mu^2}\bigg|_{\mu_0}\delta\mu^2+\cdots .
\label{eq:F_remaining}
\end{align}
The last two terms are both proportional to $n_{\rm osc}^2$; combining them we obtain
\begin{align}
\mathcal{F}(\beta,B,n)={}&\mathcal F_0(\beta,n)+\Omega^{\rm non}(B)
\nonumber\\
&+\Omega^{\rm osc}(\beta,B,\mu_0)+\Delta\mathcal F_N.
\label{eq:Fn}
\end{align}
Here
\begin{equation}
\mathcal F_0(T,n)=\Omega_0(T,\mu_0)+\mu_0 n,
\label{eq:F0}
\end{equation}
and
\begin{equation}
\Delta\mathcal F_N=\frac{\pi}{m}n_{\rm osc}^2(\beta,B,\mu_0).
\label{eq:DeltaFN}
\end{equation}
Here and below $\mu_0=2\pi n/m$. Equation~\eqref{eq:DeltaFN} is the leading correction generated by imposing the fixed-density constraint, through the quadratic density correction.

\paragraph{Harmonic spectrum of the canonical correction.}
The quadratic term in Eq.~\eqref{eq:DeltaFN} mixes the harmonics already present in $n_{\rm osc}$ and can be reorganized as a single Fourier series. To make this explicit, we define
\begin{equation}
\theta=\frac{2\pi\mu_0}{\omega_c},
\qquad
\chi=\frac{2\pi^2}{\beta\omega_c},
\end{equation}
so that
\begin{equation}
R_T(p\chi)=\frac{p\chi}{\sinh(p\chi)}.
\end{equation}
Equation~\eqref{eq:nosc} can then be written as
\begin{equation}
 n_{\rm osc}=\frac{m\omega_c\chi}{2\pi^2}
 \sum_{p=1}^{\infty}(-1)^p
 \frac{\sin(p\theta)}{\sinh(p\chi)}.
\label{eq:nosc-harmonic}
\end{equation}
Using
\begin{equation}
\sin(p\theta)\sin(j\theta)
=\frac{1}{2}\left[
\cos((p-j)\theta)-\cos((p+j)\theta)
\right],
\end{equation}
we obtain
\begin{align}
\Delta\mathcal F_N
=&\frac{m\omega_c^2\chi^2}{8\pi^3}
\sum_{p,j\geq1}
\frac{(-1)^{p+j}}{\sinh(p\chi)\sinh(j\chi)}
\nonumber\\
&\times\left[
\cos((p-j)\theta)-\cos((p+j)\theta)
\right].
\label{eq:Fcan-double-sum}
\end{align}
Thus the fixed-density constraint mixes two harmonics $p$ and $j$ into the difference and sum channels $|p-j|$ and $p+j$. Since all the original frequencies are integer multiples of the same fundamental frequency, the result can still be reorganized as a single harmonic series,
\begin{align}
\Delta\mathcal F_N
=&\Delta\mathcal F_N^{(0)}
+\frac{m\omega_c^2}{4\pi^3}
\sum_{\ell=1}^{\infty}(-1)^\ell
K_\ell(\chi)\cos(\ell\theta),
\label{eq:Fcan-spectrum}
\end{align}
where the phase-independent contribution is
\begin{equation}
\Delta\mathcal F_N^{(0)}
=\frac{m\omega_c^2\chi^2}{8\pi^3}
\sum_{p=1}^{\infty}\frac{1}{\sinh^2(p\chi)},
\label{eq:Fcan-DC}
\end{equation}
and, for $\ell\geq1$,
\begin{align}
K_\ell(\chi)=\chi^2\Bigg[
&\sum_{p=1}^{\infty}
\frac{1}{\sinh(p\chi)\sinh((p+\ell)\chi)}
\nonumber\\
&-\frac{1}{2}\sum_{p=1}^{\ell-1}
\frac{1}{\sinh(p\chi)\sinh((\ell-p)\chi)}
\Bigg].
\label{eq:Kell-convolution}
\end{align}
The first term in Eq.~\eqref{eq:Kell-convolution} comes from the difference channel, whereas the second comes from the sum channel. Both convolutions can be evaluated exactly. Using
\begin{equation}
\frac{1}{\sinh x\sinh y}
=\frac{\coth x-\coth y}{\sinh(y-x)}
\end{equation}
in the first sum and
\begin{equation}
\frac{1}{\sinh x\sinh y}
=\frac{\coth x+\coth y}{\sinh(x+y)}
\end{equation}
in the second one, the sums telescope and give the compact result
\begin{equation}
K_\ell(\chi)=
\chi^2\,
\frac{\coth(\ell\chi)-\ell}{\sinh(\ell\chi)}.
\label{eq:Kell-closed}
\end{equation}
Equation~\eqref{eq:Fcan-DC} shows that the quadratic correction contains both a phase-independent part and a genuine harmonic part. We therefore define, consistently at this order,
\begin{align}
\mathcal F^{\rm non}(\beta,B,n)
&=\Omega^{\rm non}(B)+\Delta\mathcal F_N^{(0)},
\label{eq:Fnon}\\[4pt]
\mathcal F^{\rm osc}(\beta,B,n)
&=\Omega^{\rm osc}(\beta,B,\mu_0)
+\Delta\mathcal F_{N,\rm harm},
\label{eq:Fosc-canonical}
\end{align}
where $\Delta\mathcal F_{N,\rm harm}$ denotes the $\ell\geq1$ part of
Eq.~\eqref{eq:Fcan-spectrum}.  Thus
$\mathcal F=\mathcal F_0+\mathcal F^{\rm non}+\mathcal F^{\rm osc}$, and
$\mathcal F^{\rm osc}$ has no zero-frequency component.

Combining Eq.~\eqref{eq:Fcan-spectrum} with the grand-canonical contribution, the oscillatory canonical free energy can be written as
\begin{equation}
\mathcal F^{\rm osc}
=\frac{m\omega_c^2}{4\pi^3}
\sum_{\ell=1}^{\infty}(-1)^\ell
{\cal C}^{(N)}_\ell(\chi)\cos(\ell\theta),
\label{eq:Fcan-total-spectrum}
\end{equation}
where the signed harmonic coefficient is
\begin{equation}
\begin{aligned}
{\cal C}^{(N)}_\ell(\chi)
&=\frac{R_T(\ell\chi)}{\ell^2}+K_\ell(\chi)\\
&=\frac{\chi}{\ell\sinh(\ell\chi)}
+\chi^2\frac{\coth(\ell\chi)-\ell}{\sinh(\ell\chi)}.
\end{aligned}
\label{eq:canonical-harmonic-coefficient}
\end{equation}
Equivalently,
\begin{equation}
{\cal C}^{(N)}_\ell(\chi)
=\frac{\chi}{\ell\sinh(\ell\chi)}
\left[1+\ell\chi\coth(\ell\chi)-\ell^2\chi\right].
\end{equation}
The coefficient ${\cal C}^{(N)}_\ell$ can have either sign; a sign change corresponds to a phase shift of $\pi$ for that harmonic. Thus the passage from fixed chemical potential to fixed density does not merely rescale an overall amplitude: it redistributes spectral weight among the harmonics through a nonlinear mixing process.

In the thermally damped limit $\chi\gg1$, the correction to the fundamental harmonic is exponentially small,
\begin{equation}
K_1(\chi)\simeq4\chi^2e^{-3\chi},
\end{equation}
whereas, for $\ell\geq2$,
\begin{equation}
K_\ell(\chi)\simeq-2(\ell-1)\chi^2e^{-\ell\chi}.
\end{equation}
In particular, keeping only the leading fundamental harmonic in $n_{\rm osc}$ gives, to leading exponential order,
\begin{equation}
{\cal C}^{(N)}_2
\simeq \frac{R_T(2\chi)}{4}-\frac{R_T^2(\chi)}{2}.
\label{eq:second-harmonic-canonical}
\end{equation}
The second term in Eq.~\eqref{eq:second-harmonic-canonical} is entirely generated by the fixed-density constraint. Relative to the direct Lifshitz--Kosevich second harmonic, its magnitude is
\begin{equation}
\frac{R_T^2(\chi)/2}{R_T(2\chi)/4}
=2\chi\coth\chi,
\end{equation}
showing that the canonical correction can strongly modify higher harmonics even when the oscillation of the chemical potential remains perturbatively small.

The spectral representation above follows from the perturbative solution of the fixed-density condition and must therefore be restricted to the thermally damped regime in which $|\delta\mu|/\omega_c\ll1$. It should not be extrapolated to $\chi\to0$, where the perturbative Legendre expansion breaks down and the exact low-temperature canonical treatment below is required.

As a formal consistency check on the phase-independent part only, one may nevertheless take $\chi\to0$ in Eq.~\eqref{eq:Fcan-DC}. Since
\begin{equation}
\chi^2\sum_{p=1}^{\infty}\csch^2(p\chi)\longrightarrow
\sum_{p=1}^{\infty}\frac{1}{p^2}=\frac{\pi^2}{6},
\end{equation}
we obtain $\Delta\mathcal F_N^{(0)}\to m\omega_c^2/(48\pi)$. Together with $\Omega^{\rm non}=m\omega_c^2/(48\pi)$, this gives $m\omega_c^2/(24\pi)$, in agreement with the exact $T=0$ smooth contribution derived below. This agreement is only a consistency check and does not restore the validity of the full perturbative expansion at $\chi=0$.

\subsection{Weak thermal damping and the $T=0$ limit}
In the opposite low-temperature regime, $\chi_1\ll1$ (equivalently $\beta\omega_c\gg 2\pi^2$), the Lifshitz--Kosevich damping becomes weak and the oscillatory correction to the density can no longer be treated perturbatively. The limit $T\to0$ requires some care because the grand-canonical density becomes a staircase function of the chemical potential.

At strictly zero temperature, $R_T\to1$, and for values of $\mu$ lying between Landau levels Eq.~\eqref{eq:n_expansion} gives
\begin{align}
\frac{n(\mu)}{{\cal D}_B}
&= \frac{\mu}{\omega_c} + \sum_{p=1}^\infty \frac{(-1)^p}{\pi p}\,
\sin\!\left(\frac{2\pi p\mu}{\omega_c}\right) \nonumber\\
&= \left\lfloor \frac{\mu}{\omega_c}+\frac{1}{2}\right\rfloor,
\end{align}
where ${\cal D}_B=B/(2\pi)$ is the Landau-level degeneracy per unit area.
Here $\lfloor x\rfloor$ denotes the greatest integer not exceeding $x$, while
$\{x\}\equiv x-\lfloor x\rfloor$ denotes its fractional part.  Thus, in the grand-canonical ensemble at $T=0$, the density takes only the values corresponding to an integer number of completely filled Landau levels. A generic fixed density, however, corresponds to a partially filled Landau level, for which the chemical potential lies precisely at a discontinuity of this staircase. It is therefore easier to obtain the canonical result directly by counting the occupied Landau levels.

We define the filling factor
\begin{align}
\nu&\equiv\frac{n}{{\cal D}_B}
=\frac{\mu_0}{\omega_c}=p+f,
\end{align}
where 
\begin{align}
p&=\left\lfloor\frac{\mu_0}{\omega_c}\right\rfloor,
\\
f&=\left\{\frac{\mu_0}{\omega_c}\right\},
\qquad 0\le f<1.
\end{align}
Here, $p$ Landau levels are completely filled and the level with index $p$ has filling fraction $f$. For a partially filled level, $0<f<1$, the chemical potential, defined as the energy required to add one particle, is
\begin{equation}
\mu=E_p=\omega_c\left(p+\frac12\right).
\end{equation}
Consequently,
\begin{equation}
\delta\mu = \mu - \mu_0 = \omega_c\left(\frac{1}{2} - \left\{\frac{\mu_0}{\omega_c}\right\}\right),
\qquad 0<f<1.
\label{eq:dmu_exact_T0}
\end{equation}
At integer filling ($f=0$), the canonical free energy has a cusp and the chemical potential is not uniquely defined: its left and right derivatives are the energies of the highest occupied and lowest empty Landau levels, respectively. Equation~\eqref{eq:dmu_exact_T0} corresponds to the right derivative, i.e. to the energy required to add one particle.

Direct level counting gives the Helmholtz free-energy density. At $T=0$, with $p$ completely filled Landau levels and a fraction $f$ of the next one occupied,
\begin{align}
\mathcal F
&={\cal D}_B\sum_{\lambda=0}^{p-1}E_\lambda+f{\cal D}_B E_p \nonumber \\
&=\rho_0\omega_c^2\left[\frac{p^2}{2}+f\left(p+\frac12\right)\right],
\label{eq:F_piecewise}
\end{align}
where ${\cal D}_B=\rho_0\omega_c$. The smooth zero-field contribution at the same density is
\begin{equation}
\mathcal F_0=\frac{\rho_0\mu_0^2}{2}
=\frac{\rho_0\omega_c^2}{2}(p+f)^2.
\end{equation}
Thus, the field-dependent canonical correction is
\begin{equation}
\Delta\mathcal F_{T=0}
=\mathcal F-\mathcal F_0
=\frac{\rho_0\omega_c^2}{2}(f-f^2).
\label{eq:Fosc_T0_boxed}
\end{equation}
The function $f-f^2$ contains both a zero-frequency component and a harmonic part. Using the Fourier series
\begin{equation}
f-f^2=\frac16-\frac{1}{\pi^2}
\sum_{\ell=1}^{\infty}\frac{\cos(2\pi\ell f)}{\ell^2},
\qquad 0\le f<1,
\end{equation}
and $2\pi f$ equivalent to $\theta=2\pi\mu_0/\omega_c$ modulo $2\pi$, we obtain
\begin{equation}
\Delta\mathcal F_{T=0}
=\frac{m\omega_c^2}{24\pi}
-\frac{m\omega_c^2}{4\pi^3}
\sum_{\ell=1}^{\infty}\frac{\cos(\ell\theta)}{\ell^2}.
\label{eq:F-T0-Fourier}
\end{equation}
Hence the exact canonical $T=0$ decomposition is
\begin{align}
\mathcal F^{\rm non}_{T=0}&=\frac{m\omega_c^2}{24\pi},
\\
\mathcal F^{\rm osc}_{T=0}
&=-\frac{m\omega_c^2}{4\pi^3}
\sum_{\ell=1}^{\infty}\frac{\cos(\ell\theta)}{\ell^2}.
\end{align}
Writing $F=A\mathcal F$ and $N=An$, with $A=L^2$ the  area, one
has $\partial F/\partial N=\partial\mathcal F/\partial n=E_p$, which is
constant while a given Landau level is being filled.  The total canonical
free energy is therefore piecewise linear in the filling fraction $f$.  The
field-dependent correction $\Delta\mathcal F_{T=0}$ is a sequence of
parabolic arcs, while its strictly oscillatory component has the signed
harmonic spectrum $-1/\ell^2$.  This spectrum differs qualitatively from the
grand-canonical $T=0$ spectrum, which carries the factor $(-1)^\ell$.  This
exact $T=0$ free-gas result complements the finite-temperature
canonical calculation in Sec.~\ref{sec:phase-diagram}. The numerical
coefficients there use the full parabolic density constraint and the
leading dispersion correction; the thermally damped harmonic expansion
provides their analytical interpretation.

\section{Hubbard--Stratonovich transformation}
\label{app:HS-transformation}
This appendix records the functional-integral steps leading to
Eq.~\eqref{eq:Seff}.

At inverse temperature $\beta$, the grand-canonical partition function is
\begin{align}
Z&=\Tr e^{-\beta(H-\mu N)}\nonumber\\
&=\int{\cal D}\psi^\dagger{\cal D}\psi\;
e^{-S[\psi^\dagger,\psi;{\bf A}]},
\end{align}
with Euclidean action $S=S_0+S_I$,
\begin{align}
S_0&=\int_0^\beta d\tau\int d^2x\;
\psi^\dagger[\partial_\tau+\varepsilon({\bf D})-\mu]\psi,
\label{eq:S0}\\
S_I&=\frac{f_2}{4}\int_0^\beta d\tau\int d^2x\;
\Tr\!\left[\hat Q^2\right].
\label{eq:SI}
\end{align}
The Grassmann fields satisfy the antiperiodic boundary conditions
$\psi(\beta,{\bf x})=-\psi(0,{\bf x})$ and
$\psi^\dagger(\beta,{\bf x})=-\psi^\dagger(0,{\bf x})$.

We linearize the quadrupolar interaction by means of a
Hubbard--Stratonovich transformation.  To this end, we introduce a real,
symmetric, traceless auxiliary field,
\begin{equation}
\Phi=\begin{pmatrix}\phi^+&\phi^-\\\phi^-&-\phi^+\end{pmatrix}.
\end{equation}
For the attractive channel, $f_2=-|f_2|$, the corresponding Gaussian identity is
\begin{align}
&\exp\left\{\frac{|f_2|}{4}\int_x\Tr(\hat Q^2)\right\}
\label{eq:I}\\
&={\cal N}\int{\cal D}\Phi\;
\exp\left\{-\int_x\left[
\frac{1}{4|f_2|}\Tr(\Phi^2)-\frac{1}{2}\Tr(\Phi\hat Q)
\right]\right\},
\nonumber 
\end{align}
where $\int_x\equiv\int_0^\beta d\tau\int d^2x$,
${\cal D}\Phi={\cal D}\phi^+{\cal D}\phi^-$, and
$\Phi(\beta,{\bf x})=\Phi(0,{\bf x})$.  Here ${\cal N}$ is the
$\hat Q$-independent Gaussian normalization of the functional measure,
chosen so that Eq.~\eqref{eq:I} holds exactly; it absorbs the
measure-dependent determinant of the quadratic kernel.  The partition
function therefore becomes
\begin{align}
Z=\int{\cal D}\Phi{\cal D}\psi^\dagger{\cal D}\psi\;
e^{-S[\psi^\dagger,\psi,\Phi]},
\end{align}
where
\begin{align}
S[\psi^\dagger,\psi,\Phi]
&=\int_x\Bigl\{\psi^\dagger
[\partial_\tau+\varepsilon({\bf D})-\mu]\psi
\nonumber\\
&\hspace{1. cm}-\frac{1}{2}\Tr(\Phi\hat Q)
+\frac{1}{4|f_2|}\Tr(\Phi^2)\Bigr\}.
\label{eq:SPhi}
\end{align}
In components,
\begin{align}
\Tr(\Phi^2)&=2[(\phi^+)^2+(\phi^-)^2],\\
\Tr(\Phi\hat Q)&=\phi^+\psi^\dagger(D_x^2-D_y^2)\psi
+\phi^-\psi^\dagger\{D_x,D_y\}\psi.
\end{align}

The transformed action is quadratic in the fermionic fields and may be written as
\begin{align}
S=\int_x\left\{\psi^\dagger\hat O[\Phi]\psi
+\frac{1}{4|f_2|}\Tr(\Phi^2)\right\},
\label{eq:Squadraric}
\end{align}
where 
\begin{equation}
\hat O[\Phi]=\partial_\tau-\mu+{\cal H}_{\rm eff}[\Phi],
\end{equation}
and, for the parabolic dispersion,
\begin{align}
{\cal H}_{\rm eff}[\Phi]
&=\frac{1}{2m}G_{ij}\frac{1}{2}\{D_i,D_j\}, \\
\qquad G_{ij}&=\delta_{ij}-m\Phi_{ij}.
\end{align}
These expressions retain the right-acting convention of
Eq.~\eqref{eq:Quadrupole}. A local Hermitian, gauge-invariant extension is
\begin{align}
\mathcal T_{ij}&=\frac12\{D_i,D_j\}-\frac{\delta_{ij}}2{\bf D}^2,\nonumber\\
\hat Q^H_{ij}&=\frac12\left[\psi^\dagger\mathcal T_{ij}\psi
+(\mathcal T_{ij}\psi)^\dagger\psi\right].
\label{app:Hermitian-quadrupole}
\end{align}
For real symmetric traceless $\Phi$, the coupling
$-\frac12\int d^2x\,\Phi_{ij}\hat Q^H_{ij}$ gives
\begin{equation}
\mathcal H^H_{\rm eff}[\Phi]
=\varepsilon({\bf D})-\frac14\{\Phi_{ij}(\mathbf x),\mathcal T_{ij}\},
\label{app:Hermitian-Heff}
\end{equation}
with repeated indices summed and boundary terms taken to vanish. The
anticommutator here is between multiplication by $\Phi_{ij}$ and the
differential operator $\mathcal T_{ij}$. For uniform $\Phi$ and parabolic
$\varepsilon$, this reduces to Eq.~\eqref{eq:Heff}; spatially varying
$\Phi$ generates the gradient terms fixed by this ordering.  This
prescription specifies a Hermitian extension of the nonuniform interaction.
If both the interaction and the Hubbard--Stratonovich decoupling are defined
using $\hat Q^H$, it provides the gradient vertices required for a future
finite-field fluctuation analysis.

The Gaussian integral over the Grassmann fields can now be carried out exactly,
\begin{align}
Z=\int{\cal D}\Phi\;e^{-S_{\rm eff}[\Phi]},
\end{align}
with
\begin{align}
S_{\rm eff}[\Phi]
=\frac{1}{4|f_2|}\int_x\Tr(\Phi^2)
-\Tr\ln\!\left(\beta\hat O[\Phi]\right).
\end{align}
Here $\hat O[\Phi]$ is given by Eq.~\eqref{eq:O}.
The functional trace runs over space, imaginary time (or, equivalently, Matsubara frequency), and the one-particle Hilbert space in the magnetic field. Equation~\eqref{eq:Seff} is the formal fermionic determinant for the
right-acting convention above. For the Hermitian extension, the same
determinant construction uses $\hat O^H=\partial_\tau-\mu+\mathcal H^H_{\rm eff}$
and the consistently redefined quadrupolar interaction.

The stationary configurations of the remaining functional integral over
$\Phi$ determine the possible uniform nematic states, while an expansion
about them would generate nematic fluctuations.  The main text addresses
the first problem by restricting $\Phi$ to a static, spatially uniform
saddle.

\section{Landau spectrum in a uniform nematic background}
\label{app:spectrum-uniform}
For a static and spatially uniform saddle the quadratic Hamiltonian is
\begin{equation}
{\cal H}_{\rm eff}=\frac{1}{2m}{\bf D}^{T}G{\bf D},
\end{equation}
where $G$ is a real symmetric positive-definite $2\times2$ matrix with
$\det G=\Delta^2$.  Because $G$ is constant, operator-ordering issues do not
arise in this expression.

For $B>0$, we choose an orientation-preserving real matrix $S$ such that
\begin{equation}
S^{T}GS={\bf 1},
\end{equation}
and define ${\bf y}=S^T{\bf x}$ and
$\widetilde{\bf D}=S^{-1}{\bf D}$.  Then
\begin{equation}
{\cal H}_{\rm eff}
=\frac{1}{2m}\widetilde{\bf D}^{T}\widetilde{\bf D}.
\end{equation}
Since $[D_x,D_y]=-iB$,
\begin{equation}
[\widetilde D_1,\widetilde D_2]
=-iB\det(S^{-1}).
\end{equation}
Taking determinants of $S^TGS={\bf1}$ gives
$\det S=1/\sqrt{\det G}=1/\Delta$, and therefore
\begin{equation}
[\widetilde D_1,\widetilde D_2]=-iB\Delta.
\end{equation}
The level spacing is set by the magnitude of this commutator.  The spectrum is consequently
\begin{equation}
E_\lambda=\left(\lambda+\frac12\right)\frac{B\Delta}{m}.
\end{equation}

Denoting the physical area by $A=L^2$, the transformed area is
$A_y=A/\Delta$, whereas the effective magnetic field
in the transformed problem is $B\Delta$.  Hence the number of states in each
Landau level is
\begin{equation}
\frac{B\Delta}{2\pi}A_y
=\frac{BA}{2\pi}.
\end{equation}
Thus the degeneracy per unit physical area is exactly $B/(2\pi)$, independent
of the nematic deformation.

For the traceless saddle introduced in
Sec.~\ref{sec:nematic-model-spectrum},
\begin{equation}
G={\bf1}-m\bar\Phi,
\end{equation}
and the identity valid for any real traceless $2\times2$ matrix gives
\begin{equation}
\det G=1-\frac{m^2}{2}\Tr(\bar\Phi^2)=1-q^2.
\end{equation}
This proves Eqs.~\eqref{eq:ciclotron} and \eqref{eq:degeneracy}.

\subsection{Self-consistent spectrum and Landau-level crossings}
\label{app:selfconsistent-spectrum}
Substituting the self-consistent saddle into Eq.~\eqref{eq:Fosc-MF} gives the
mean-field oscillatory grand potential,
\begin{align}
\Omega_{\rm MF}^{\rm osc}
={}&a\left(\frac{\omega_c}{\mu}\right)^2
\Delta_{\rm sp}
\sum_{n=1}^{\infty}\frac{(-1)^n}{\pi^2n^2}
\nonumber\\
&\times R_T\!\left(\frac{\chi_n}{\Delta_{\rm sp}}\right)
\cos\!\left(\frac{2\pi n\mu}
{\omega_c\Delta_{\rm sp}}\right),
\label{eq:Omegaosc-selfconsistent}
\end{align}
where
\begin{equation}
\Delta_{\rm sp}=\sqrt{1-q_{\rm sp}^2(T,B)}.
\end{equation}
If $q_{\rm sp}$ were field independent, Eq.~\eqref{eq:Omegaosc-selfconsistent}
would reduce to the free-gas expression with the renormalized cyclotron mass
$m_c=m/\Delta_{\rm sp}$.  In equilibrium, however, $q_{\rm sp}(T,B)$ inherits
the oscillatory field dependence of Eq.~\eqref{eq:qsp-TB}.  Nematic
self-consistency therefore modulates both the dHvA phase and the thermal
damping.  This feedback is the central mean-field mechanism analyzed in the
quantum-oscillation spectra in Sec.~\ref{sec:dhva-nematic}.

The self-consistent saddle makes the Landau-level spacing field dependent
through $\Delta_{\rm sp}(B)$, and therefore shifts the fields at which
individual levels cross the fixed chemical potential.  As long as the
quadratic saddle remains positive definite, $q_{\rm sp}<1$, the level
spacing in Eq.~\eqref{eq:ciclotron} remains positive and the mutual ordering
of the Landau ladder is preserved.  In the numerical parameter range
examined here, we find that the field dependence of the saddle is too weak
to make the Landau-level trajectories nonmonotonic.  Within this range,
each level therefore crosses $\mu$ at most once, so self-consistency shifts
the crossing sequence without producing additional crossings or reordering
the levels.

The preceding discussion applies at fixed chemical potential.  At fixed
density, \(\mu\) adjusts with the field and the parabolic model exhibits the
phase locking derived in Sec.~\ref{sec:MF-fixed-density}.  The canonical constraint must therefore be
imposed by a Legendre transformation rather than by reinterpreting the
fixed-\(\mu\) crossing fields.  In the thermally damped regime this procedure
is controlled and generates nonlinear mixing of the dHvA harmonics, in direct
parallel with the free-gas analysis of
Sec.~\ref{sec:free-benchmarks}.

\section{Landau expansion at fixed chemical potential}
\label{app:Landau-expansion}
The purpose of this appendix is to derive the parabolic contributions to
the quadratic and quartic Landau coefficients used in
Sec.~\ref{sec:mean-field}.  In
particular, we derive the susceptibility
${\cal X}_{2,{\rm par}}(T,B)$ quoted in
Eqs.~\eqref{eq:X2-total} and \eqref{eq:X2-osc}, and the quartic
coefficient $u_{f,{\rm par}}(T,B)$ given in
Eqs.~\eqref{eq:u-smooth}--\eqref{eq:u-par-osc}.  We also explain why
the local expansion becomes nonuniform in the low-temperature limit.

The calculation starts from the fixed-$\mu$ parabolic grand potential in
Eqs.~\eqref{eq:calF0}--\eqref{eq:Fosc-MF}.  We expand only in the
nematic amplitude $q$ around the isotropic state, while keeping the full
dependence on $T/\omega_c$ in the Lifshitz--Kosevich factor.  The
contribution of the radial irrelevant operator is treated separately in
Appendix~\ref{app:irrelevant-term}.

The only elementary expansion needed below is
\begin{align}
\Delta(q)=\sqrt{1-q^2}
&=1-\frac{q^2}{2}-\frac{q^4}{8}+O(q^6),
\label{app:Delta-expansion1}\\
\frac{1}{\Delta(q)}=\frac{1}{\sqrt{1-q^2}}
&=1+\frac{q^2}{2}+\frac{3q^4}{8}+O(q^6).
\label{app:Delta-expansion2}
\end{align}

\subsection{Quadratic coefficient}
We first determine the coefficient of $q^2$ in the Landau expansion
\begin{equation}
\Delta\Omega_{\rm MF}
=
\frac{a}{2}r(T,B)q^2
+\frac{a}{4}u(T,B)q^4+\cdots .
\end{equation}
For the parabolic theory, the smooth fermionic terms are
Eqs.~\eqref{eq:calF0} and \eqref{eq:calFnon}.  Using
Eqs.~\eqref{app:Delta-expansion1} and \eqref{app:Delta-expansion2} their changes relative to the isotropic
state are
\begin{align}
\Omega_f^0(q)-\Omega_f^0(0)
=&
-a\left[
1+\frac{\pi^2}{3}\left(\frac{T}{\mu}\right)^2
\right] \times\nonumber \\
&\times \left(
\frac{q^2}{2}+\frac{3q^4}{8}
\right)
+O(q^6),
\label{app:smooth-zero-expansion}
\\
\Omega_f^{\rm non}(q)-\Omega_f^{\rm non}(0)
=&
-\frac{a}{12}
\left(\frac{\omega_c}{\mu}\right)^2
\left(
\frac{q^2}{2}+\frac{q^4}{8}
\right)
+O(q^6).
\label{app:smooth-B-expansion}
\end{align}
Combining their quadratic terms with the Hubbard--Stratonovich
contribution $a q^2/(2|F_2|)$ gives
\begin{align}
r_{\rm par}(T,B)
={}&\frac{1}{|F_2|}-1
-\frac{\pi^2}{3}\left(\frac{T}{\mu}\right)^2
\nonumber\\
&-\frac{1}{12}\left(\frac{\omega_c}{\mu}\right)^2
-{\cal X}_{2,{\rm par}}^{\rm osc}(T,B),
\end{align}
which is Eq.~\eqref{eq:X2-total}.  It remains to derive the oscillatory
term ${\cal X}_{2,{\rm par}}^{\rm osc}$.

The $q$ dependence of the $n$th harmonic in
Eq.~\eqref{eq:Fosc-MF} is entirely contained in
\begin{equation}
\Delta\,
R_T\!\left(\frac{\chi_n}{\Delta}\right)
\cos\!\left(\frac{\theta_n}{\Delta}\right),
\label{app:harmonic-factor}
\end{equation}
where $\theta_n=2\pi n\mu/\omega_c$ and
$\chi_n=2\pi^2nT/\omega_c$.  Using
\begin{equation}
\frac{d\ln R_T(z)}{dz}
=-L(z),
\qquad
L(z)=\coth z-\frac{1}{z},
\end{equation}
the derivative of Eq.~\eqref{app:harmonic-factor} at the isotropic point is
\begin{align}
&\left.
\frac{d}{d\Delta}
\left[
\Delta
R_T\!\left(\frac{\chi_n}{\Delta}\right)
\cos\!\left(\frac{\theta_n}{\Delta}\right)
\right]
\right|_{\Delta=1}
= 
\label{app:first-harmonic-derivative} \\
&R_T(\chi_n)
\Big\{
[1+\chi_n L(\chi_n)]\cos\theta_n
+\theta_n\sin\theta_n
\Big\}.
\nonumber
\end{align}

Since $\Delta(q)-1=-q^2/2+O(q^4)$, the corresponding quadratic
contribution to the oscillatory grand potential is
\begin{align}
\left.\Delta\Omega_f^{\rm osc}\right|_{q^2}
={}&
-\frac{a q^2}{2}
\left(\frac{\omega_c}{\mu}\right)^2
\sum_{n=1}^{\infty}
\frac{(-1)^n}{\pi^2n^2}
R_T(\chi_n)
\\
&\times
\Big\{
[1+\chi_nL(\chi_n)]\cos\theta_n
+\theta_n\sin\theta_n
\Big\}.
\nonumber
\end{align}
Comparison with $(a/2)r q^2$, together with
$r_{\rm par}=1/|F_2|-{\cal X}_{2,{\rm par}}$, therefore gives
\begin{align}
{\cal X}_{2,{\rm par}}^{\rm osc}
={}&
\left(\frac{\omega_c}{\mu}\right)^2
\sum_{n=1}^{\infty}
\frac{(-1)^n}{\pi^2n^2}
R_T(\chi_n)
\nonumber\\
&\times
\Big\{
[1+\chi_nL(\chi_n)]\cos\theta_n
+\theta_n\sin\theta_n
\Big\},
\end{align}
which is Eq.~\eqref{eq:X2-osc}.

Because
$\theta_n\sim\mu/\omega_c$, the term
$\theta_n\sin\theta_n$ compensates one power of the explicit
$(\omega_c/\mu)^2$ prefactor.  The leading oscillation of the quadratic
coefficient is therefore of order $\omega_c/\mu$, whereas the cosine
quadrature starts at order $(\omega_c/\mu)^2$.

\subsection{Quartic coefficient}
We next extract the coefficient of $q^4$.  The smooth contribution follows
immediately from Eqs.~\eqref{app:smooth-zero-expansion} and
\eqref{app:smooth-B-expansion}.  Comparing their quartic terms with
$(a/4)u q^4$ gives
\begin{equation}
u_{f,{\rm par}}^{\rm sm}
=
-\frac{3}{2}
\left[
1+\frac{\pi^2}{3}\left(\frac{T}{\mu}\right)^2
\right]
-\frac{1}{24}
\left(\frac{\omega_c}{\mu}\right)^2,
\end{equation}
which is Eq.~\eqref{eq:u-smooth}.

The oscillatory part requires the expansion of
Eq.~\eqref{app:harmonic-factor} through $q^4$.  For any function
$F(\Delta)$, Eqs.~\eqref{app:Delta-expansion1} and \eqref{app:Delta-expansion2}  imply
\begin{equation}
F[\Delta(q)]
=
F(1)
-\frac{q^2}{2}F'(1)
+\frac{q^4}{8}
\left[F''(1)-F'(1)\right]
+O(q^6).
\label{app:generic-Delta-expansion}
\end{equation}
Thus the quartic coefficient is determined by the combination of first
and second derivatives of the harmonic factor.

Using the same LK identity as above, and denoting
$L_n=L(\chi_n)$ and
$L'_n=L'(\chi_n)$ as in Sec.~\ref{sec:mean-field}, one obtains
\begin{align}
&\left.
\left(
\frac{d^2}{d\Delta^2}
-\frac{d}{d\Delta}
\right)
\left[
\Delta
R_T\!\left(\frac{\chi_n}{\Delta}\right)
\cos\!\left(\frac{\theta_n}{\Delta}\right)
\right]
\right|_{\Delta=1}
\label{app:second-harmonic-derivative}
\\
&\qquad =
R_T(\chi_n)
\Big\{
\left[
\chi_n^2(L_n^2-L'_n)
-\theta_n^2
-1-\chi_nL_n
\right]\cos\theta_n
\nonumber\\
&\qquad
+\theta_n(2\chi_nL_n-1)\sin\theta_n
\Big\}.
\nonumber
\end{align}

Substituting this result into the oscillatory grand potential and comparing
the coefficient of $q^4$ with $(a/4)u q^4$ gives
\begin{align}
u_{f,{\rm par}}^{\rm osc}
={}&
\frac{1}{2}
\left(\frac{\omega_c}{\mu}\right)^2
\sum_{n=1}^{\infty}
\frac{(-1)^n}{\pi^2n^2}
R_T(\chi_n)
\nonumber\\
&\times
\Big\{
\left[
\chi_n^2(L_n^2-L'_n)
-\theta_n^2
-1-\chi_nL_n
\right]\cos\theta_n
\nonumber\\
&
+\theta_n(2\chi_nL_n-1)\sin\theta_n
\Big\}.
\label{app:uosc-before-theta}
\end{align}

Using $\theta_n=2\pi n\mu/\omega_c$, this can be written as
\begin{align}
u_{f,{\rm par}}^{\rm osc}
={}&
\sum_{n=1}^{\infty}(-1)^nR_T(\chi_n)
\Bigg\{
-2\cos\theta_n
\\
&+
\frac{\omega_c/\mu}{\pi n}
(2\chi_nL_n-1)\sin\theta_n
\nonumber\\
&+
\frac{(\omega_c/\mu)^2}{2\pi^2n^2}
\left[
\chi_n^2(L_n^2-L'_n)
-1-\chi_nL_n
\right]\cos\theta_n
\Bigg\},
\nonumber
\end{align}
which is Eq.~\eqref{eq:u-par-osc}.

This result also explains the different parametric size of the
oscillations in the quadratic and quartic coefficients.  The term
$-\theta_n^2\cos\theta_n$ in
Eq.~\eqref{app:uosc-before-theta} produces two powers of
$\mu/\omega_c$ and therefore cancels the explicit
$(\omega_c/\mu)^2$ prefactor.  Consequently,
\begin{equation}
u_{f,{\rm par}}^{\rm osc}
=
-2\sum_{n=1}^{\infty}
(-1)^nR_T(\chi_n)\cos\theta_n
+O\!\left(\frac{\omega_c}{\mu}\right),
\end{equation}
reproducing Eq.~\eqref{eq:uosc-leading}.  The quartic coefficient can
therefore oscillate by an amount of order unity even though the oscillatory
grand potential itself carries an explicit weak-field factor
$(\omega_c/\mu)^2$.

\subsection{Range of validity and the low-temperature limit}
The local expansion above is an expansion not only of the smooth
thermodynamic terms but also of the rapidly varying dHvA phase.  For the
$n$th harmonic,
\begin{equation}
\frac{\theta_n}{\Delta(q)}
=
\theta_n
\left[
1+\frac{q^2}{2}+O(q^4)
\right],
\end{equation}
so that the phase change generated by the nematic deformation is of order
$nq^2\mu/\omega_c$.  The expansion of an individual thermally active
harmonic therefore requires
\begin{equation}
n q^2\frac{\mu}{\omega_c}\ll1.
\end{equation}
Since the LK factor suppresses harmonics only for
$nT/\omega_c\gtrsim1$, the highest relevant harmonics are parametrically
$n\sim\omega_c/T$.  The most restrictive condition is consequently
\begin{equation}
\mu q^2\ll T,
\end{equation}
which is the low-temperature form of the phase-expansion criterion discussed
around Eq.~\eqref{eq:Landau-phase-control}.  At any finite temperature this
condition is satisfied sufficiently close to a continuous transition, but
the allowed interval in $q$ shrinks as $T\to0$.

This nonuniformity can also be seen directly at zero temperature.  For a
finite field away from a Landau-level crossing, an infinitesimal variation
of $q$ does not change the number of occupied levels,
\begin{equation}
N_{\rm LL}
=
\left\lfloor
\frac{\mu}{\omega_c}+\frac12
\right\rfloor .
\end{equation}
The parabolic grand potential can then be summed exactly,
\begin{equation}
\Omega_{f,{\rm par}}(q)
=
-{\cal D}_B
\left[
N_{\rm LL}\mu
-\frac{\omega_c\Delta(q)}{2}N_{\rm LL}^2
\right].
\end{equation}
Expanding only $\Delta(q)$ gives
\begin{equation}
u_{f,{\rm par}}(T=0,B)
=
-\frac12
\left(
N_{\rm LL}\frac{\omega_c}{\mu}
\right)^2,
\label{eq:u-T0-exact}
\end{equation}
provided that $\mu$ does not coincide with a Landau level.  Along a
weak-field sequence away from the crossings,
$N_{\rm LL}=\mu/\omega_c+O(1)$, so that
$u_{f,{\rm par}}\to-1/2$.  This differs from the strictly zero-field result
$u_{f,{\rm par}}=-3/2$, obtained by setting $B=0$ before performing the
Landau expansion.  The difference reflects the noncommutativity of the
$B\to0$ limit and the local expansion of the rapidly oscillating
finite-field grand potential; it is not a thermodynamic discontinuity.

The singular character of the low-temperature limit is even more apparent
at a Landau-level crossing.  There,
$\theta_1=\pi$ modulo $2\pi$, and all terms in the leading harmonic series
of Eq.~\eqref{eq:uosc-leading} add with the same sign,
\begin{equation}
u_{f,{\rm par}}^{\rm osc}
=
-2\sum_{n=1}^{\infty}R_T(\chi_n)
+O\!\left(\frac{\omega_c}{\mu}\right).
\end{equation}
For $T\ll\omega_c$,
\begin{equation}
\sum_{n=1}^{\infty}
\frac{n\chi_1}{\sinh(n\chi_1)}
\simeq
\frac{\pi^2}{4\chi_1},
\end{equation}
and therefore
\begin{equation}
u_{f,{\rm par}}^{\rm osc}
\sim
-\frac{\omega_c}{4T}
\qquad
(\mu\ \hbox{at a Landau level}).
\label{eq:u-crossing-asymptotic}
\end{equation}
The growing negative quartic response is the finite-temperature precursor
of the nonanalytic zero-temperature level crossing.  In this regime the
local Landau expansion is not uniform, and the phase structure must instead
be determined from the unexpanded thermodynamic potential used in
Sec.~\ref{sec:full-potential}.

\section{Fixed-density Landau coefficients and harmonic mixing}
\label{app:fixedN-nematic}
The purpose of this appendix is to derive the fixed-density results used in
Sec.~\ref{sec:MF-fixed-density}.  We first show how the Legendre transformation
produces the canonical harmonic mixing summarized in
Eqs.~\eqref{eq:FN-spectrum-nematic}--\eqref{eq:Cell-nematic-fixedN}.  We then
expand the canonical free energy around the isotropic state and recover the
parabolic Landau coefficients in Eqs.~\eqref{eq:rN-par-smooth}--
\eqref{eq:uN-par-osc-explicit}.  Finally, we derive the parabolic coefficients directly from constrained
Landau-level sums.  Radial corrections and finite-amplitude tests are
deferred to Appendix~\ref{app:irrelevant-term}.

The central simplification at fixed density is the phase locking already
established in Sec.~\ref{sec:MF-fixed-density}.  The smooth density constraint
gives $\bar\mu(q)=\mu_0\Delta(q)$, so that
$2\pi p\bar\mu/[\omega_c\Delta]=p\theta_0$ is independent of the nematic
amplitude.  Consequently, derivatives with respect to $q$ act on the smooth
prefactor and on the Lifshitz--Kosevich factor, but not on the rapidly varying
dHvA phase.  This is the origin of the weaker canonical modulation of the
Landau coefficients.

\subsection{Canonical harmonic mixing}
We begin with the origin of the harmonic spectrum in
Eq.~\eqref{eq:FN-spectrum-nematic}.  At fixed nematic amplitude, the Legendre
transformation about the smooth chemical potential gives
Eq.~\eqref{eq:FN-nematic-fixedN},
\begin{equation}
{\cal F}_N
={\cal F}_{N,{\rm sm}}
+\Omega_{\rm osc}(q,\bar\mu)
+\frac{n_{\rm osc}^2(q)}{2\kappa(q)}+
O(n_{\rm osc}^3).
\label{app:FN-Legendre-summary}
\end{equation}
The first oscillatory term is the direct dHvA contribution.  The second is
the leading canonical correction generated by the oscillation of the
chemical potential.

Substituting Eqs.~\eqref{eq:nosc-nematic-fixedN} and
\eqref{eq:kappa-nematic-fixedN} into the quadratic term in
Eq.~\eqref{app:FN-Legendre-summary} and using
\begin{equation*}
\sin(p\theta_0)\sin(j\theta_0)
=\frac12\left[
\cos((p-j)\theta_0)-\cos((p+j)\theta_0)
\right]
\end{equation*}
shows directly that the Legendre correction mixes the original harmonics
through difference and sum channels.  Collecting all terms with the same
harmonic index $\ell$ gives the kernel
\begin{equation}
K_\ell(z)
=z^2\frac{\coth(\ell z)-\ell}{\sinh(\ell z)},
\end{equation}
where $z=\widetilde\chi=\chi/\Delta(q)$.  Adding the direct dHvA contribution
therefore gives
\begin{equation}
{\cal C}^{(N)}_\ell(z)
=\frac{R_T(\ell z)}{\ell^2}+K_\ell(z),
\end{equation}
which is Eq.~\eqref{eq:Cell-nematic-fixedN} and leads immediately to the
canonical spectrum in Eq.~\eqref{eq:FN-spectrum-nematic}.

The product of equal harmonics also generates a zero-frequency contribution.
In the notation of Sec.~\ref{sec:MF-fixed-density}, its $q$-dependent part is
\begin{equation}
{\cal F}^{(0)}_{2,{\rm can}}(q)
=\frac{m\omega_c^2}{8\pi^3}\,\Delta(q)
\left[\frac{\chi}{\Delta(q)}\right]^2
\sum_{p=1}^{\infty}
\frac{1}{\sinh^2[p\chi/\Delta(q)]}.
\label{app:canonical-DC-term}
\end{equation}
This term is independent of the rapid phase and is therefore not part of the
nonzero-harmonic sum in Eq.~\eqref{eq:FN-spectrum-nematic}.  It is retained
automatically in the direct level-sum evaluation used below.

\subsection{Local parabolic Landau coefficients}
We next derive the coefficients of the local expansion
Eq.~\eqref{eq:Landau-fixedN-explicit}.  Using the notation
$A_0$, $\epsilon_0$, $\chi$, and $\theta_0$ introduced in
Sec.~\ref{sec:MF-fixed-density}, the direct smooth canonical contribution can
be written, up to a $q$-independent constant, as
\begin{align}
{\cal F}_{N,{\rm par}}^{\rm sm}(q)-{\cal F}_{N,{\rm par}}^{\rm sm}(0)
={}&A_0\left(1+\frac{\epsilon_0^2}{12}\right)
[\Delta(q)-1]
\nonumber\\
&-A_0\frac{\pi^2T^2}{3\mu_0^2}
\left[\frac{1}{\Delta(q)}-1\right].
\label{app:fixedN-smooth-potential}
\end{align}
Expanding
$\Delta(q)=1-q^2/2-q^4/8+O(q^6)$ and adding the
Hubbard--Stratonovich term gives directly
Eqs.~\eqref{eq:rN-par-smooth} and \eqref{eq:uN-par-smooth}.
In particular, the Legendre constraint changes the zero-field parabolic
quartic coefficient from $-3/2$ at fixed chemical potential to $-1/2$ at
fixed density.

The oscillatory coefficients follow just as directly from
Eq.~\eqref{eq:FN-spectrum-nematic}.  With
$C_\ell(z)\equiv{\cal C}^{(N)}_\ell(z)$ as in
Eq.~\eqref{eq:Cell-explicit-fixedN}, the only quantity that must be expanded
is
$\Delta(q)C_\ell[\chi/\Delta(q)]$.  To quartic order,
\begin{align}
\Delta(q)C_\ell\!\left(\frac{\chi}{\Delta(q)}\right)
=&C_\ell(\chi)
-\frac{q^2}{2}\left[C_\ell-\chi C_\ell'\right]
\label{app:fixedN-harmonic-expansion}
\\
&+\frac{q^4}{8}
\left[\chi^2C_\ell''-C_\ell+\chi C_\ell'\right]
+O(q^6),
\nonumber
\end{align}
where all functions and derivatives on the right-hand side are evaluated at
$z=\chi$.  Comparing the coefficients of $q^2$ and $q^4$ with
Eq.~\eqref{eq:Landau-fixedN-explicit} gives precisely
Eqs.~\eqref{eq:rN-par-osc-explicit} and
\eqref{eq:uN-par-osc-explicit}.

Because this phase is locked, the local expansion generates no derivatives
of $\theta_0$.  Both canonical oscillatory coefficients therefore retain the
explicit prefactor $(\omega_c/\mu_0)^2$, in contrast with the order-one
quartic oscillation at fixed chemical potential.

\subsection{Direct parabolic level-sum construction}
The harmonic expansion above is useful for interpretation in the thermally
damped regime.  The direct construction uses finite-temperature Landau-level sums,
with the density constraint imposed explicitly.  This keeps the complete parabolic thermal occupations and also
includes the phase-independent term in Eq.~\eqref{app:canonical-DC-term}
without separating it from the other contributions.

At each $q$, the parabolic chemical potential is determined by
\begin{equation}
n={\cal D}_B\sum_{\lambda\geq0}
 n_F[E_\lambda(q)-\mu_{\rm par}(q)],
\label{eq:canonical-full-density}
\end{equation}
where
$E_\lambda(q)=e_\lambda\Delta(q)$ and
$e_\lambda=(\lambda+1/2)\omega_c$.  For the local expansion we write
\begin{align}
E_\lambda(q)
&=e_\lambda+E_{\lambda,2}q^2+E_{\lambda,4}q^4+O(q^6),
\nonumber\\
\mu_{\rm par}(q)
&=\mu_{\rm par}(0)+\mu_2q^2+\mu_4q^4+O(q^6),
\end{align}
with
\begin{equation}
E_{\lambda,2}=-\frac{e_\lambda}{2},
\qquad
E_{\lambda,4}=-\frac{e_\lambda}{8}.
\end{equation}
Defining
$f_\lambda=n_F[e_\lambda-\mu_{\rm par}(0)]$, with primes denoting
derivatives with respect to the energy argument, the fixed-density condition
gives
\begin{align}
\mu_2
&=\frac{\sum_\lambda f_\lambda' E_{\lambda,2}}
        {\sum_\lambda f_\lambda'},
\label{app:mu2-fixedN}\\
\mu_4
&=\frac{\sum_\lambda\left[
 f_\lambda' E_{\lambda,4}
 +\frac12 f_\lambda''(E_{\lambda,2}-\mu_2)^2
 \right]}
 {\sum_\lambda f_\lambda'}.
\label{app:mu4-fixedN}
\end{align}
The parabolic coefficients obtained from the same constrained level sum are
then
\begin{align}
r_{N,{\rm par}}^{\rm levels}
&=\frac1{|F_2|}
+\frac{2{\cal D}_B}{A_0}
\sum_\lambda f_\lambda E_{\lambda,2},
\label{app:rN-par-levels}\\
u_{N,{\rm par}}^{\rm levels}
&=\frac{4{\cal D}_B}{A_0}
\sum_\lambda\left[
 f_\lambda E_{\lambda,4}
 +\frac12 f_\lambda'(E_{\lambda,2}-\mu_2)E_{\lambda,2}
\right].
\label{app:uN-par-levels}
\end{align}

The radial corrections and their zero-field limit are derived in
Appendix~\ref{app:canonical-radial}.  The finite-amplitude canonical
construction and its numerical checks are given in
Appendices~\ref{app:canonical-continuation} and~\ref{app:completion-convergence}.

\section{Oscillation-induced shift of the nematic saddle}
\label{app:saddle-readjustment}
The purpose of this appendix is to determine the correction that arises
because the nematic order parameter itself responds to the dHvA
oscillations.  At fixed density, Eq.~\eqref{eq:FN-nematic-fixedN}
separates the canonical free energy, at fixed nematic amplitude, into a
smooth part, a direct oscillatory contribution, and the second-order
Legendre correction $n_{\rm osc}^2/(2\kappa)$.  The harmonic spectrum in
Eq.~\eqref{eq:FN-spectrum-nematic} is obtained at a fixed nematic
background.  Here we determine the additional contribution produced by the
oscillation-induced shift of the nematic minimum itself.

To make this effect explicit, we write the canonical free energy to first
order in the oscillatory contribution as
\begin{equation}
{\cal F}_N(q)
=
{\cal F}_{N,{\rm sm}}(q)
+
{\cal F}_{N,{\rm osc}}^{(1)}(q)
+\cdots .
\end{equation}
We denote by $q_{\rm sm}$ a minimum of the smooth free energy,
\begin{equation}
\left.
\partial_q{\cal F}_{N,{\rm sm}}
\right|_{q=q_{\rm sm}}
=0,
\qquad
\left.
\partial_q^2{\cal F}_{N,{\rm sm}}
\right|_{q=q_{\rm sm}}
>0.
\end{equation}
The oscillatory contribution shifts this minimum to
\begin{equation}
q_{\rm sp}=q_{\rm sm}+\delta q_{\rm sp}.
\end{equation}
Expanding the stationarity condition
$\partial_q{\cal F}_N(q_{\rm sp})=0$ to first order gives
\begin{equation}
0=
\left.
\partial_q^2{\cal F}_{N,{\rm sm}}
\right|_{q_{\rm sm}}
\delta q_{\rm sp}
+
\left.
\partial_q{\cal F}_{N,{\rm osc}}^{(1)}
\right|_{q_{\rm sm}},
\end{equation}
and therefore
\begin{equation}
\delta q_{\rm sp}
=
-
\left.
\frac{
\partial_q{\cal F}_{N,{\rm osc}}^{(1)}
}{
\partial_q^2{\cal F}_{N,{\rm sm}}
}
\right|_{q=q_{\rm sm}} .
\label{app:dq-saddle-fixedN}
\end{equation}
Thus the shift is determined by the oscillatory thermodynamic force divided
by the curvature of the smooth nematic minimum.

Expanding the free energy itself to second order in this displacement and
using Eq.~\eqref{app:dq-saddle-fixedN} gives
\begin{equation}
\Delta{\cal F}_{N,{\rm shift}}^{(2)}
=
-
\left.
\frac{
\left(\partial_q{\cal F}_{N,{\rm osc}}^{(1)}\right)^2
}{
2\,\partial_q^2{\cal F}_{N,{\rm sm}}
}
\right|_{q=q_{\rm sm}} .
\label{app:FN-saddle-shift}
\end{equation}
This static correction arises from the displacement of the equilibrium
minimum and is distinct from the canonical term $n_{\rm osc}^2/(2\kappa)$
in Eq.~\eqref{eq:FN-correction-nematic}, which follows directly from the
density constraint.

At fixed density, phase locking fixes the harmonic structure of
Eq.~\eqref{app:FN-saddle-shift}.  Using Eq.~\eqref{eq:phase-lock-fixedN}, the direct oscillatory
contribution to the canonical free energy is
\begin{align}
{\cal F}_{N,{\rm osc}}^{(1)}(q)
={}&
\frac{m\omega_c^2}{4\pi^3}\Delta(q)
\sum_{p=1}^{\infty}
\frac{(-1)^p}{p^2}
R_T\!\left(p\widetilde\chi\right)
\cos(p\theta_0),
\label{app:Fosc-saddle-fixedN}
\end{align}
where
\begin{equation}
\widetilde\chi
=
\frac{2\pi^2}{\beta\omega_c\Delta(q)},
\qquad
\theta_0=\frac{2\pi\mu_0}{\omega_c}.
\end{equation}
Since $\theta_0$ is independent of $q$, differentiation acts only on the
prefactor and the LK damping factor:
\begin{align}
\partial_q{\cal F}_{N,{\rm osc}}^{(1)}
={}&
\frac{m\omega_c^2}{4\pi^3}
\sum_{p=1}^{\infty}
\frac{(-1)^p}{p^2}
\frac{d}{dq}
\left[
\Delta(q)
R_T\!\left(p\widetilde\chi\right)
\right]
\cos(p\theta_0).
\label{app:dFosc-dq-fixedN}
\end{align}
It therefore generates no large factors of $\mu_0/\omega_c$, consistently
with the phase-locking suppression of the canonical Landau-coefficient
oscillations in Eqs.~\eqref{eq:rN-par-osc-explicit} and
\eqref{eq:uN-par-osc-explicit}.

Equation~\eqref{app:FN-saddle-shift} also shows that the shift of the
nematic minimum generates harmonic mixing.  Since
$\partial_q{\cal F}_{N,{\rm osc}}^{(1)}$ is a cosine series, its square
contains
\begin{equation}
\cos(p\theta_0)\cos(j\theta_0)
=
\frac{1}{2}
\left[
\cos((p-j)\theta_0)
+
\cos((p+j)\theta_0)
\right].
\end{equation}
Thus the saddle shift generates sum and difference frequencies, as does
the Legendre correction, through the distinct mechanism identified above.

The relative size of the two effects follows from their weak-field scales.
The direct oscillatory free energy is of order $m\omega_c^2$, while the
curvature of a regular smooth nematic minimum is of order $m\mu_0^2$.
Consequently, the saddle-shift contribution is smaller than the leading
canonical second-order mixing by an additional factor of order
$
\left(\omega_c/\mu_0\right)^2,
$
provided that the smooth curvature remains finite.  This is the hierarchy
used in Sec.~\ref{sec:MF-fixed-density}.  If the smooth curvature becomes
anomalously small, the perturbative estimate in
Eq.~\eqref{app:dq-saddle-fixedN} is no longer uniform and the full
thermodynamic potential must be minimized instead.

To second order in the oscillatory amplitude, $\Delta(q)$ in the Legendre
term may be evaluated at $q_{\rm sm}$.  Since $n_{\rm osc}^2$ is already
second order whereas the saddle displacement is first order, their product
contributes only at third order.  The Legendre term therefore gives the
leading canonical harmonic reorganization.

\section{Fermi-surface irrelevant correction and finite-amplitude continuation}
\label{app:irrelevant-term}
The radial dispersion correction introduced in Sec.~\ref{sec:mean-field}
serves two related but distinct purposes.  First, it supplies the local
quadratic and quartic coefficients of the nematic Landau expansion, including
the corrections entering Eqs.~\eqref{eq:X2-complete-lowenergy},
\eqref{eq:u-gamma-smooth}, and \eqref{eq:u-gamma-osc}.  Second, a global
minimization requires a potential defined at finite nematic amplitude, beyond
the information contained in these coefficients.  The distinction concerns
the range of validity of the radial operator, which is an effective
Fermi-surface correction rather than a microscopic dispersion over the entire
band.

An unrestricted sum over occupied Landau levels includes states far below
the Fermi surface even in the limit $q\to0$.  Small $q$ alone does not remove
those states.  Instead, the local coefficients are extracted by expanding
the sum and specifying how analytic contributions from the remote band bottom
are treated.  At $B=0$, a cancellation in the deformation-dependent part of
the grand potential can be exhibited directly.  At finite $B$, a lower-edge
contribution remains in the discrete level sum and is explicitly subtracted
in the low-energy prescription adopted here.  Once the local coefficients
are fixed, a separately regulated continuation supplies the finite-$q$
potential.  Matching counterterms preserve those coefficients, leaving the
regulator dependence in terms of order $q^6$ and higher.  We keep the
zero-temperature stabilization mechanism of
Refs.~\cite{OgKiFr2001,BaOx2003,Lawler2006} and extend the local calculation
to finite temperature and magnetic field.

\subsection{Zero-field reference}
We start at $B=0$, where the cancellation of the deep-state contribution to
the deformation-dependent energy can be seen explicitly.  The radial operator
introduced in Sec.~\ref{sec:mean-field} is
\begin{equation}
\delta\varepsilon_\gamma({\bf D})
=\frac{\gamma}{8mk_F^4}({\bf D}^2-k_F^2)^3,
\label{app:gamma-op}
\end{equation}
where $k_F^2=2m\mu$.  Without a magnetic field it becomes
\begin{equation}
\delta\varepsilon_\gamma({\bf \nabla})
=\frac{\gamma}{8mk_F^4}(-{\bf \nabla}^2-k_F^2)^3,
\label{app:gamma-op-B0}
\end{equation}
or, in momentum space,
\begin{equation}
\delta\varepsilon_\gamma({\bf k})
=\frac{\gamma}{8mk_F^4}({\bf k}^2-k_F^2)^3.
\label{app:gamma-op-k}
\end{equation}
We choose the nematic director as the $x$ principal axis and let $\vartheta$
denote the polar angle of ${\bf k}$ measured from it.  Introducing
\begin{equation}
\rho=\frac{k^2}{k_F^2},
\end{equation}
we write the dispersion as
\begin{equation}
\frac{\varepsilon(\rho,\vartheta)}{\mu}
=\rho(1+q\cos2\vartheta)+\frac{\gamma}{4}(\rho-1)^3.
\label{app:dispersion-y}
\end{equation}
The Fermi surface satisfies
$\varepsilon[\rho_F(\vartheta),\vartheta]=\mu$.  Solving perturbatively,
with $\rho_F=1+a_1q\cos(2\vartheta)+\cdots$, gives
\begin{align}
\rho_F={}&1-q\cos2\vartheta+q^2\cos^2(2\vartheta)
+\frac{\gamma-4}{4}q^3\cos^3(2\vartheta)
\nonumber\\
&+(1-\gamma)q^4\cos^4(2\vartheta)+O(q^5).
\label{app:yF-series}
\end{align}

At zero temperature, the fermionic grand potential is the integral of
$\varepsilon-\mu$ over occupied states.  Since
$k\,dk=(k_F^2/2)d\rho=m\mu\,d\rho$, subtracting the isotropic value gives
\begin{align}
\Delta\Omega_f
=-\frac{m\mu^2}{4\pi^2}
\int_0^{2\pi}d\vartheta\Bigg\{&
\int_0^{\rho_F}d\rho\,[1-\varepsilon_q(\rho,\vartheta)/\mu]
\nonumber\\
&-\int_0^1d\rho\,[1-\varepsilon_0(\rho)/\mu]\Bigg\}.
\label{app:Omega-gamma-integral}
\end{align}
Inserting Eq.~\eqref{app:yF-series} and using
$\langle\cos^2(2\vartheta)\rangle=1/2$ and
$\langle\cos^4(2\vartheta)\rangle=3/8$ yields
\begin{equation}
\Delta\Omega_f
=-\frac{m\mu^2}{8\pi}q^2
+\frac{3m\mu^2}{256\pi}(\gamma-8)q^4
+O(q^6).
\label{eq:gamma-Landau-zero}
\end{equation}
Thus the radial term leaves the quadratic coefficient unchanged at
$T=B=0$ and contributes $3\gamma/16$ to the dimensionless quartic
coefficient.  Adding the Hubbard--Stratonovich term reproduces
Eq.~\eqref{eq:r0-u0}, $u_0=3(\gamma-8)/16$, so the local fixed-$\mu$
saddle is stabilized for $\gamma>8$.

The fact that the original integral extends to $\rho=0$ does not, by itself,
invalidate this local result.  To see the cancellation explicitly, treat the
radial term to first order in $\gamma$.  The unperturbed parabolic Fermi
surface is
\begin{equation}
\rho_F^{(0)}(\vartheta)=\frac{1}{1+q\cos2\vartheta}.
\end{equation}
Because the radial operator is independent of $q$ in momentum space, all
states occupied in both configurations cancel in the difference
$\delta\Omega_\gamma(q)-\delta\Omega_\gamma(0)$.  The remaining contribution
is
\begin{align}
\delta\Omega_\gamma^{(1)}(q)
-\delta\Omega_\gamma^{(1)}(0)
={}&\frac{\gamma m\mu^2}{16\pi^2}
\int_0^{2\pi}d\vartheta
\int_1^{\rho_F^{(0)}(\vartheta)}d\rho\,(\rho-1)^3
\nonumber\\
={}&\frac{\gamma m\mu^2}{64\pi^2}
\int_0^{2\pi}d\vartheta\,
[\rho_F^{(0)}(\vartheta)-1]^4.
\label{app:gamma-FS-cancellation}
\end{align}
Here the superscript $(1)$ denotes first order in the radial perturbation.
For $q\ll1$, $\rho_F^{(0)}-1=-q\cos2\vartheta+O(q^2)$, and
Eq.~\eqref{app:gamma-FS-cancellation} gives
$3\gamma m\mu^2q^4/(256\pi)+O(q^6)$, as in
Eq.~\eqref{eq:gamma-Landau-zero}.  Only the thin region between the
isotropic and distorted Fermi surfaces contributes to this difference;
its characteristic energy width is of order $\mu|q|$.  The cancellation
relies on the zero-field momentum representation and on the perturbative
comparison at fixed microscopic dispersion.  It should not be assumed to
remove every analytic deep-band contribution in the Landau-level sums at
finite magnetic field.

\subsection{Landau-level shifts}
\label{app:radial-level-shifts}
At finite magnetic field, the quadratic nematic Landau problem has been
solved in Sec.~\ref{sec:nematic-model-spectrum}.  Its unperturbed levels are
\[
E_\lambda^{(0)}(q)
=\left(\lambda+\frac12\right)\omega_c\Delta(q),
\qquad \Delta(q)=\sqrt{1-q^2},
\]
as in Eq.~\eqref{eq:ciclotron}.  We use the anisotropic Landau states as
the unperturbed basis and treat Eq.~\eqref{app:gamma-op} to first order in
perturbation theory.  The level shift is
\begin{equation}
\delta E_\lambda(q)
=\langle\lambda,q|\delta\varepsilon_\gamma({\bf D})|\lambda,q\rangle.
\label{app:deltaE}
\end{equation}
Expanding the cubic operator gives
\begin{align}
\delta E_\lambda(q)
=\frac{\gamma}{8mk_F^4}\Big[&
\langle({\bf D}^2)^3\rangle_\lambda
-3k_F^2\langle({\bf D}^2)^2\rangle_\lambda
\nonumber\\
&+3k_F^4\langle{\bf D}^2\rangle_\lambda-k_F^6\Big].
\label{app:gamma-levelshift}
\end{align}

It is important here to distinguish the original operator
$D^2=D_x^2+D_y^2$ from the quadratic nematic Hamiltonian.  In its
principal-axis frame, set
$\widetilde D_x=\sqrt{1+q}\,D_x$,
$\widetilde D_y=\sqrt{1-q}\,D_y$, and $s_\lambda=2\lambda+1$.
Since $[\widetilde D_x,\widetilde D_y]=-iB\Delta$, the oscillator satisfies
\[
\langle\widetilde D_x^2\rangle_\lambda
=\langle\widetilde D_y^2\rangle_\lambda
=\frac{B s_\lambda\Delta}{2}.
\]
Transforming back and evaluating the higher oscillator moments yields
\begin{align}
\langle {\bf D}^2\rangle_\lambda
&=\frac{B s_\lambda}{\Delta},
\label{app:D2-moment}\\
\langle ({\bf D}^2)^2\rangle_\lambda
&=B^2\left[
 s_\lambda^2
+\frac{3q^2}{2(1-q^2)}(s_\lambda^2+1)
\right],
\label{app:D4-moment}\\
\langle ({\bf D}^2)^3\rangle_\lambda
&=\frac{B^3s_\lambda}{2(1-q^2)^{3/2}}
\left[2s_\lambda^2+q^2(3s_\lambda^2+25)\right].
\label{app:D6-moment}
\end{align}
At $q=0$, these reduce to
$\langle({\bf D}^2)^n\rangle_\lambda=[B(2\lambda+1)]^n$.
The isotropic shift is accordingly
\begin{equation}
\delta E_\lambda(0)
=\frac{\gamma}{8mk_F^4}
\left[B(2\lambda+1)-k_F^2\right]^3.
\label{app:gamma-levelshift-iso}
\end{equation}
It is cubic in the distance of the Landau-level kinetic energy from the
reference Fermi surface.  Nevertheless, evaluating it in an unrestricted
occupied-level sum samples the entire filled band.  The low-energy
interpretation must therefore be imposed on the thermodynamic quantities,
not inferred from the form of the level shift alone.

\subsection{Local coefficients at fixed chemical potential and low-energy prescription}
\label{app:radial-local-mu}
We next calculate the radial corrections to the local quadratic and
quartic coefficients, denoted by $\delta{\cal X}_{2,\gamma}$ and
$\delta u_\gamma$ in Sec.~\ref{sec:mean-field}.  We first expand the
unrestricted level sum, and then identify its low-energy continuum and
oscillatory contributions.  Analytic terms arising from the lower band edge
are separated explicitly.  In particular, these subtractions are already
needed when $q\to0$; they are not introduced only for finite-amplitude
minimization.

Expanding the level shift in even powers of $q$,
\begin{equation}
\delta E_\lambda(q)
=\delta E_\lambda^{(0)}+q^2\delta E_\lambda^{(2)}+O(q^4),
\end{equation}
and introducing
\begin{equation}
\epsilon_B=\frac{\omega_c}{\mu},\qquad
x_\lambda=\frac{E_\lambda}{\mu}
=\frac{B}{k_F^2}(2\lambda+1)
=\epsilon_B\left(\lambda+\frac12\right),
\end{equation}
we find
\begin{align}
\delta E_\lambda^{(0)}
&=\frac{\gamma\mu}{4}(x_\lambda-1)^3,
\label{app:gamma-E0}\\
\delta E_\lambda^{(2)}
&=\frac{\gamma\mu}{4}
\Bigg[\frac32x_\lambda(x_\lambda-1)(2x_\lambda-1)
\nonumber\\
&\hspace{2.0cm}
+\frac{B^2}{2k_F^4}(25x_\lambda-9)\Bigg].
\label{app:gamma-E2}
\end{align}

To first order in the radial operator, its grand-potential contribution is
\begin{equation}
\delta\Omega_\gamma(q)
={\cal D}_B\sum_{\lambda=0}^{\infty}
 n_F[E_\lambda^{(0)}(q)]\,\delta E_\lambda(q).
\label{app:gamma-Omega}
\end{equation}
In this appendix, $n_F(E)=[e^{\beta(E-\mu)}+1]^{-1}$ takes the absolute
one-particle energy as its argument; in the relative-energy convention of
Sec.~\ref{sec:mean-field} and Appendices~\ref{app:fixedN-nematic} and
\ref{app:magnetization}, the same function is written $n_F(E-\mu)$.
Primes on $n_F$ below denote derivatives with respect to $E$ at fixed $\mu$.

Since $E_\lambda^{(0)}(q)=E_\lambda\Delta(q)$,
\[
E_\lambda^{(0)}(q)=E_\lambda
\left(1-\frac{q^2}{2}+O(q^4)\right),
\]
so that
\[
n_F[E_\lambda^{(0)}(q)]
=n_F(E_\lambda)-\frac{E_\lambda}{2}n_F'(E_\lambda)q^2
+O(q^4).
\]
Comparing the coefficient of $q^2$ in
Eq.~\eqref{app:gamma-Omega} with $(a/2)r q^2$, where
$r=1/|F_2|-{\cal X}_2$, gives
\begin{align}
\delta{\cal X}_{2,\gamma}(T,B)
=-\frac{2{\cal D}_B}{a}\sum_{\lambda=0}^{\infty}
\Bigg\{&n_F(E_\lambda)\,\delta E_\lambda^{(2)}
\label{app:deltaX2-gamma}\\
&-\frac{E_\lambda}{2}n_F'(E_\lambda)
\,\delta E_\lambda^{(0)}\Bigg\}.
\nonumber
\end{align}
Here $E_\lambda=(\lambda+1/2)\omega_c$.
This formula is exact at the retained perturbative order in $\gamma$
\emph{for the unrestricted level sum}.  It does not yet implement the
low-energy subtraction at the bottom of the band.

Introducing $t=T/\mu$ and $f(x)=[e^{(x-1)/t}+1]^{-1}$, we can write
\begin{equation}
\delta{\cal X}_{2,\gamma}
=-\gamma\epsilon_B\sum_{\lambda=0}^{\infty}{\cal F}(x_\lambda),
\label{app:deltaX-Fsum}
\end{equation}
where
\begin{align}
{\cal F}(x)={}&f(x)\left[
\frac32x(x-1)(2x-1)
+\frac{\epsilon_B^2}{8}(25x-9)\right]
\nonumber\\
&-\frac{x}{2}f'(x)(x-1)^3.
\label{app:Fgamma-y}
\end{align}
The half-integer Poisson formula separates the continuum integral from
nonzero harmonics:
\begin{align}
\epsilon_B\sum_{\lambda=0}^{\infty}{\cal F}(x_\lambda)
={}&\int_0^\infty dx\,{\cal F}(x)
\label{app:Poisson-gamma}\\
&+2\sum_{n=1}^{\infty}(-1)^n
\int_0^\infty dx\,{\cal F}(x)
\cos\left(\frac{2\pi nx}{\epsilon_B}\right).
\nonumber
\end{align}
A Sommerfeld expansion of the continuum contribution gives
\begin{equation}
\int_0^\infty dx\,{\cal F}(x)
=\frac{\pi^2}{4}t^2+\frac{7}{16}\epsilon_B^2
+O(t^4,\epsilon_B^2t^2).
\label{app:gamma-smooth-integral}
\end{equation}

\paragraph*{Lower-band-edge subtraction at small deformation.}
The continuum approximation to a Landau-level sum contains analytic
corrections from the lower endpoint $x=0$.  For a sufficiently smooth test
function $h(x)$, the midpoint Euler--Maclaurin expansion reads
\begin{equation}
\epsilon_B\sum_{\lambda=0}^{\infty}
h\!\left[\epsilon_B\left(\lambda+\frac12\right)\right]
=\int_0^\infty dx\,h(x)
+\frac{\epsilon_B^2}{24}h'(0)+O(\epsilon_B^4),
\label{app:midpoint-EM-edge}
\end{equation}
where the displayed correction arises from the \emph{bottom} of the band.
For the degenerate Fermi function, the Fermi-edge oscillations are treated
separately by the nonzero Poisson harmonics in
Eq.~\eqref{app:Poisson-gamma}; Eq.~\eqref{app:midpoint-EM-edge}
identifies the smooth lower-endpoint contribution.

From Eq.~\eqref{app:Fgamma-y},
${\cal F}'(0)=3/2+O(\epsilon_B^2)$, up to corrections exponentially
small in $\mu/T$.  Thus the \emph{unsubtracted} quadratic susceptibility
contains
\begin{equation}
\delta{\cal X}_{2,\gamma}^{\rm edge}
=-\gamma\frac{\epsilon_B^2}{24}{\cal F}'(0)
=-\frac{\gamma\epsilon_B^2}{16}
+O(\gamma\epsilon_B^4).
\label{app:gamma-edge-quadratic}
\end{equation}
This term comes from the extrapolation of the radial operator to the
remote band bottom, even though it is extracted from the $q^2$ coefficient
at $q=0$.  It is not a new Fermi-surface response, nor does taking
$q\to0$ eliminate it.  The effective low-energy prescription adopted in
this work \emph{subtracts} this analytic lower-edge contribution.
Equations~\eqref{app:deltaX2-gamma} and \eqref{app:deltaX-Fsum}
therefore refer to the unrestricted expressions \emph{before} subtraction;
the low-energy coefficient used in the main text is defined \emph{after}
it.  With this convention,
\begin{align}
\delta{\cal X}_{2,\gamma}^{\rm sm}
={}&-\gamma\left[
\frac{\pi^2}{4}\left(\frac{T}{\mu}\right)^2
+\frac{7}{16}\left(\frac{\omega_c}{\mu}\right)^2
\right]
\nonumber\\
&+O\!\left(\gamma\frac{T^4}{\mu^4},
\gamma\frac{\omega_c^2T^2}{\mu^4}\right).
\label{eq:deltaX-gamma-smooth}
\end{align}
The subtraction is a choice of effective-theory matching, not a cancellation
that follows automatically from expanding in the nematic amplitude.

The leading Fermi-surface harmonic comes from the terms near $x=1$.
Using the Fourier transform of the thermal Fermi edge,
\begin{equation}
\int dx\,[-f'(x)]e^{i\theta_nx}
=e^{i\theta_n}R_T(\chi_n),
\end{equation}
where $\theta_n=2\pi n/\epsilon_B$, and integrating by parts gives
\begin{align}
\left.\int dx\,f(x)(x-1)e^{i\theta_nx}\right|_{\rm FS}
={}&\frac{e^{i\theta_n}}{\theta_n^2}R_T(\chi_n)
\left[1+\chi_nL(\chi_n)\right].
\label{app:linear-Fourier}
\end{align}
Here $|_{\rm FS}$ indicates the Fermi-edge contribution, with analytic
band-bottom terms excluded by the stated prescription.  The resulting
oscillatory susceptibility is
\begin{align}
\delta{\cal X}_{2,\gamma}^{\rm osc}
={}&-\frac{3\gamma}{4}
\left(\frac{\omega_c}{\mu}\right)^2
\sum_{n=1}^{\infty}
\frac{(-1)^n}{\pi^2n^2}R_T(\chi_n)
\nonumber\\
&\times\left[1+\chi_nL(\chi_n)\right]\cos\theta_n
+O\!\left(\gamma\left(\frac{\omega_c}{\mu}\right)^3\right).
\label{eq:deltaX-gamma-osc}
\end{align}
Together with the parabolic result,
Eqs.~\eqref{eq:deltaX-gamma-smooth} and
\eqref{eq:deltaX-gamma-osc} give
Eq.~\eqref{eq:X2-complete-lowenergy}.  The radial term changes the leading
oscillatory quadratic susceptibility only at order
$\gamma\epsilon_B^2$; the $O(\epsilon_B)\sin\theta_n$ contribution
remains that of the parabolic theory.

For the quartic coefficient we continue the expansion to $q^4$:
\begin{equation}
\delta E_\lambda(q)
=\delta E_\lambda^{(0)}+q^2\delta E_\lambda^{(2)}
+q^4\delta E_\lambda^{(4)}+O(q^6),
\end{equation}
where
\begin{align}
\delta E_\lambda^{(4)}
=\frac{\gamma\mu}{4}\Bigg[&
\frac{33}{8}x_\lambda^3-\frac92x_\lambda^2
+\frac98x_\lambda
\nonumber\\
&+\frac{\epsilon_B^2}{4}
\left(\frac{75}{4}x_\lambda-\frac92\right)\Bigg].
\label{app:gamma-E4}
\end{align}
The occupation factor becomes
\begin{align}
n_F(E_\lambda\Delta)
={}&n_F(E_\lambda)-\frac{E_\lambda}{2}n_F'(E_\lambda)q^2
\label{app:nF}\\
&+\frac18\left[E_\lambda^2n_F''(E_\lambda)
-E_\lambda n_F'(E_\lambda)\right]q^4
+O(q^6).
\nonumber
\end{align}
Multiplying these expansions and comparing with
$(a/4)\delta u_\gamma q^4$ gives the unrestricted level sum
\begin{align}
\delta u_\gamma(T,B)
={}&\frac{4{\cal D}_B}{a}\sum_{\lambda=0}^{\infty}\Bigg\{
 n_F(E_\lambda)\delta E_\lambda^{(4)}
\nonumber\\
&-\frac{E_\lambda}{2}n_F'(E_\lambda)
\delta E_\lambda^{(2)}
\nonumber\\
&+\frac18\left[E_\lambda^2n_F''(E_\lambda)
-E_\lambda n_F'(E_\lambda)\right]
\delta E_\lambda^{(0)}\Bigg\}.
\label{app:delta-u-gamma-levelsum}
\end{align}
This expression is also useful for numerical checks on the expansion.

The same lower-band-edge prescription applies to the quartic coefficient.
To make its magnitude explicit, write the dimensionless summand in
Eq.~\eqref{app:delta-u-gamma-levelsum} as
$\delta u_\gamma=2\gamma\epsilon_B\sum_\lambda{\cal U}(x_\lambda)$.
For $T\ll\mu$, terms involving derivatives of the Fermi function are
exponentially small at $x=0$, and Eq.~\eqref{app:gamma-E4} gives
${\cal U}'(0)=9/8+O(\epsilon_B^2)$.  The midpoint contribution to the
unsubtracted quartic coefficient is therefore
\begin{equation}
\delta u_\gamma^{\rm edge}
=2\gamma\frac{\epsilon_B^2}{24}{\cal U}'(0)
=\frac{3\gamma\epsilon_B^2}{32}
+O(\gamma\epsilon_B^4).
\label{app:gamma-edge-quartic}
\end{equation}
As for the quadratic term, this is an analytic band-bottom contribution
present already in the small-$q$ expansion.  We subtract it when defining
the local low-energy quartic coefficient.  The continuum contribution,
after integrations by parts and a Sommerfeld expansion, is
\begin{align}
u_\gamma^{\rm sm}
={}&\frac{3\gamma}{16}
+\frac{15\pi^2\gamma}{4}\left(\frac{T}{\mu}\right)^2
+\frac{71\gamma}{16}\epsilon_B^2
\nonumber\\
&+O\!\left(\gamma\frac{T^4}{\mu^4},
\gamma\epsilon_B^2\frac{T^2}{\mu^2},
\gamma\epsilon_B^4\right).
\end{align}
This is Eq.~\eqref{eq:u-gamma-smooth} of the main text.  Its constant
contribution $3\gamma/16$ agrees with the independent zero-field
calculation in Eq.~\eqref{eq:gamma-Landau-zero}.

The leading oscillatory quartic correction is obtained by applying the
half-integer Poisson formula to
Eq.~\eqref{app:delta-u-gamma-levelsum} and retaining the Fermi-edge
contribution:
\begin{align}
\delta u_\gamma^{\rm osc}
={}&\frac{3\gamma\epsilon_B}{2\pi}
\sum_{n=1}^{\infty}\frac{(-1)^n}{n}R_T(\chi_n)
\left[1-\chi_nL(\chi_n)\right]\sin\theta_n
\nonumber\\
&+O(\gamma\epsilon_B^2).
\end{align}
This is Eq.~\eqref{eq:u-gamma-osc} of the main text.  The radial
correction therefore does not change the leading $O(1)$ parabolic
oscillation of the quartic coefficient; its first oscillatory correction
is smaller by one power of $\omega_c/\mu$.

We can summarize the local control parameter directly from
Eq.~\eqref{app:gamma-op-k}.  With
\begin{equation}
\xi=\frac{k^2-k_F^2}{2m},
\end{equation}
the relative size of the radial correction is
\begin{equation}
\frac{|\delta\varepsilon_\gamma(\xi)|}{|\xi|}
=\frac{\gamma}{4}\left(\frac{\xi}{\mu}\right)^2.
\end{equation}
For the Fermi-edge terms at finite temperature and weak magnetic field,
$|\xi|\lesssim\max(T,\omega_c)$, so the perturbative expansion is
controlled when
\begin{equation}
\frac{\gamma}{4}
\left[\frac{\max(T,\omega_c)}{\mu}\right]^2\ll1.
\end{equation}
This criterion applies to the low-energy contributions, not to the
unrestricted sum over all occupied levels.  The analytic edge terms in
Eqs.~\eqref{app:gamma-edge-quadratic} and
\eqref{app:gamma-edge-quartic} show explicitly why a small-$q$
expansion still requires a prescription for the band bottom.  The
subtraction fixes the local coefficients used in the main text.

\subsection{Radial corrections at fixed density}
\label{app:canonical-radial}
We now add the radial dispersion correction to first order in $\gamma$ to the coefficients of the Landau 
expansion at fixed density.
The microscopic reference momentum is held fixed at
$k_{F0}^2=2m\mu_0$ during the Legendre transformation.  By stationarity of
the parabolic Legendre functional, the $O(\gamma)$ correction to the
canonical free energy is simply
\begin{equation}
\delta{\cal F}_{N,\gamma}(q)
=\delta\Omega_\gamma
[q,\mu_{\rm par}(q);k_{F0}]+O(\gamma^2);
\label{app:gamma-envelope-fixedN}
\end{equation}
the $O(\gamma)$ shift of the chemical potential produces no additional
first-order term.

We use the level-shift expansion
\begin{equation}
\delta E_\lambda(q)
=\delta E_\lambda^{(0)}
+q^2\delta E_\lambda^{(2)}
+q^4\delta E_\lambda^{(4)}+O(q^6),
\end{equation}
with Eqs.~\eqref{app:gamma-E0}, \eqref{app:gamma-E2}, and
\eqref{app:gamma-E4} evaluated using the fixed reference
$k_{F0}^2=2m\mu_0$.  Combining this expansion with
Eqs.~\eqref{app:mu2-fixedN} and \eqref{app:mu4-fixedN} gives
\begin{align}
\delta r_{N,\gamma}
={}&\frac{2{\cal D}_B}{A_0}\sum_\lambda
\left[
 f_\lambda\delta E_\lambda^{(2)}
 +f_\lambda'(E_{\lambda,2}-\mu_2)\delta E_\lambda^{(0)}
\right],
\label{eq:rN-gamma-explicit}\\
\delta u_{N,\gamma}
={}&\frac{4{\cal D}_B}{A_0}\sum_\lambda
\Bigg\{
 f_\lambda\delta E_\lambda^{(4)}
 +f_\lambda'(E_{\lambda,2}-\mu_2)\delta E_\lambda^{(2)}
\nonumber\\
&
+\left[
 f_\lambda'(E_{\lambda,4}-\mu_4)
 +\frac12 f_\lambda''(E_{\lambda,2}-\mu_2)^2
 \right]\delta E_\lambda^{(0)}
\Bigg\}.
\label{eq:uN-gamma-explicit}
\end{align}
These expressions are the fixed-density counterparts of the local
radial-correction formulas derived in the previous subsection.
The coefficients used in the numerical phase diagrams are therefore
\begin{equation}
r_N=r_{N,{\rm par}}^{\rm levels}+\delta r_{N,\gamma},
\qquad
u_N=u_{N,{\rm par}}^{\rm levels}+\delta u_{N,\gamma}.
\label{eq:rN-uN-complete}
\end{equation}
They are exact in the parabolic thermal occupations at finite temperature
and retain the leading contribution of the radial dispersion correction.
They remain, of course, a local expansion in the nematic amplitude.

The zero-field limit provides a simple check.  As $B,T\to0$,
${\cal D}_B\sum_\lambda\to(m/2\pi)\int_0^\infty d\epsilon$.  The Fermi-edge terms
from $\delta E_\lambda^{(0)}$ and $\delta E_\lambda^{(2)}$ vanish, and the
remaining integral gives
\begin{equation}
\delta r_{N,\gamma}(0,0)=0,
\qquad
\delta u_{N,\gamma}(0,0)=\frac{3\gamma}{16}.
\end{equation}
Together with $u_{N,{\rm par}}(0,0)=-1/2$, this reproduces
Eq.~\eqref{eq:rN-uN-zero},
\begin{equation*}
r_N(0,0)=\frac1{|F_2|}-1,
\qquad
u_N(0,0)=\frac{3\gamma-8}{16}.
\end{equation*}
Thus the local fixed-density saddle is stabilized for $\gamma>8/3$, a weaker
condition than the fixed-chemical-potential criterion $\gamma>8$ in
Eq.~\eqref{eq:r0-u0}.

\subsection{Regulated finite-amplitude continuation and matching}
\label{app:numerical-completion}
The local coefficients determine the expansion of the nematic potential
through $q^4$, but do not determine its global behavior.  Although the
unrestricted sum in Eq.~\eqref{app:gamma-Omega} already samples deep levels
for $q\ll1$, the finite-amplitude problem raises an additional issue:
the deformation changes which states contribute, and terms of order $q^6$
and higher depend on the continuation of the radial operator away from the
low-energy region.  For this reason, we restrict the explicit radial
correction to a smooth energy shell centered on the isotropic Fermi surface.

For the isotropic reference level, define
$\xi_\lambda=(\lambda+1/2)\omega_c-\mu$.  The shell profile is
\begin{equation}
W_{\Lambda,p}(\xi)=\exp[-|\xi/\Lambda|^p],
\label{app:global-regulator-cutoff}
\end{equation}
where $\Lambda$ is the shell width and $p$ its profile exponent.
The regulated correction is
\begin{equation}
\delta\Omega_{\gamma,W}(q)
={\cal D}_B\sum_{\lambda\geq0}
W_{\Lambda,p}(\xi_\lambda)
 n_F[E_\lambda^{(0)}(q)]\,\delta E_\lambda(q).
\label{app:global-regulator}
\end{equation}
The regulator is defined relative to the isotropic energy and does not
depend explicitly on $q$.  The dependence on nematic order enters through
the occupations and level shifts.  The parameters $\Lambda$ and $p$ define
a continuation of the low-energy operator. A useful
diagnostic is $\eta_\Lambda\sim(\gamma/4)(\Lambda/\mu)^2$, which estimates
the size of the radial correction at the edge of the retained shell.

We work with the dimensionless correction
\begin{equation}
G_W(q)\equiv\frac{\delta\Omega_{\gamma,W}(q)}{a},
\end{equation}
and expand
\begin{equation}
G_W(q)-G_W(0)=g_2q^2+g_4q^4+O(q^6).
\label{app:GW-expansion}
\end{equation}
The regulated coefficients $g_2$ and $g_4$ generally differ from the
analytic low-energy coefficients, because the regulator changes the
level sum.  We therefore subtract these regulator-dependent terms and
replace them with the previously defined values
$\delta r_\gamma=-\delta{\cal X}_{2,\gamma}$ and
$\delta u_\gamma=u_\gamma^{\rm sm}+\delta u_\gamma^{\rm osc}$.
The matched finite-amplitude correction used in
Eq.~\eqref{eq:full-potential-mu} is
\begin{align}
\delta\omega_\gamma^{\rm ct}(q)
={}&G_W(q)-G_W(0)
+\left(\frac{\delta r_\gamma}{2}-g_2\right)q^2
\nonumber\\
&+\left(\frac{\delta u_\gamma}{4}-g_4\right)q^4
+c_6q^6+c_8q^8.
\label{eq:completion-counterterms}
\end{align}
The optional dimensionless continuation coefficients $c_6$ and $c_8$
vanish in the reference calculation.  The matching counterterms are
recomputed at each $(T,B)$ and do not change the target local coefficients.
By construction,
\[
\delta\omega_\gamma^{\rm ct}(q)
=\frac{\delta r_\gamma}{2}q^2
+\frac{\delta u_\gamma}{4}q^4+O(q^6),
\]
independently of $\Lambda$ and $p$ through quartic order.  The
regulator dependence therefore first appears at $q^6$.  When taking
field derivatives for magnetic response, both the regulator and the
counterterms must be differentiated along with the rest of the potential;
see Appendix~\ref{app:magnetization}.

\subsubsection{Canonical continuation}
\label{app:canonical-continuation}

At fixed density we construct a finite-amplitude canonical free energy by
solving the full parabolic density constraint at every $q$ and adding the
regulated radial correction of Eqs.~\eqref{app:global-regulator} and  \eqref{app:global-regulator-cutoff}, now
referenced to $\mu_0$.  Here $\Lambda$ is the width of the retained energy
shell and $p$ is the regulator exponent defined in
Eq.~\eqref{app:global-regulator-cutoff}.  We denote this quantity by
${\cal F}_{N,{\rm reg}}(q)$
and write
\begin{equation}
\frac{{\cal F}_{N,{\rm reg}}(q)-{\cal F}_{N,{\rm reg}}(0)}{A_0}
=g_2q^2+g_4q^4+O(q^6).
\end{equation}
Counterterms restore the local coefficients in
Eq.~\eqref{eq:rN-uN-complete}.  The potential used in the test is therefore
augmented by optional coefficients $c_6$ and $c_8$ multiplying $q^6$ and
$q^8$, respectively, only to test sensitivity to the finite-amplitude
continuation:
\begin{align}
V_N(q)={}&
\frac{{\cal F}_{N,{\rm reg}}(q)-{\cal F}_{N,{\rm reg}}(0)}{A_0}
+\left(\frac{r_N}{2}-g_2\right)q^2
\nonumber\\
&+\left(\frac{u_N}{4}-g_4\right)q^4
+c_6q^6+c_8q^8.
\label{app:canonical-counterterm-potential}
\end{align}
The matching also separates local and global conclusions.  If the local
quartic coefficient becomes negative while the potential is stabilized at
larger amplitude, a continuous instability of the isotropic state can be
preempted by first-order coexistence.  The precise coexistence field and
order-parameter jump depend on higher-order terms.  When $u>0$, the
existence of a more distant competing minimum is itself sensitive to those
terms.  These distinctions motivate the regulator-variation tests below.

\subsection{Global minimization and endpoint structure}
\label{app:global-endpoints}
The local expansion determines stability near $q=0$, whereas the global
transition requires comparison of all candidate minima.  We minimize the
full potential of Eq.~\eqref{eq:full-potential-mu}, using the matched
radial term in Eq.~\eqref{eq:completion-counterterms}.

For the equilibrium boundaries in
Figs.~\ref{fig:full-potential-first-order}(b) and
\ref{fig:ensemble-comparison-cuts}(a), we evaluate $\mathcal R(q)$ at each
$(T,B)$, refine candidate finite-amplitude maxima, and compare their
values with $\mathcal R(0)={\cal X}_2$.  The largest value determines
$F_{2,c}$ through Eq.~\eqref{eq:full-potential-boundary}: a maximum at
$q=0$ corresponds to a continuous transition, whereas a dominant
finite-$q$ maximum gives first-order coexistence with a finite jump.
For specified $(T,B,F_2)$, we locate and refine local minima of
$V_\mu(q)$, always retaining the isotropic state $q=0$, where
$V_\mu(0)=0$.  Comparing their energies gives $q_{{\rm sp},\mu}$,
used in Fig.~\ref{fig:phase-maps-ensembles}(a).

Both searches sample a broad but finite interval below $q=1$, followed
by local refinement.  Quantitative results are quoted only for small
deformations, where the effective description is controlled.  In the
figures, $q\leq0.14$; for $\gamma=10$ this implies
$\gamma q^2\leq0.196$.  No lower large-$q$ minimum was found in the
boundary scans, but the behavior at large deformation is not interpreted
as a quantitative prediction of the low-energy theory.

The reference continuation uses $\Lambda/\mu=0.3$, $p=4$, and
$c_6=c_8=0$.

\paragraph{Representative first-order segment and endpoint structure.}

For $F_2=-1.0078$, the reference continuation gives coexistence at
\begin{equation}
\frac{\mu}{\omega_c}=18.54657836,
\qquad q_{\rm jump}=0.098599,
\end{equation}
with
\begin{equation}
r_\mu=0.00164484>0,
\qquad u_\mu=-0.663339.
\end{equation}
The positive quadratic coefficient means that the isotropic solution is
still locally stable when the finite-$q$ nematic minimum becomes
degenerate with it.  The transition is therefore first order: global
coexistence occurs before the local instability at $r_\mu=0$.  The
coexistence point lies between the appearance of a metastable nematic
minimum at $\mu/\omega_c\simeq18.5409981$ and the loss of local
isotropic stability at $\mu/\omega_c=18.5593725$.

The two solutions of $r=u=0$ illustrate the role of global minimization
in classifying endpoints.  Here $v$ denotes the coefficient of $q^6/6$
in the dimensionless local expansion of Eq.~\eqref{eq:phase-V-Landau}
for the reference continuation.  At the right candidate,
\[
\frac{\mu}{\omega_c}=18.65029362,
\]
we find $v=58.8336>0$, and the order-parameter jump vanishes on
approach.  The global potential thus confirms a tricritical endpoint
within this continuation.

At the left candidate,
\[
\frac{\mu}{\omega_c}=18.36085737,
\]
we instead obtain $v=-4.31455$.  A finite-amplitude minimum becomes
favorable before the local condition $r=u=0$ is reached, so that the
candidate is preempted.  The first-order and continuous boundaries meet
near $\mu/\omega_c=18.3600650$, with a coexisting deformation
$q\simeq0.06419$.

This asymmetry matters physically.  Negative local quartic coefficients
lead to first-order preemption in a stabilized potential, while the
classification of an endpoint reached with $u>0$ depends more directly
on higher-order terms.  Varying the regulator preserves the signs of the
sixth-order coefficients at the two reference candidates, whereas
adding allowed $q^6$ and $q^8$ terms can remove the finite-$q$
preemption of the left candidate.  The existence of the tested
first-order segment is therefore more robust than the detailed endpoint
locations and classifications.

The radial corrections in Eqs.~\eqref{eq:gamma-Landau-zero} and
\eqref{eq:deltaX-gamma-smooth}--\eqref{eq:deltaX-gamma-osc}
are those used in Sec.~\ref{sec:mean-field};
Eq.~\eqref{eq:completion-counterterms} defines the finite-amplitude
continuation used in Sec.~\ref{sec:full-potential}.  The level shift
\eqref{app:gamma-levelshift} also enters the magnetic response in
Appendix~\ref{app:magnetization}.

\subsection{Bounded uniform canonical preemption test}
\label{app:canonical-preemption-check}
The conditions $r_N=0$ and $u_N>0$ establish only the local
continuous-instability boundary.
To support the fixed-density phase boundaries used in
Sec.~\ref{sec:phase-diagram}, we also test whether a distinct uniform minimum
at finite $q$ can preempt the local transition.

Because $F_2$ enters this construction only through $r_N$, it is sufficient
to test the boundary $r_N=0$: if $V_N(q)>0$ for every $q>0$ there, no
finite-amplitude minimum can preempt the transition on the $r_N>0$ side.
For numerical stability we monitor $4V_N(q)/q^4$, whose $q\to0$ limit is
$u_N$.

For $\gamma=10$ we tested $T/\mu_0=0.006,0.009,0.012$,
$20\leq\mu_0/\omega_c\leq32$, and $0.005\leq q\leq0.99$.  Across the regulator
choices $(\Lambda/\mu_0,p)=(0.2,2),(0.3,4),(0.5,4),(0.3,6)$ and the
higher-order terms $(c_6,c_8)=(0,0),(1,0),(0,1),(0,-1)$, the minimum of
$4V_N/q^4$ at $r_N=0$ remains larger than $1.4046$.  Refining the reference
scan in both field and amplitude gives the same classification.  No uniform
first-order preemption is therefore detected within the sampled domain and
tested continuation family.  This numerical check supports the use of the local
canonical condition in Eq.~\eqref{eq:phase-V-canonical-condition}; it does
not assign quantitative significance to large deformations outside the
small-$q$ regime of the effective theory.

\subsection{Sensitivity to the finite-amplitude continuation}
\label{app:continuation-sensitivity}
The reference continuation uses $\Lambda/\mu=0.3$, $p=4$, and
$c_6=c_8=0$.  We also tested $(\Lambda/\mu,p)=(0.2,2)$,
$(0.5,4)$, and $(0.3,6)$.  For $T/\mu=0.006$, $\gamma=10$, and
$F_2=-1.0078$, all four prescriptions exhibit the same first-order
segment, within
\begin{align}
18.54618&\lesssim\frac{\mu}{\omega_c}\lesssim18.54673,
\nonumber\\
0.09804&\lesssim q_{\rm jump}\lesssim0.10005.
\end{align}
The first-order segment persists under these variations; its precise
position and jump retain the expected dependence on the chosen
finite-amplitude continuation.

\subsection{Numerical convergence and validation}
\label{app:completion-convergence}
For the numerical implementation, the scalar density equation
Eq.~\eqref{eq:canonical-full-density} is solved by bracketing and the analytic
coefficients above are then evaluated directly.  Fitting the independently
Legendre-transformed level-sum potential at small $q^2$ gives the same
coefficients to better than $3.2\times10^{-7}$ for the representative tests
with $\gamma=0$ and $10$; the relative density residual is below
$8\times10^{-15}$, and increasing the thermal level cutoff from $45T$ to
$60T$ changes the coefficients by less than $10^{-10}$.

Finally, convergence was checked separately from regulator dependence.
For the harmonic cutoff $n_{\max}$ of
Eq.~\eqref{eq:full-potential-S}, the coexistence inverse field is
$18.54675125$, $18.54657908$, and $18.54657836$ for
$n_{\max}=3,5,10$, respectively; $n_{\max}=20$ and $40$ agree with the
last value to the reported precision.  The jump changes by less than
$10^{-6}$ between $10$ and $20$ harmonics.  The regulated Landau-level
sum is truncated once $|\xi_\lambda/\Lambda|^p>40$; thresholds $25$,
$60$, and $90$ give the same sixth-order check at the shown precision.
Independent 55-digit differentiation verifies the counterterm matching,
while mesh refinement and equality-of-potential tests check the global
minimization.  The energy difference between the representative
coexisting minima is below $10^{-14}a$.

\section{Magnetic response}
\label{app:magnetization}
This appendix derives the magnetic response formulas used in
Sec.~\ref{sec:dhva-nematic}, including the thermodynamic constraint,
the motion of the nematic saddle, and the harmonic representation.
The numerical field sweeps and spectral analysis are described in
Appendix~\ref{app:numerical-spectral}.

The central thermodynamic point is Eq.~\eqref{eq:VI-envelope-M}.  Because the
equilibrium saddle satisfies \(\partial_q{\cal V}_X=0\), the magnetization is
obtained by differentiating the thermodynamic potential with respect to the
field at fixed \(q\), and only afterwards evaluating the result at
\(q=q_{{\rm sp},X}(B)\).  There is therefore no separate explicit
\(dq_{{\rm sp},X}/dB\) contribution to the magnetization.  The field dependence
of the saddle nevertheless enters through the value of \(q\) at which the
partial derivative is evaluated.  The susceptibility behaves differently:
a second field derivative probes the motion of the saddle explicitly, as
shown below.

\subsection{Magnetization at fixed chemical potential}
\label{app:fixedmu-magnetization}
At fixed chemical potential, the parabolic contribution follows directly by
differentiating the smooth and oscillatory grand potentials of
Eqs.~\eqref{eq:calFnon} and \eqref{eq:Fosc-MF} at fixed \(q\).  This gives the
smooth diamagnetic term in Eq.~\eqref{eq:VI-Mmu-diamagnetic} and the dHvA
response in Eq.~\eqref{eq:VI-Mmu-par}.  The derivative of the rapidly varying
phase produces the dominant sine contribution, while derivatives of the
prefactor and of the LK factor generate the smaller cosine quadrature.  At
\(q=0\) the result reduces to the free-gas expression
Eq.~\eqref{eq:Mosc_full}.

It remains to include the leading correction from the radial dispersion
operator.  Before introducing any finite-amplitude regulator, its
contribution to the grand potential is
\begin{equation}
\delta\Omega_\gamma(q)
={\cal D}_B\sum_{\lambda\geq0}
 f_\lambda^{(\mu)}\,\delta E_\lambda(q),
\end{equation}
where $f_\lambda^{(\mu)}=n_F[E_\lambda(q)-\mu]$.
The magnetic field enters this expression through three places: the Landau
level degeneracy, the occupation factors, and the level shift itself.
Using \(\partial_B{\cal D}_B={\cal D}_B/B\) and
\(\partial_BE_\lambda=E_\lambda/B\) at fixed \(q\), we obtain
\begin{align}
\delta M_{\mu,\gamma}(q)
={}&-{\cal D}_B\sum_{\lambda\geq0}\Bigg[
\frac{f_\lambda^{(\mu)}\delta E_\lambda}{B}
+\frac{E_\lambda}{B}(f_\lambda^{(\mu)})'\delta E_\lambda
\label{eq:VI-Mmu-gamma}\\
&\hspace{3.1cm}
+f_\lambda^{(\mu)}\partial_B\delta E_\lambda
\Bigg]_{q,\mu,k_F}.
\nonumber
\end{align}
Here \(\delta E_\lambda(q)\) is the level shift derived in
Eq.~\eqref{app:gamma-levelshift}, the prime denotes differentiation of the
Fermi function with respect to its energy argument, and
\(k_F^2=2m\mu\) is held fixed.  Equation~\eqref{eq:VI-Mmu-gamma} is the
unregulated low-energy response.  When the finite-amplitude potential of
Appendix~\ref{app:irrelevant-term} is used, the field derivatives of the
regulator and of the matching counterterms must be included as part of the
same thermodynamic derivative.

Together with Eqs.~\eqref{eq:VI-Mmu-diamagnetic} and
\eqref{eq:VI-Mmu-par}, this gives the fixed-\(\mu\) magnetization quoted in
Eq.~\eqref{eq:VI-Mmu-final}.

\subsection{Magnetization at fixed density}
\label{app:canonical-magnetization}
At fixed density, the chemical potential is not an external parameter: it
must adjust with the magnetic field so that the density remains constant.
The Legendre transformation makes the thermodynamic structure particularly
simple.  If \(\mu_*(B)\) is the chemical potential that satisfies the density
constraint, then
\begin{align}
M_N
&=-\left.\frac{d}{dB}
\left[\Omega(\mu_*,B)+\mu_*n\right]\right|_n
\nonumber\\
&=-\left.\frac{\partial\Omega}{\partial B}\right|_{\mu=\mu_*},
\label{eq:VI-Legendre-M}
\end{align}
because the terms proportional to \(d\mu_*/dB\) cancel by the density
constraint \(\partial_\mu\Omega+n=0\).  Thus the canonical magnetization is
the grand-canonical field derivative evaluated at the chemical potential
that enforces the fixed density.

For the parabolic Landau levels, the density constraint at fixed \(q\) is
\begin{equation}
n={\cal D}_B\sum_{\lambda\geq0}f_\lambda^{(N)},
\end{equation}
where
$f_\lambda^{(N)}=n_F[E_\lambda(q)-\mu_{\rm par}(q,B)]$.
Differentiating this equation with respect to \(B\) gives the field dependence
of the constraint chemical potential,
\begin{equation}
\partial_B\mu_{\rm par}
=
\frac{\sum_\lambda
\left[f_\lambda^{(N)}+E_\lambda(f_\lambda^{(N)})'\right]}
{B\sum_\lambda(f_\lambda^{(N)})'}.
\label{app:muB-full-density}
\end{equation}
The derivative of the Landau degeneracy is essential in this relation.

For the parabolic part, the dependence on \(\partial_B\mu_{\rm par}\) cancels
from the Legendre transform.  The result can be
written directly as
\begin{equation}
M_{N,{\rm par}}^{\rm levels}
=-\frac{{\cal D}_B}{B}\sum_\lambda
\left[g_\lambda+E_\lambda f_\lambda^{(N)}\right].
\label{app:MN-par-levels}
\end{equation}
Here we define
\begin{equation}
g_\lambda=-T\ln[1+\exp((\mu_{\rm par}-E_\lambda)/T)].
\end{equation}
This expression automatically contains the full finite-temperature
canonical level occupation and does not require a harmonic expansion.

The radial dispersion correction is treated to first order in \(\gamma\),
with the microscopic reference momentum \(k_{F0}^2=2m\mu_0\) held fixed.
Because the correction is evaluated along the field-dependent density
constraint, its derivative retains the factor \(\partial_B\mu_{\rm par}\):
\begin{align}
\delta M_{N,\gamma}(q)
={}&-{\cal D}_B\sum_{\lambda\geq0}\Bigg[
\frac{f_\lambda^{(N)}\delta E_\lambda}{B}
+f_\lambda^{(N)}\partial_B\delta E_\lambda
\label{eq:VI-MN-gamma}\\
&
+(f_\lambda^{(N)})'
\left(\frac{E_\lambda}{B}-\partial_B\mu_{\rm par}\right)
\delta E_\lambda
\Bigg]_{q,n,k_{F0}}.
\nonumber
\end{align}
Equations~\eqref{app:MN-par-levels} and \eqref{eq:VI-MN-gamma}, evaluated at
the equilibrium saddle, give Eq.~\eqref{eq:VI-MN-final}.  In the thermally
damped regime they reduce to the harmonic representation discussed in
Sec.~\ref{sec:dhva-nematic}; for the numerical equilibrium response we use the
full finite-temperature level sums.

\subsection{Susceptibility and the motion of the nematic saddle}
\label{app:equilibrium-susceptibility}
The susceptibility contains an additional piece of physics that is absent
from the magnetization.  The magnetization uses stationarity to remove the
explicit derivative of the saddle.  When the field derivative is taken once
more, however, the field-induced motion of the minimum contributes to the
response.

Differentiating the stationarity condition
\(\partial_q{\cal V}_X(q_{{\rm sp},X},B)=0\) gives
\begin{equation}
\frac{dq_{{\rm sp},X}}{dB}
=-\left.
\frac{\partial_B\partial_q{\cal V}_X}
{\partial_q^2{\cal V}_X}
\right|_{q=q_{{\rm sp},X}}.
\end{equation}
Substitution into \(\chi_X=dM_X/dB\) yields
\begin{equation}
\chi_X
=
-\partial_B^2{\cal V}_X
+\frac{(\partial_B\partial_q{\cal V}_X)^2}
{\partial_q^2{\cal V}_X}
\bigg|_{q=q_{{\rm sp},X}},
\label{eq:VI-chi-decomposition}
\end{equation}
which is Eq.~\eqref{eq:VI-chi-envelope} of the main text.  The first term is
the response at a fixed nematic background.  The second is the additional
static response produced by the field-induced shift of the equilibrium
minimum.  On a stable branch, where
\(\partial_q^2{\cal V}_X>0\), this second contribution is nonnegative.  It
vanishes in the isotropic phase because the potential is even in \(q\).

Equation~\eqref{eq:VI-chi-decomposition} must be evaluated from the same
thermodynamic potential that determines the equilibrium saddle.  In
particular, when the finite-amplitude continuation of
Appendix~\ref{app:irrelevant-term} is used, the regulator and matching
counterterms contribute consistently to the field derivatives.

Equivalently, one may minimize the complete potential at each magnetic field
and differentiate the resulting equilibrium free energy.  On a smooth stable
branch, this procedure gives the same susceptibility as
Eq.~\eqref{eq:VI-chi-decomposition}.

\subsection{Harmonic interpretation of the magnetic response}
The expressions above describe the complete equilibrium magnetic response.
We now turn to a simpler harmonic representation whose purpose is to
identify the phase and amplitude of the dHvA oscillations and to clarify the
difference between the two thermodynamic constraints.
We first consider a fixed nematic background.  The field dependence of the
equilibrium saddle is already included in the complete thermodynamic
response discussed in the preceding subsection.

In the thermally damped regime, the parabolic magnetization can be written
harmonic by harmonic as
\begin{equation}
M_X^{\rm osc}
=
\sum_{\ell=1}^{\infty}(-1)^\ell
\left[
S_{X,\ell}\sin\Phi_{X,\ell}
+
Q_{X,\ell}\cos\Phi_{X,\ell}
\right].
\label{app:M-harmonic-decomposition}
\end{equation}
The coefficient $S_{X,\ell}$ contains the leading dHvA oscillation.
The smaller coefficient $Q_{X,\ell}$ comes from the field dependence of the
slow prefactor and of the LK damping factor.  This decomposition is useful
for interpreting the waveform, but it is not needed to define the complete
equilibrium magnetization.

At fixed chemical potential,
\begin{align}
\Phi_{\mu,\ell}&=\frac{\theta_\ell}{\Delta},
&
z_\ell&=\frac{\chi_\ell}{\Delta},
\nonumber\\
S_{\mu,\ell}
&=-\frac{\mu}{2\pi^2\ell}R_T(z_\ell),
&
Q_{\mu,\ell}
&=S_{\mu,\ell}\frac{\epsilon_q}{\pi\ell}
\left[
1+\frac{z_\ell}{2}L(z_\ell)
\right],
\label{app:Mmu-harmonic-coefficients}
\end{align}
where
\[
\epsilon_q=\frac{\omega_c\Delta}{\mu}.
\]
The important feature is that the phase itself depends on the nematic
deformation through $1/\Delta(q)$.  A field-dependent nematic saddle can
therefore modulate both the amplitude and the phase of the oscillations.

At fixed density,
\begin{align}
\Phi_{N,\ell}&=\ell\theta_0,
\qquad z=\widetilde\chi,
\nonumber\\
S_{N,\ell}
&=-\frac{\omega_c\Delta}{4\pi^3}
\ell\theta_0 C_\ell(z),
\nonumber\\
Q_{N,\ell}
&=-\frac{\omega_c\Delta}{4\pi^3}
\left[
2C_\ell(z)-zC_\ell'(z)
\right].
\label{app:MN-harmonic-coefficients}
\end{align}
In this case the rapid phase $\ell\theta_0$ is independent of $q$.
The nematic deformation changes the thermal damping and the harmonic
amplitudes, but does not shift the parabolic Onsager phase.  In addition,
the coefficient
\[
C_\ell(z)=\frac{R_T(\ell z)}{\ell^2}+K_\ell(z)
\]
contains the harmonic mixing produced by the fixed-density constraint.
The appropriate isotropic reference is therefore the canonical $q=0$
spectrum, not the grand-canonical LK spectrum.

The magnitude of each magnetization harmonic is
\begin{equation}
{\cal A}_{X,\ell}
=
\sqrt{
S_{X,\ell}^2+Q_{X,\ell}^2
}.
\end{equation}
Since $Q_{X,\ell}$ is subleading in the weak-field regime, neglecting it
gives the leading expressions
Eqs.~\eqref{eq:VI-Mmu-leading-comparison} and
\eqref{eq:VI-MN-leading-comparison} of the main text.

The same harmonic representation also gives a simple interpretation of the
susceptibility at fixed nematic background.  Differentiating one harmonic of
Eq.~\eqref{app:M-harmonic-decomposition} with respect to the magnetic field
gives
\begin{align}
\chi_{X,\ell}^{\rm osc,rig}(q)
=
\frac{(-1)^\ell}{B}\Big\{&
\left[
B\partial_B S_{X,\ell}
+\Phi_{X,\ell}Q_{X,\ell}
\right]\sin\Phi_{X,\ell}
\nonumber\\
&+
\left[
B\partial_B Q_{X,\ell}
-\Phi_{X,\ell}S_{X,\ell}
\right]\cos\Phi_{X,\ell}
\Big\}.
\label{eq:VI-chi-harmonic-general}
\end{align}
Here all field derivatives are taken at fixed $q$.  The dominant contribution
is
\[
-\frac{(-1)^\ell}{B}
\Phi_{X,\ell}S_{X,\ell}\cos\Phi_{X,\ell}.
\]
Thus an additional field derivative approximately shifts the leading dHvA
oscillation by $\pi/2$ and enhances it by the large semiclassical phase
$\Phi_{X,\ell}$.  This is the origin of the stronger harmonic structure of
the susceptibility discussed in Sec.~\ref{sec:dhva-nematic}.

For completeness, the slow field dependence of the coefficients at fixed
chemical potential is
\begin{align}
B\partial_BS_{\mu,\ell}
&=
z_\ell L(z_\ell)S_{\mu,\ell},
\nonumber\\
B\partial_BQ_{\mu,\ell}
&=
z_\ell L(z_\ell)Q_{\mu,\ell}
+
S_{\mu,\ell}\frac{\epsilon_q}{\pi\ell}
\left[
1-\frac{z_\ell^2}{2}L'(z_\ell)
\right],
\label{app:chi-mu-derivatives}
\end{align}
while at fixed density
\begin{align}
B\partial_BS_{N,\ell}
&=
\frac{\omega_c\Delta}{4\pi^3}
\ell\theta_0\,zC_\ell'(z),
\nonumber\\
B\partial_BQ_{N,\ell}
&=
-\frac{\omega_c\Delta}{4\pi^3}
\left[
2C_\ell(z)-2zC_\ell'(z)+z^2C_\ell''(z)
\right].
\label{app:chi-N-derivatives}
\end{align}

The fixed-$q$ formulas in this subsection provide the rigid-background
reference used throughout Sec.~\ref{sec:dhva-nematic}. They isolate the
kinematic effect of a nematic deformation from the additional field
dependence generated by the self-consistent saddle. At fixed chemical
potential, the deformation changes both the oscillation phase and the LK
mass through $1/\Delta(q)$; at fixed density, the phase remains locked while
the LK mass and harmonic amplitudes change. The equilibrium calculations in Appendix~\ref{app:equilibrium-sweeps}
include, in addition, the field dependence of
$q_{{\rm sp},X}(B)$.

\section{Numerical procedures and spectral analysis}
\label{app:numerical-spectral}
This appendix specifies the equilibrium field sweeps, isotropic references,
and finite-window spectral analysis used in Sec.~\ref{sec:dhva-nematic}.
The response is evaluated from the same potentials used in minimization,
with the thermodynamic derivatives derived in Appendix~\ref{app:magnetization}.

\subsection{Construction of the equilibrium field sweeps}
\label{app:equilibrium-sweeps}
We now turn from the analytic response formulas to the numerical evaluation
of the equilibrium curves shown in Sec.~\ref{sec:dhva-nematic}. At fixed
chemical potential, each field sweep minimizes the same unexpanded
finite-temperature potential used for the global phase boundary. At fixed
density, the sweeps minimize the finite-amplitude canonical potential with
the density constraint imposed at each trial amplitude; this calculation is
distinct from the displayed canonical phase boundary, which uses local
coefficients from finite-temperature Landau-level sums supported by bounded
preemption tests, rather than a full global canonical phase diagram. The
magnetization and susceptibility are obtained from field derivatives of the
respective minimized potentials, with the corresponding isotropic system
evaluated at the same external variables and at the same perturbative order.

At fixed chemical potential, the potential contains the parabolic
contribution together with the matched finite-amplitude radial correction of
Appendix~\ref{app:numerical-completion}.  At fixed density, the chemical potential
is obtained from the full density constraint at each field and trial value of
\(q\) before the canonical free energy is minimized.  In both ensembles the
regulator and counterterm contributions are differentiated together with the
rest of the potential.  This guarantees that the magnetic response and the
equilibrium saddle follow the same thermodynamic functional.

The isotropic reference is treated independently rather than obtained by
subtracting a fitted background from the nematic signal.  The pointwise
nematic corrections displayed in Figs.~\ref{fig:VI-spectral-transfer} and~\ref{fig:VI-selfconsistent-susceptibility} are therefore direct
equilibrium-minus-isotropic differences.  They vanish identically in the
isotropic phase.  Near a phase boundary, field derivatives are evaluated
within each stable branch so that a differentiation stencil does not mix the
ordered and isotropic solutions across the transition.

For the percentage panels in Figs.~\ref{fig:VI-spectral-transfer}
and~\ref{fig:VI-selfconsistent-susceptibility}, let $S_X=M_X/M_0$ for the
magnetization and $S_X=\chi_X/(2\pi x_X^2M_0)$ for the scaled
susceptibility.  The isotropic reference entering the denominator is
$S_{X,{\rm ref}}^{\rm iso}=S_X^{\rm iso}-P_{3,X}^{\rm iso}$ for the
magnetization, where $P_{3,X}^{\rm iso}$ is fitted over the full interval,
and $S_{X,{\rm ref}}^{\rm iso}=S_X^{\rm iso}$ for the susceptibility, with
no background subtraction.  On the common inverse-field grid
$x_i=20+8(i-1)/(N_x-1)$, with $N_x=4097$, the root-mean-square (RMS)
normalization is
\begin{equation}
S_{X,{\rm RMS}}^{\rm iso}
 =\left[\frac{1}{N_x}\sum_{i=1}^{N_x}
\bigl(S_{X,{\rm ref}}^{\rm iso}(x_i)\bigr)^2\right]^{1/2}.
\label{app:isotropic-rms}
\end{equation}
The plotted percentage is
\begin{equation}
\delta S_X^{(\%)}(x_i)
=100\,
\frac{S_X^{\rm eq}(x_i)-S_X^{\rm iso}(x_i)}
{S_{X,{\rm RMS}}^{\rm iso}},
\qquad X=\mu,N.
\label{app:percent-rms}
\end{equation}
Equation~\eqref{app:isotropic-rms} is the uniform-grid implementation of
$[(x_{\max}-x_{\min})^{-1}\int_{x_{\min}}^{x_{\max}}dx\,
S^2(x)]^{1/2}$.
The numerator in Eq.~\eqref{app:percent-rms} is nevertheless the direct
equilibrium-minus-isotropic difference formed before background subtraction.
The factor $100$ produces the displayed percentages.  The same inverse-field
interval and prescription
are used for both ensembles, with a separate isotropic RMS for each ensemble
and response.  This normalization measures the nematic change in units of the
typical isotropic amplitude and therefore permits a relative comparison even
when the isotropic amplitudes differ.

\paragraph*{Numerical implementation and convergence.}
The minimization uses a broad amplitude scan followed by one-dimensional
refinement of every candidate minimum, with the isotropic state always kept
as a competing solution.  The equilibrium field sweeps use 4097 uniformly
spaced points in the inverse-field variable.  Repeating the calculation with
2049, 8193, and 16385 points leaves the harmonic ratios quoted below unchanged
at the reported precision.  Magnetization and susceptibility are obtained
from centered high-order finite differences on smooth portions of the
branches; one-sided branch-preserving stencils are used only near the ends of
an ordered interval.  These numerical choices affect the discretization
error but not the thermodynamic definition of the response.

\subsection{Finite-window spectra and harmonic weights}
\label{app:spectral-weights}
The Fourier analysis of a field sweep requires some care because the signal
is known only over a finite interval of inverse magnetic field.  This is
different from the ideal Fourier transform of an infinitely long periodic
signal.

We use
\begin{equation}
E_{0,\mu}=\mu,
\qquad
E_{0,N}=\mu_0,
\qquad
x_X=\frac{E_{0,X}}{\omega_c},
\label{eq:VI-spectral-variables}
\end{equation}
so that \(x_X\propto 1/B\).  The isotropic fundamental frequency is
\(\nu_{B,0}^{(X)}=mE_{0,X}\), and Fourier frequencies are expressed in
units of this reference frequency,
\begin{equation}
\widehat\nu_B
=
\frac{\nu_B}{\nu_{B,0}^{(X)}}.
\label{eq:VI-normalized-frequency}
\end{equation}

The signal is sampled over the finite interval
\[
x_{\min}\leq x_X\leq x_{\max}.
\]
Its length
$
L=x_{\max}-x_{\min}
$
limits the frequency resolution to
order $1/L$.  To suppress spectral leakage from the interval endpoints,
we use the Hann window~\cite{Harris1978},
\begin{equation}
w(x_X)
=
\frac{1}{2}
\left[
1-\cos\left(
2\pi\frac{x_X-x_{\min}}{L}
\right)
\right],
\end{equation}
where $x_{\min}\leq x_X\leq x_{\max}$.
The Hann window smoothly brings the signal to zero at both ends of the field
interval.  This strongly suppresses the artificial side lobes produced by a
sharp cutoff, at the price of a modest additional broadening of the main
Fourier peak.  We apply exactly the same window to the equilibrium and
isotropic signals.

Before performing the Fourier transform, we remove the slowly varying
nonoscillatory background by fitting the magnetization over the full
inverse-field interval with a third-degree polynomial in
$x_X=E_{0,X}/\omega_c\propto 1/B$,
\begin{equation}
P_{3,X}(x_X)
=a_0+a_1x_X+a_2x_X^2+a_3x_X^3,
\end{equation}
and defining the background-subtracted signal as
\begin{equation}
s_X(x_X)=M_X(x_X)-P_{3,X}(x_X).
\end{equation}
This subtraction removes only the slowly varying component used in the
finite-window Fourier analysis; it does not modify the thermodynamic
magnetization.  The windowed transform is then
\begin{align}
\widetilde M_X(\widehat\nu_B)
&=
\int_{x_{\min}}^{x_{\max}}dx_X\,
w(x_X)s_X(x_X)
e^{-2\pi i\widehat\nu_Bx_X},
\end{align}
and its power spectrum is
 \begin{align}
{\cal P}_{M,X}(\widehat\nu_B)
&=
\frac{
|\widetilde M_X(\widehat\nu_B)|^2
}{
{\cal N}_w
},
\label{eq:VI-physical-spectrum}
\end{align}
with the normalization
\begin{equation}
{\cal N}_w
=
\int_{x_{\min}}^{x_{\max}}dx_X\,w^2(x_X).
\end{equation}
The finite-window character of the transform is especially important for the
equilibrium nematic signal.  If \(q_{{\rm sp},X}\) varies with magnetic field,
the oscillation amplitude changes across the window and, at fixed chemical
potential, the phase itself is no longer exactly linear in \(1/B\).
Consequently, spectral weight can broaden or move within a Fourier peak.
The position of the maximum of such a finite-window peak should therefore
not be identified with the local oscillation frequency defined from the
derivative of the phase.

For this reason, we characterize each harmonic by its integrated spectral
weight rather than by the height of a single Fourier bin.  For the
\(\ell\)th harmonic we define
\begin{equation}
W_{\ell,X}
=
\int_{\ell-0.25}^{\ell+0.25}
d\widehat\nu_B\,
{\cal P}_{M,X}(\widehat\nu_B).
\label{eq:VI-integrated-weight}
\end{equation}
The integration interval collects the broadened spectral weight around the
corresponding isotropic harmonic while keeping neighboring harmonics
separate.  We then use the ratios
\[
\frac{W_{\ell,X}}{W_{1,X}}
\]
to compare harmonic content.  The overall normalization of the Hann window
cancels from these ratios.

It is useful to distinguish the physical frequency resolution from the
frequency grid used to draw the spectrum.  A denser Fourier grid gives a
smoother representation of the peak, but it cannot improve the intrinsic
resolution set by the finite field interval \(L\).  The integrated weights
are therefore evaluated directly over their frequency intervals and checked
for numerical convergence independently of the plotting grid.

The susceptibility spectra are treated in exactly the same way.  We first
form the scaled susceptibility displayed in
Fig.~\ref{fig:VI-selfconsistent-susceptibility}, and only then perform the
background subtraction, Hann windowing, and Fourier transform.  The reported
susceptibility harmonic ratios consequently refer to this scaled signal.

As numerical checks, replacing the cubic background by a quintic polynomial
changes the representative equilibrium harmonic ratios by at most
$0.14\%$, and refinement of the frequency integration leaves the quoted
weights unchanged at the displayed precision.  These tests concern only the
spectral analysis; they do not modify the underlying equilibrium
magnetization or susceptibility.

For Fig.~\ref{fig:VI-spectral-transfer}, the equilibrium magnetization ratios
\[
\left(
\frac{W_2}{W_1},
\frac{W_3}{W_1},
\frac{W_4}{W_1}
\right)
\]
are
\[
(1.53\times10^{-4},
 2.43\times10^{-5},
 1.55\times10^{-5})
\]
at fixed chemical potential and
\[
(6.57\times10^{-3},
 1.05\times10^{-4},
 2.59\times10^{-6})
\]
at fixed density.  In the latter case, most of the third- and fourth-harmonic
background is already present in the isotropic canonical spectrum.  The
additional nematic changes in those two ratios are only
\(1.46\times10^{-8}\) and \(5.35\times10^{-10}\), respectively.

\subsection{Temperature evolution of the spectral redistribution}
\label{app:temperature-harmonic-redistribution}
We next use the same finite-window analysis to determine how the change in harmonic content produced by nematic order evolves with temperature.
At every temperature we calculate the equilibrium spectrum and the
corresponding isotropic spectrum using exactly the same field interval,
background subtraction, Hann window, and frequency integration.  The direct
nematic change in the \(\ell\)th harmonic ratio is then
\begin{equation}
\delta\!\left(\frac{W_\ell}{W_1}\right)
=
\left(\frac{W_\ell}{W_1}\right)_{\rm nem}
-
\left(\frac{W_\ell}{W_1}\right)_{\rm iso}.
\label{eq:VII-experimental-harmonic-difference}
\end{equation}
 The isotropic
canonical system already contains harmonic mixing through the density
constraint, so Eq.~\eqref{eq:VII-experimental-harmonic-difference} isolates
the additional redistribution produced by nematic order.

Because the Fourier transform is taken over a finite field interval, a small
change of temperature can slightly shift or broaden spectral weight within
one of the integration windows.  The signed quantity above can therefore
show small fluctuations that reflect the finite-window analysis rather than
a new thermodynamic structure.

\begin{samepage}
To display the magnitude of the temperature-dependent spectral
redistribution, we define the Gaussian-smoothed local RMS magnitude
\begin{equation}
{\cal D}_\ell(T)
=
\left\{
\int dT'\,
G_{\sigma_T}(T-T')
\left[
\delta\!\left(\frac{W_\ell}{W_1}\right)(T')
\right]^2
\right\}^{1/2}.
\label{eq:VII-experimental-thermal-envelope}
\end{equation}
\end{samepage}
The normalized continuous kernel in this notation is
\begin{align}
G_{\sigma_T}(\tau)
&=\frac{1}{\sqrt{2\pi}\sigma_T}
\exp\!\left[-\frac{\tau^2}{2\sigma_T^2}\right],
\nonumber\\
\int_{-\infty}^{\infty}d\tau\,G_{\sigma_T}(\tau)&=1.
\label{app:temperature-gaussian-kernel}
\end{align}
Here $\sigma_T$ is the smoothing width in temperature.  For the plotted
curves, Eq.~\eqref{eq:VII-experimental-thermal-envelope} is implemented on
the 81-point uniform grid
$T_i/E_0=0.009+i(1.375\times10^{-4})$, $i=0,\ldots,80$, with
$\sigma_T/E_0=2.5\times10^{-4}$.  We define
$\Delta T=T_{i+1}-T_i$, so that
$\Delta T/E_0=1.375\times10^{-4}$, and set
$i_j=\operatorname{clip}(i+j,0,80)$.  Writing
$\Delta_\ell(T)=\delta(W_\ell/W_1)(T)$, the discrete convolution is
\begin{align}
g_j&=
\frac{\exp[-(j\Delta T)^2/(2\sigma_T^2)]}
{\displaystyle\sum_{k=-J}^{J}
\exp[-(k\Delta T)^2/(2\sigma_T^2)]},
\qquad -J\leq j\leq J,
\nonumber\\
J&=\left\lceil\frac{4\sigma_T}{\Delta T}\right\rceil=8,
\nonumber\\
{\cal D}_\ell(T_i)
&=\left[
\sum_{j=-J}^{J}g_j\,
\Delta_\ell^2\!\left(T_{i_j}\right)
\right]^{1/2}.
\label{app:temperature-gaussian-discrete}
\end{align}
Thus the truncated discrete kernel is normalized once before convolution,
and indices outside the sampled interval are replaced by the nearest endpoint
(constant edge padding); there is no position-dependent edge
renormalization.  Only the controlled portion of this grid within the
displayed interval $0.009\leq T/E_0\leq0.012$ is shown.
Squaring removes the sign, the Gaussian kernel averages over nearby
temperatures, and the square root restores the scale of the original ratio.
The signed harmonic-ratio difference retains the direction of spectral transfer.
This RMS envelope is a numerical diagnostic of the finite-window Fourier
data, not an additional thermodynamic observable.
We retain only temperatures for which the chosen field window contains a
controlled ordered region.  A curve is terminated when this condition is no
longer satisfied or when the corresponding spectral change falls below the
numerical display threshold.
\begin{figure}[t]
\centering
\includegraphics[width=\columnwidth]
{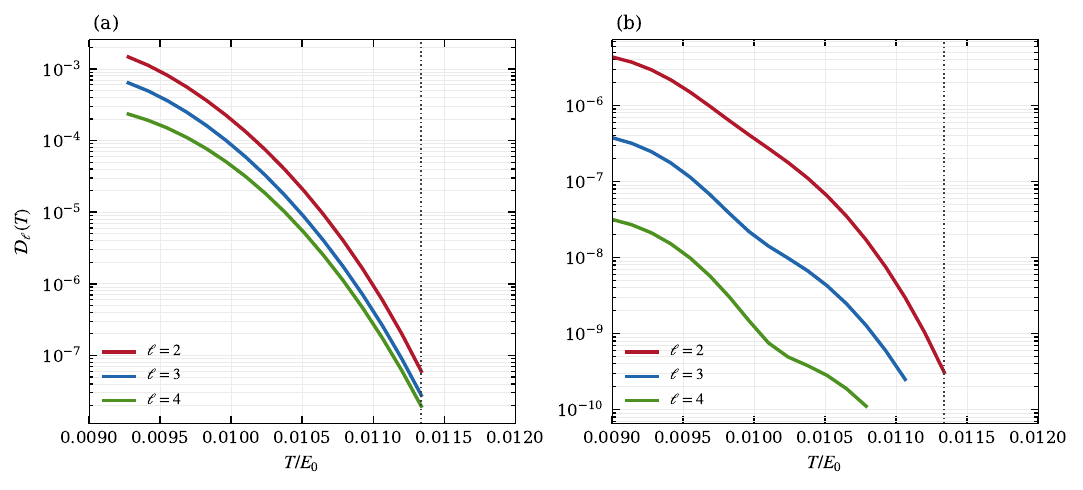}
\caption{Temperature dependence of the nematic redistribution of harmonic
weight.  We show the Gaussian-smoothed RMS magnitude
${\cal D}_\ell(T)$ defined in
Eq.~\eqref{eq:VII-experimental-thermal-envelope} for
$\ell=2,3,4$.  Panels (a) and (b) correspond to fixed chemical potential
and fixed density, respectively.  The field window is
$20\leq E_0/\omega_c\leq28$, with $\gamma=10$ and
$\sigma_T/E_0=2.5\times10^{-4}$.  We use $F_2=-1.0078$ at fixed $\mu$
and $F_2=-1.009$ at fixed density, over
$0.009\leq T/E_0\leq0.012$.  The equilibrium and isotropic spectra are
processed with the same finite-window convention.  Dotted lines indicate
the last sampled temperature for which the field window contains a
controlled ordered region.  Values below $10^{-10}$ are omitted from the
display.}
\label{fig:VII-experimental-harmonic-difference}
\end{figure}

Figure~\ref{fig:VII-experimental-harmonic-difference} shows that the nematic
redistribution of the second through fourth harmonics decreases as the
temperature is raised.  For the displayed parameters the effect is also much
smaller at fixed density, consistent with phase locking and with the larger
canonical harmonic background already present in the isotropic reference.
The termination of a curve is a consequence of the control window or of the
display threshold.

\noindent\textit{AI-assisted code development.---}
Parts of the Python code used for the numerical calculations were developed
with assistance from OpenAI ChatGPT (GPT-5.6 Sol). The AI-assisted code was
reviewed by the authors, executed locally, and validated through convergence
tests, analytic limits, and internal consistency checks. The authors take
full responsibility for the numerical implementation and reported results.


%

\end{document}